\documentclass[a4paper,10pt,twoside,floatfix]{cpc-hepnp}

\usepackage{multicol}
\usepackage{graphicx}
\usepackage{booktabs}
\usepackage{amssymb,bm,mathrsfs,bbm,amscd}
\usepackage[tbtags]{amsmath}
\usepackage{lastpage}
\usepackage{CJK}

\begin{document}
\begin{CJK*}{GB}{gbsn}

\title{Spectral fluctuations in the interacting boson model  \thanks{Supported by National Natural Science
Foundation of China (11875158,11875171)}}

\author{Yu-Qing Wu$^{1}$
\quad Wei Teng$^{1}$\quad Xiao-Jie Hou$^{1}$\quad Gui-Xiu
Na$^{1}$\quad Yu Zhang
$^{1;1)}$\email{dlzhangyu$_-$physics@163.com}\\ \quad Bing-Cheng He
$^{2}$ \quad Yan-An Luo$^{2}$\quad}

\maketitle

\address{
$^1$ Department of Physics, Liaoning Normal University, Dalian, 116029, China\\
$2$ School of Physics, Nankai University, Tianjin 300071, China\\}

\begin{abstract}
The energy dependence of the spectral fluctuations in the
interacting boson model (IBM) and its connections to the mean-field
structures have been analyzed through adopting two statistical
measures, the nearest neighbor level spacing distribution $P(S)$
measuring the chaoticity (regularity) in energy spectra and the
$\Delta_3(L)$ statistics of Dyson and Metha measuring the spectral
rigidity. Specifically, the statistical results as functions of the
energy cutoff have been worked out for different dynamical
situations including the U(5)-SU(3) and SU(3)-O(6) transitions as
well as those near the AW arc of regularity. It is found that most
of the changes in spectral fluctuations are triggered near the
stationary points of the classical potential especially for the
cases in the deformed region of the IBM phase diagram. The results
thus justify the stationary point effects from the point of view of
statistics. In addition, the approximate degeneracies in the $2^+$
spectrum on the AW arc is also revealed from the statistical
calculations.
\end{abstract}

\begin{multicols}{2}
\section{Introduction}
The interacting boson model (IBM)~\cite{IachelloBook87}, in addition
to being the important model for heavy and intermediate-heavy
nuclei, also provides a theoretical laboratory for studying
different many-body problems such as quantum phase
transition~\cite{CJ2009,CJC2010,Iachello2011} and quantum
chaos~\cite{Alhassid1990,Alhassid1991I,Alhassid1991II,Alhassid1992,Alhassid1993,Whelan1993}.
The IBM has three dynamical symmetries (DSs), including U(5), O(6)
and SU(3). In each dynamical symmetry limit, the system is
completely integrable and corresponds to a fully regular spectrum.
Away from the symmetry limits, the systems are expected to generate
chaotic spectra except for the cases of mixing only the U(5) and
O(6) DSs, in which the spectra are found to be still regular due to
the O(5) sub-symmetry. Here, a chaotic spectrum means that the level
distributions approximately obey the statistics predicted by the
gaussian orthogonal ensemble (GOE) of random matrices, while a
regular spectrum may approximately follow the Poisson
statistics~\cite{Whelan1993}. An important signature of chaos is
just the statistical fluctuations of energy spectra, which is the
main topic of this work.

The regular and chaotic behaviors in the context of the IBM have
been extensively
studied~\cite{Alhassid1990,Alhassid1991I,Alhassid1991II,Alhassid1992,Alhassid1993,Whelan1993,Leviatan2012}
using different theoretical tools including the energy spectral
statistics~\cite{Brody1981,Dyson1963}. The chaotic map of the IBM
triangle, which covers various situations in the IBM in terms of two
control parameters, was revealed by Whelan and Alhassid. In
particular, a nearly regular arc connecting the U(5) and SU(3) DSs
has been discovered by them~\cite{Alhassid1992,Whelan1993} thus with
the name Alhassid-Whelan arc of regularity (AW arc) given to it. A
fascinating point of this arc is that its parameter trajectory may
point to a very chaotic situation with the Hamiltonian mixing the
U(5), O(6) and SU(3) dynamical symmetries. However, the statistical
analysis~\cite{Alhassid1992,Whelan1993} indicate that the spectra
associated with the entire AW arc are unexpected regular (especially
in a large $N$ case~\cite{Karampagia2015}) in contrast to its
adjacent parameter area. The AW arc has been extensively studied
from different
angles~\cite{Cejnar1998I,Cejnar1998II,Jolie2004,Amon2007,Macek2007,Macek2009,Bonatsos2010,Bonatsos2011}.
Its regularity nature has been explained in different ways including
the SU(3) quasidynamical symmetry~\cite{Bonatsos2010,Bonatsos2011}
but remains to be clarified. In addition, the empirical signatures
of this arc were also identified~\cite{Jolie2004}.

Most of the statistical calculations in the IBM were performed on
all the excited states for a given spin. However, it is known that
the spectral properties in a system may change as a function of the
excitation energy. In particular, the excited state quantum phase
transitions
phenomena~\cite{Caprio2008,Cejnar2021,Stransky2014,Stransky2015,Macek2019,Zhang2016}
were found to widely occur in the IBM and suggested to be connected
to the stationary points of the potential
functions~\cite{Caprio2008,Cejnar2021}. It means that the excited
states in an IBM system can be divided into different ``families"
according to the stationary points~\cite{Zhang2016}. The recent
analysis~\cite{Zhang2021} has revealed that the spectral
fluctuations in the different ``families" are indeed different
especially in the chaotic region of the IBM. Such a finding was
gained from the statistical calculations by predividing the excited
states into different groups based on the mean-field
analysis~\cite{Zhang2021}. It is remained to answer whether or not
the statistical results themselves can be taken as a criterion of
distinguishing the excited states. In fact, an analysis of the
energy dependence of the spectral fluctuations has already been
given in \cite{Karampagia2015} but the study focuses on the cases in
the vicinity of the AW arc with only the states of zero spin being
discussed. In this work, we hope to give a more general examination
of the energy dependence of spectral fluctuations in the IBM to
reveal the possible connections between spectral fluctuations and
mean-field structures.

The investigations in this work will be placed on the cases
associated with the U(5)-SU(3) and SU(3)-O(6) transitions as well as
the AW arc, therefore covering most of the interesting situations in
the IBM from the point of view of spectral statistics. Particularly,
both zero- and nonzero-spin states will be involved in the
statistical analysis. Two statistical measures, the nearest neighbor
level spacing $P(S)$~\cite{Brody1981} and the $\Delta_3$ statistics
of Dyson and Mehta~\cite{Dyson1963}, will be adopted to analyze
spectral fluctuation because they were successfully applied to
reveal the quantal chaos and regularity in the
IBM~\cite{Alhassid1992,Karampagia2015,Whelan1993,Paar1992,Shu2003}.
To discuss the energy dependence of spectral fluctuation, the number
of statistical samples should be large enough in order to guarantee
the reasonability of the statistical results evolving as a function
of the excitation energy. A large-$N$ calculation of the IBM by the
IBAR code~\cite{Casperson2012} makes it possible to do such a kind
of analysis.

The article is organized as follows. In Sec.~II, the model
Hamiltonian and its mean-field structure are introduced, and two
statistical schemes for spectral fluctuation are described in
detailed. In Sec.~III, the energy dependence of the statistical
results are analyzed. A summary is given in Sec.~IV.

\section{Model and statistical method}

\subsection{The Hamiltonian and mean-field structure}

A Hamiltonian in the IBM framework is constructed from two kinds of
boson operators: $s$-boson with $J^\pi=0^+$ and $d$-boson with
$J^\pi=2^+$~\cite{IachelloBook87}. To discuss different situations
in the IBM, it is convenient to adopt the consistent-$Q$
Hamiltonian~\cite{Warner1983}, which can be written as
\begin{equation} \label{Hamiltonian-IBM}
\hat{H}(\eta,~\chi)=\varepsilon \left[ (1-\eta)\hat{n}_{d} -
\frac{\eta}{4N}\hat{Q}^{\chi} \cdot \hat{Q}^{\chi} \right] \, .
\end{equation}
In the Hamiltonian, $\hat{n}_{d}=d^\dag\cdot\tilde{d}$ is the
$d$-boson number operator, $\hat{Q}^{\chi} = (d^{\dag} s + s^{\dag}
\tilde{d})^{(2)} + \chi (d^{\dag} \tilde{d})^{(2)}$ is the
quadrupole operator, $\eta$ and $\chi$ are the control parameters
with $\eta\in[0,1]$ and $\chi\in[-\sqrt{7}/2,0]$, and $\varepsilon$
is a scale factor set as 1 for convenience. The different dynamical
situations in the IBM are then characterized by the different values
of the control parameters $\eta$ and $\chi$. Specifically, the
Hamiltonian is in the U(5) DS when $\eta=0$; it is in the O(6) DS
when $\eta=1$ and $\chi=0$; it is in the SU(3) DS when $\eta=1$ and
$\chi=-\frac{\sqrt{7}}{2}$. Three DSs in the IBM describe three
typical nuclear shapes including the spherical (U(5)), the
axially-deformed (SU(3)) and the $\gamma$-unstable (O(6)).

\begin{center}
\includegraphics[scale=0.28]{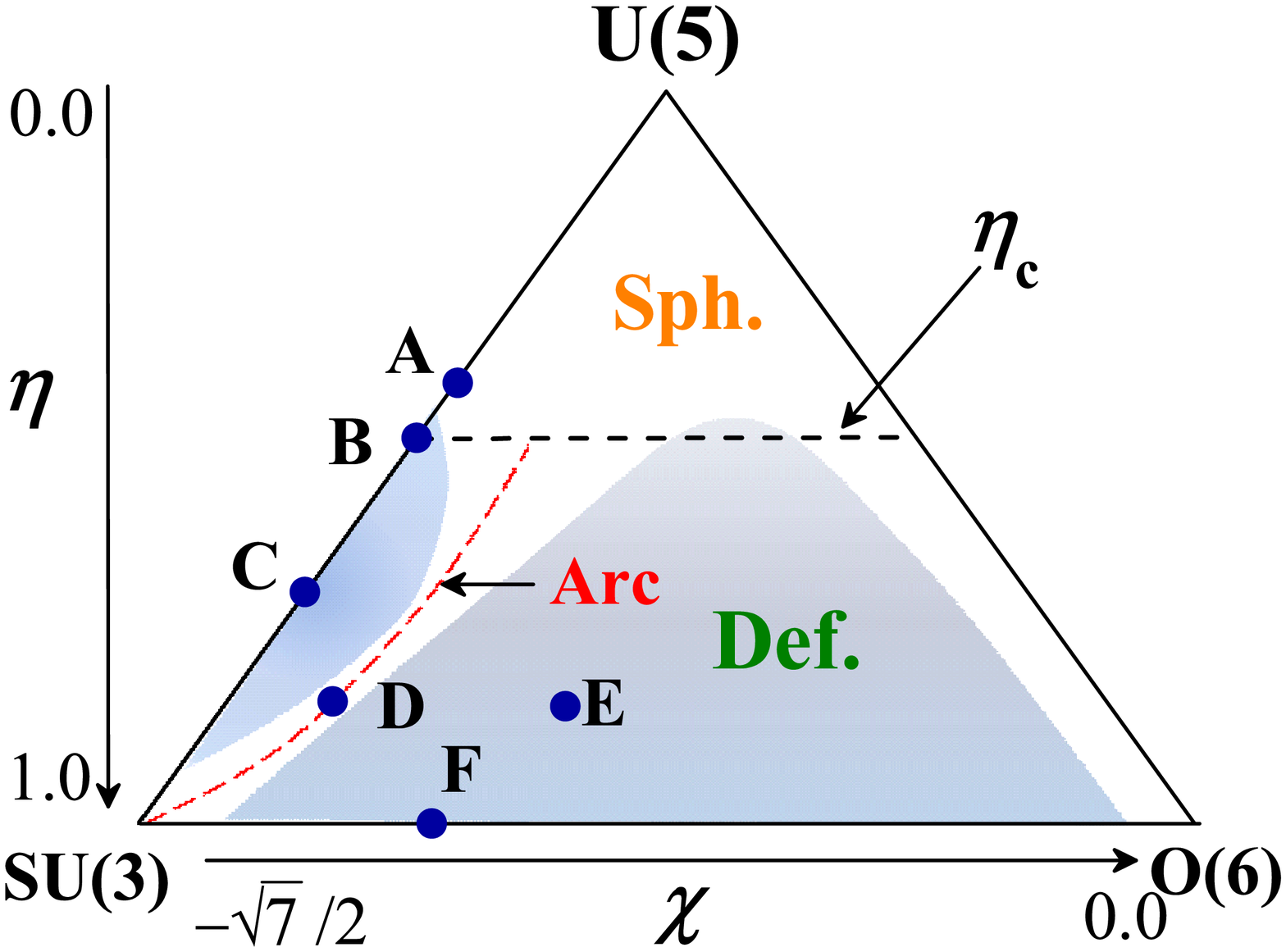}
\figcaption{The triangle phase diagram of the IBM with the dashed
black line denoting the critical points of the 1st-order GSQPTs
described by (\ref{critical}) and six parameter points A, B, C, D, E
and F with
$(\eta,~\chi)=(0.38,~-\sqrt{7}/2),~(0.47,~-\sqrt{7}/2),~(0.75,~-\sqrt{7}/2),~(0.88,~-1.032),
~(0.88,~-0.7)$ and $(1.0,~-0.9)$ being selected to analyze the
spectral fluctuations in different situations. In addition, the
chaotic region of the IBM identified previously~\cite{Whelan1993}
are schematically illustrated by two pieces of shade area with a red
curve used to signify the trajectory of the AW arc of regularity
passing through the chaotic region.}\label{F01}
\end{center}

The mean-field structure of the IBM can be established by using the
coherent state defined with~\cite{IachelloBook87}
\begin{eqnarray}\label{coherent}
|\beta, \gamma, N\rangle&=&\frac{1}{\sqrt{N! (1 + \beta^2)^N}}
[s^\dag + \beta \mathrm{cos} \gamma~ d_0^\dag\, \nonumber \\&+&
\frac{1}{\sqrt{2}} \beta \mathrm{sin} \gamma (d_2^\dag + d_{ -
2}^\dag)]^N |0\rangle\, .
\end{eqnarray}
The scaled potential surface corresponding to the Hamiltonian
(\ref{Hamiltonian-IBM}) in the large-$N$ limit is then given
as~\cite{Iachello2004}
\begin{eqnarray}\label{V}\nonumber
V(\beta, \gamma)&=&\frac{1}{N}
\langle\beta, \gamma, N | \hat{H}(\eta,~\chi) | \beta, \gamma, N\rangle|_{N\rightarrow\infty} \\
\nonumber &=& (1-\eta) \frac{\beta^2}{1+\beta^2} - \frac{\eta}{4(1 +
\beta^2)^2}\\ &\times&[4\beta^2 - 4\sqrt{\frac{2}{7}}\chi\beta^3
\mathrm{cos3} \gamma + \frac{2}{7}\chi^2\beta^4]\, .
\end{eqnarray}
By minimizing this potential with respect to $\beta$ and $\gamma$,
one can prove that the 1st-order ground state quantum phase
transitions (GSQPTs) may occur at the parameter points
with~\cite{Iachello2004}
\begin{equation}\label{critical}\eta_\mathrm{c}=\frac{14}{28+\chi^2},~~~\chi\in[-\frac{\sqrt{7}}{2},~0)\,
\end{equation} and the 2nd-order GSQPT takes place only at the point $\eta_\mathrm{c}=1/2$ with $\chi=0$ on the U(5)-O(6) transitional line.

The two-dimensional phase diagram of the IBM can be mapped into a
triangle. As shown in Fig.~\ref{F01}, each vertex of the triangle
corresponds to a given DS and the whole area is cut into two regions
by the 1st-order transitional line, the spherical and deformed. It
is further shown in Fig.~\ref{F01} that the AW arc denoted by the
dotted curve extends its trajectory from the SU(3) vertex to the
interior of the triangle with the parameter trajectory being
approximately described by
\begin{equation}\label{AW}
\chi=\frac{4+(\sqrt{7}-4)\eta}{6\eta-8}\, ,
\end{equation}
which can be determined either from a minimal fraction of the
chaotic phase-space volume~\cite{Whelan1993} or from the minimal
values of the entropy-ratio product~\cite{Cejnar1998II}. The
fraction of chaotic-space volume~\cite{Whelan1993}, which is defined
as an phase-space integral under given conditions, is a measure of
classical chaos and therefore applied to test the chaocity generated
by the classical limit of the Hamiltonian. Its calculations are
usually done by Monte Carlo methods. The smaller the fraction, the
more regular the system is. In contrast, the entropy-ratio product,
which is defined based on the Shannon information entropy of wave
function~\cite{Cejnar1998II}, may be considered as a quantum measure
of chaos. The wave functions and all the reference basis need to be
known to calculate the entropy-ratio product. Similarly, the smaller
the entropy-ration product, the more regular the system is. In fact,
the two methods agree with each other very well in characterizing
the trajectory of the regular arc in the IBM (see, for example,
Fig.8 in \cite{Cejnar1998II}), which embodies the consistency
between classical and quantum chaos appearing in an IBM system. More
details of the two methods can be read from
\cite{Alhassid1991I,Whelan1993,Cejnar1998I,Cejnar1998II}. On the
other hand, another approximate parametrization of the arc can be
obtained as
\begin{equation}\label{qsu3}
\chi=\frac{2\sqrt{2}-(2\sqrt{2}+\sqrt{7})\eta}{2\eta}\, ,
\end{equation} by connecting the approximate SU(3) symmetry with the AW
arc of regularity~\cite{Bonatsos2011}. One can check that
Eq.~(\ref{qsu3}) describes the parameter trajectory being very close
to the one described by (\ref{AW}) for $\eta\in[0.5,~1.0]$.

As schematically illustrated in Fig.~\ref{F01}, the trajectory of
the AW arc may pass through two chaotic regions in the IBM phase
diagram described by the parameters
$(\eta,~\chi)$~\cite{Whelan1993}, while the two chaotic regions may
cover most of the deformed area of the phase diagram. To analyze the
spectral fluctuations in different situations, six parameter points
are selected to do statistical calculations. These parameter points
correspond to $(\eta,~\chi)=(0.38,~-\frac{\sqrt{7}}{2})$,
$(0.47,~-\frac{\sqrt{7}}{2})$, $(0.75,~-\frac{\sqrt{7}}{2})$,
$(0.88,~-1.032)$, $(0.88,~-0.7)$, and $(1.0,~-0.9)$, which have been
denoted by A, B, C, D, E and F, respectively, as shown in
Fig.~\ref{F01}. The points A, B and C are taken to illustrate three
typical situations in the U(5)-SU(3) GSQPT, namely the spherical
phase, critical point and deformed phase; the points D and E are
chosen to indicate the situations inside the triangle but lying on
and off the AW arc; the point F represents a typical case on the
SU(3)-O(6) line, which in the large-$N$ limit corresponds to a
crossover~\cite{IachelloBook87}.

\begin{center}
\includegraphics[scale=0.28]{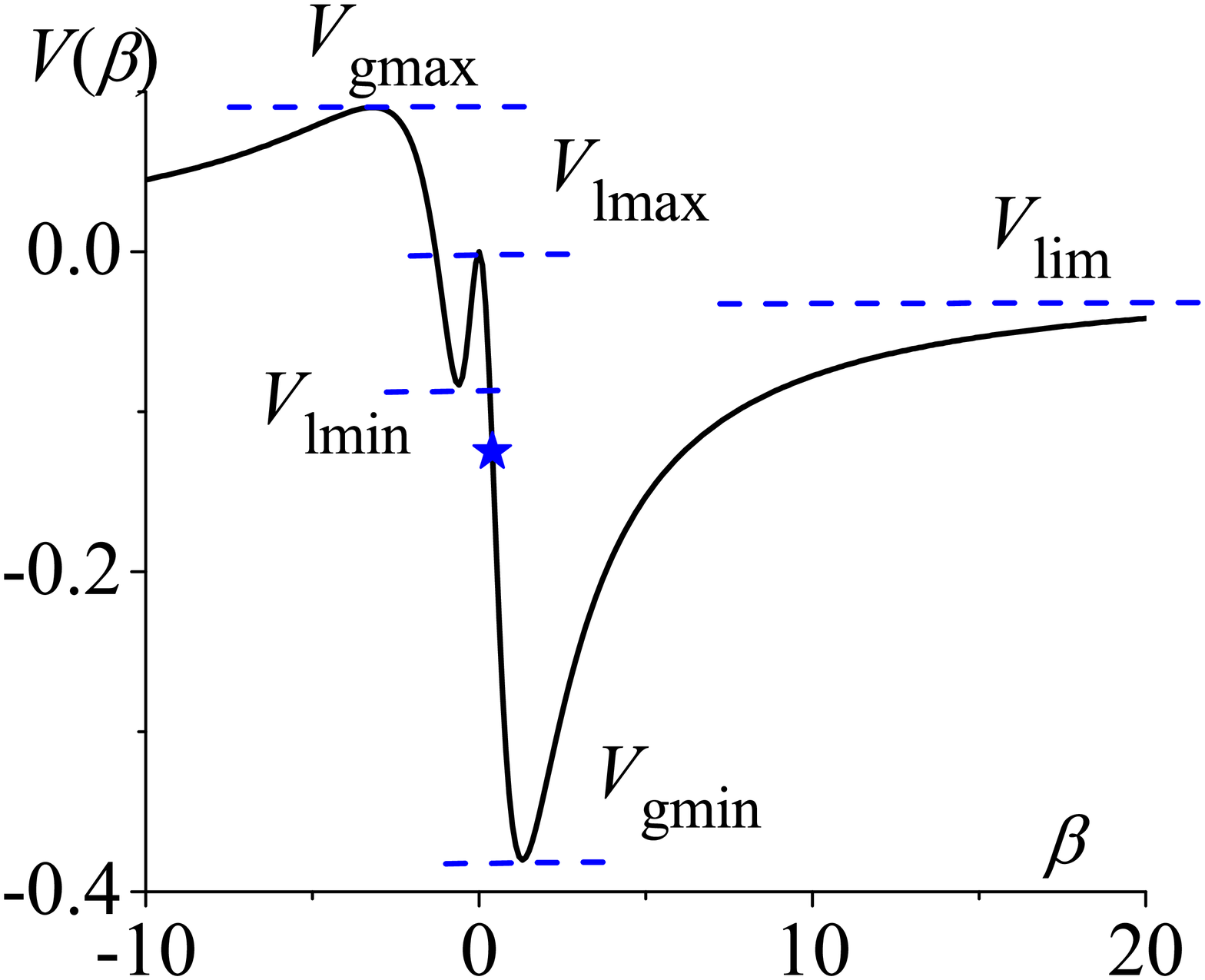}
\figcaption{The potential curve $V(\beta)$ (in any unit) solved from
(\ref{V}) with $(\eta,~\chi)=(0.9,-1.3)$. The stationary points on
the curve are signified by the shot dashed lines.}\label{F02}
\end{center}

One can prove that the extreme values of the potential function
(\ref{V}) will appear at either $\gamma=0^\circ$ or
$\gamma=60^\circ$, while the latter can be equivalently realized by
taking a negative $\beta$ value due to
$V(\beta,\gamma=60^\circ)=V(-\beta,\gamma=0^\circ)$. The stationary
points discussed here are defined as the extreme value points of the
potential function $V(\beta)\equiv V(\beta,\gamma=0^\circ)$, namely
those with $\frac{\partial V(\beta)}{\partial\beta}=0$. As an
example, the potential curve at a typical parameter point inside the
triangle is shown in Fig.~\ref{F02} to illustrate stationary point
structure. In general, the stationary points of a potential curve
include the global minimum, the global maximum, the local minimum,
the local maximum and the $|\beta|\rightarrow\infty$ limit point,
which are denoted as $V_{\mathrm{gmin}}$, $V_{\mathrm{gmax}}$,
$V_{\mathrm{lmin}}$, $V_{\mathrm{lmax}}$ and $V_{\mathrm{lim}}$,
respectively, as seen in Fig.~\ref{F02}. Among them, the local
minimal point $V_{\mathrm{lmin}}$ may correspond to a saddle point
of $V(\beta,\gamma)$ if observing it from the degree of freedom of
$\beta$ and $\gamma$. At mean-field level, $V_{\mathrm{gmin}}$ and
$V_{\mathrm{gmax}}$ just correspond to the ground-state energy and
highest excited energy, respectively. Therefore, more emphasis
should be placed on $V_{\mathrm{lmin}}$, $V_{\mathrm{lmin}}$ and
$V_{\mathrm{lmin}}$. In terms of excitation energy, the related
``critical'' energies are further given as
\begin{eqnarray}
&&E_{\mathrm{lmin}}=V_{\mathrm{lmin}}-V_{\mathrm{gmin}},\\
&&E_{\mathrm{lmax}}=V_{\mathrm{lmax}}-V_{\mathrm{gmin}},\\
&&E_{\mathrm{lim}}=V_{\mathrm{lim}}-V_{\mathrm{gmin}}\, .\
\end{eqnarray}
It is easy to know that the energy scale in the ``critical''
energies is as same as the excitation energy per boson $E/N$ (in any
unit), which means that the ``critical'' energies defined above can
be directly taken to compar the excitation energy $E/N$ solved from
the same Hamiltonian.

\subsection{Statistical measures}

To measure the spectral fluctuations in the IBM, two statistical
measures will be adopted here, including the nearest neighbor level
spacing distribution $P(S)$~\cite{Brody1981} and the $\Delta_3$
statistics of Dyson and Mehta~\cite{Dyson1963}. The spectral
statistics should be performed to the so-called unfolded spectrum in
order to be consistent with the requirements of the Gaussian
orthogonal ensemble (GOE)~\cite{Alhassid1992}. First we construct
the staircase function of the spectrum $N(E)$ defined as the number
of levels below $E$ with the level energies ${E}$ solved from the
Hamiltonian (\ref{Hamiltonian-IBM}). $N(E)$ is further separated
into average and fluctuating parts
\begin{equation}
N(E)=N_{\mathrm{av}}(E)+N_{\mathrm{fluct}}(E)\, .\
\end{equation}
The average part can be expanded as a polynomial of sixth order in
$E$~\cite{Alhassid1992,Karampagia2015}
\begin{equation}
N_{\mathrm{av}}(E)=a_0+a_1E+a_2E^2+a_3E^3+a_4E^4+a_5E^5+a_6E^6\,
\end{equation}
with the expanding parameters $a_i$ determined from the best fit to
$N(E)$. Then, the unfolded spectrum is obtained via the mapping
$\tilde{E_i} = N_{\mathrm{av}}(E_i)$. With the unfolded spectrum,
the nearest neighbor level spacings are obtained from
\begin{equation}S_i = \tilde{E}_{i+1}- \tilde{E}_i\, ,\end{equation} and the
distribution $P(S)$ is then given as the probability of two
neighboring levels to be separated by a distance $S$. Specifically,
$P(S)$ will be shown as the histogram of the normalized spacing and
the results are further fitted to the Brody
distribution~\cite{Brody1981}
\begin{equation}\label{brody}
P_\omega(S) = \alpha(1 + \omega)S^\omega \mathrm{exp}(-\alpha
S^{1+\omega})\,
\end{equation} with
$\alpha= \Gamma[(2 + \omega)/(1 + \omega)]^{1+\omega} $. The Brody
distribution interpolates between Poisson statistics $(\omega = 0)$
characterizing a fully regular system and the Wigner distribution
$(\omega = 1)$ indicating a completely chaotic
system~\cite{Alhassid1992}. As a result, the intermediate value with
$\omega\in[0,~1]$ provides a quantitative estimation of the quantum
chaos in the spectrum.

Spectral rigidity, $\Delta_3(L)$, is a measure of the deviation of
the staircase function from a straight line~\cite{Dyson1963}. It is
defined by
\begin{equation}
\Delta_3 (a,L) = \frac{1}{L} \mathrm{min}_{A,B}\int_a^{a+L}
[N(\tilde{E})-A\tilde{E}- B]^2d\tilde{E}\, ,\end{equation} where $A$
and $B$ give the best local fit to $N(\tilde{E})$ in the interval
$a\leq\tilde{E}\leq a+L$ with $L$ being the energy length of the
interval. A rigid spectrum is supposed to give a smaller $\Delta_3$
while a soft spectrum gives a larger $\Delta_3$. A smoother
$\Delta_3(L)$ can be obtained by averaging $\Delta_3(a,L)$ over
$n_a$ intervals $(a,~a + L)$,
\begin{equation} \Delta_3(L) =
\frac{1}{n_a}\sum_a\Delta_3(a,L)\, .
\end{equation} The successive intervals are taken to overlap by $L/2$. In the concrete
calculations, a useful formula~\cite{Alhassid1992}
\begin{eqnarray}\nonumber
\Delta_3(a,L)&=&\frac{n^2}{16}-\frac{1}{L^2}\big(\sum_{i=1}^n\tilde{\epsilon}_i\big)^2+\frac{3n}{2L^2}\big(\sum_{i=1}^n\tilde{\epsilon}_i^2\big)\\
&-&\frac{3}{L^4}\big(\sum_{i=1}^n\tilde{\epsilon}_i^2\big)+\frac{1}{L}\big(\sum_{i=1}^n(n-2i+1)\tilde{\epsilon}_i\big)\,
\end{eqnarray}
with $\tilde{\epsilon}_i=\tilde{E}_i-(a+L/2)$ is often adopted. For
the Poisson statistics, it is given by
\begin{equation}
\Delta_3^\mathrm{P}(L)=\frac{L}{15}\, ,
\end{equation}
while for the GOE (chaotic) case, it is approximately given by
\begin{equation}
\Delta_3^{\mathrm{GOE}}(L)=\frac{1}{\pi^2}(\mathrm{log}L-0.0687)\,
\end{equation} for $L\gg1$. The exact form of $\Delta_3$ statistics for the GOE and Poisson limit can be
solved by the integral~\cite{Alhassid1992}
\begin{equation}
\Delta_3(L)=\frac{2}{L^4}\int_0^L(L^3-2L^2r+r^3)\Sigma^2(r)dr\,
\end{equation}
with $\Sigma^2(L)=L$ given for the Poisson statistics and
\begin{eqnarray}\label{sigma}\nonumber
\Sigma^2(L)&=&\frac{2}{\pi^2}\Big[\mathrm{ln}(2\pi
L)+\bar{\gamma}+1+\frac{1}{2}\Big(\mathrm{Si}(\pi L)\Big)^2\\
\nonumber &~&-\frac{\pi}{2}\mathrm{Si}(\pi L)-\mathrm{cos}(2\pi
L)-\mathrm{Ci}(2\pi L)\\
&~&+\pi^2L\Big(1-\frac{2}{\pi}\mathrm{Si}(2\pi L)\Big)\Big]\,
\end{eqnarray}
given for the GOE~\cite{Alhassid1992,Haq1982}. In Eq.~(\ref{sigma}),
$\bar{\gamma}$ is the Euler constant and Si (Ci) is the sine
(cosine) integral. Similar to the Brody distribution, one can fit
the calculated $\Delta_3(L)$ with the
parameterization~\cite{Brody1981,Alhassid1992,Karampagia2015,Honig1989}
\begin{equation}\label{delta}
\Delta_3^{q}(L)=\Delta_3^{\mathrm{Poisson}}[(1-q)L]+\Delta_3^{\mathrm{GOE}}(qL),~~~~q\in[0,~1]\,
\end{equation}
in order to give a quantitative estimation of the possible deviation
of $\Delta_3(L)$ from the regular ($q=0$) or chaotic limit ($q=1$).

\begin{center}
\includegraphics[scale=0.165]{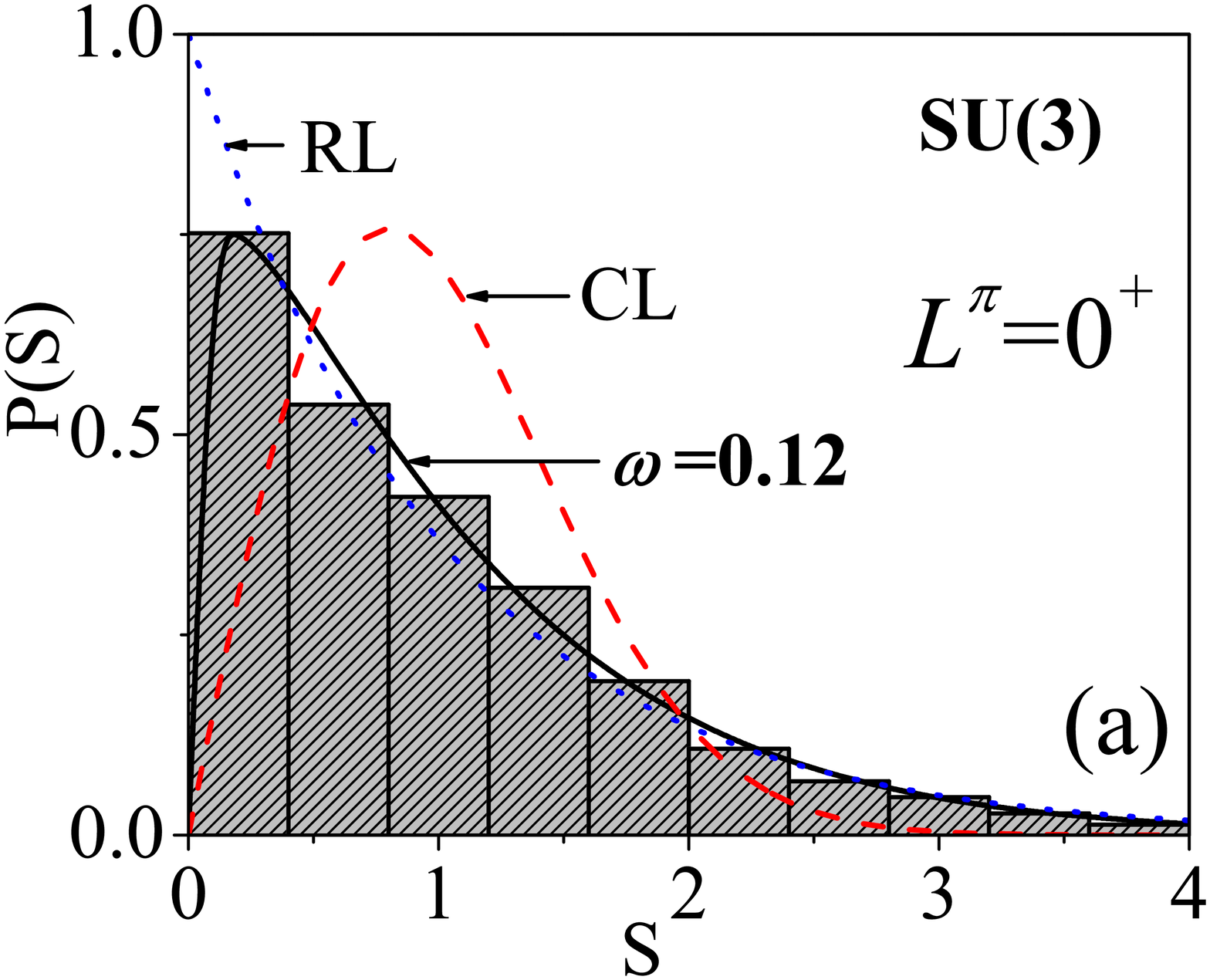}
\includegraphics[scale=0.165]{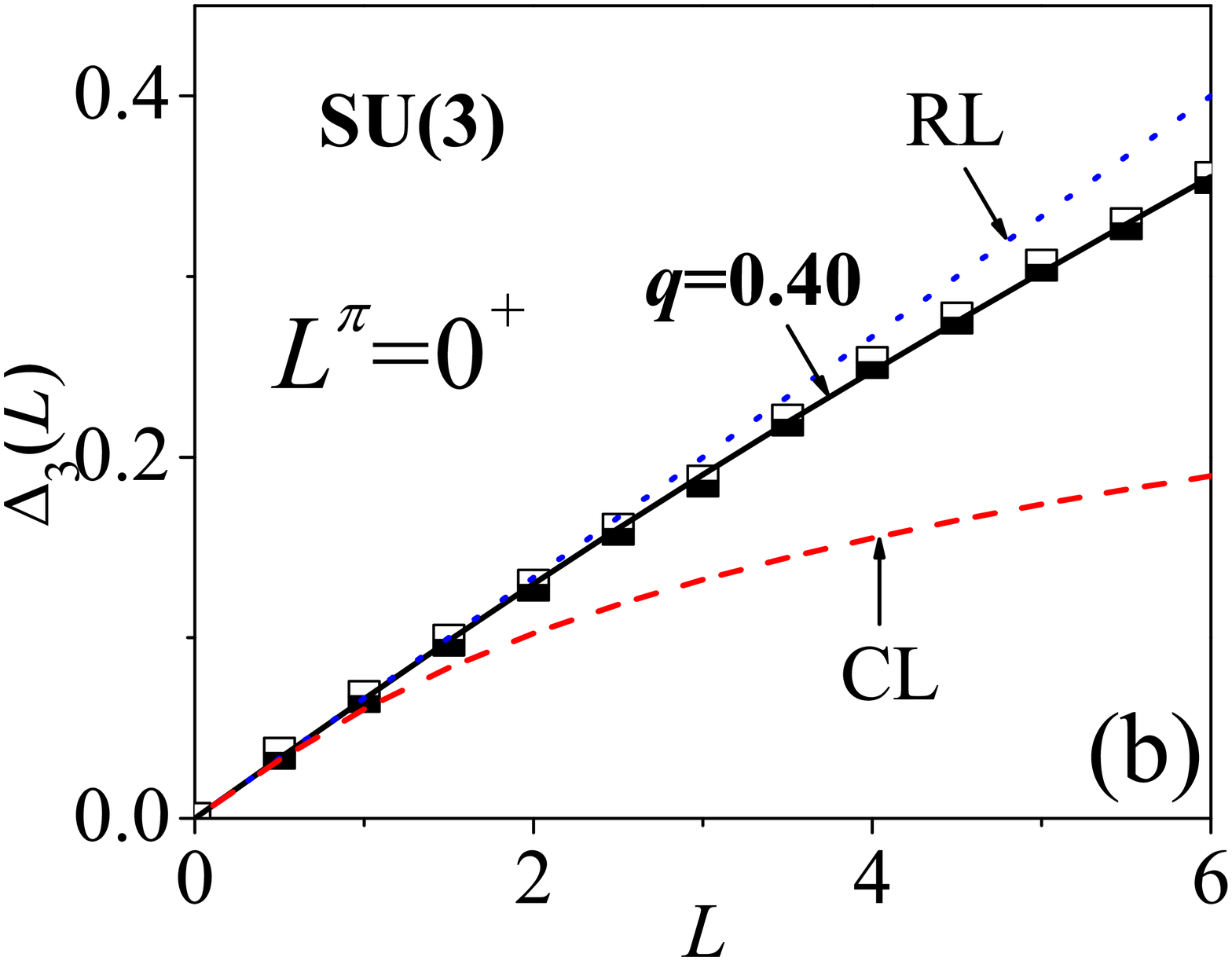}
\includegraphics[scale=0.165]{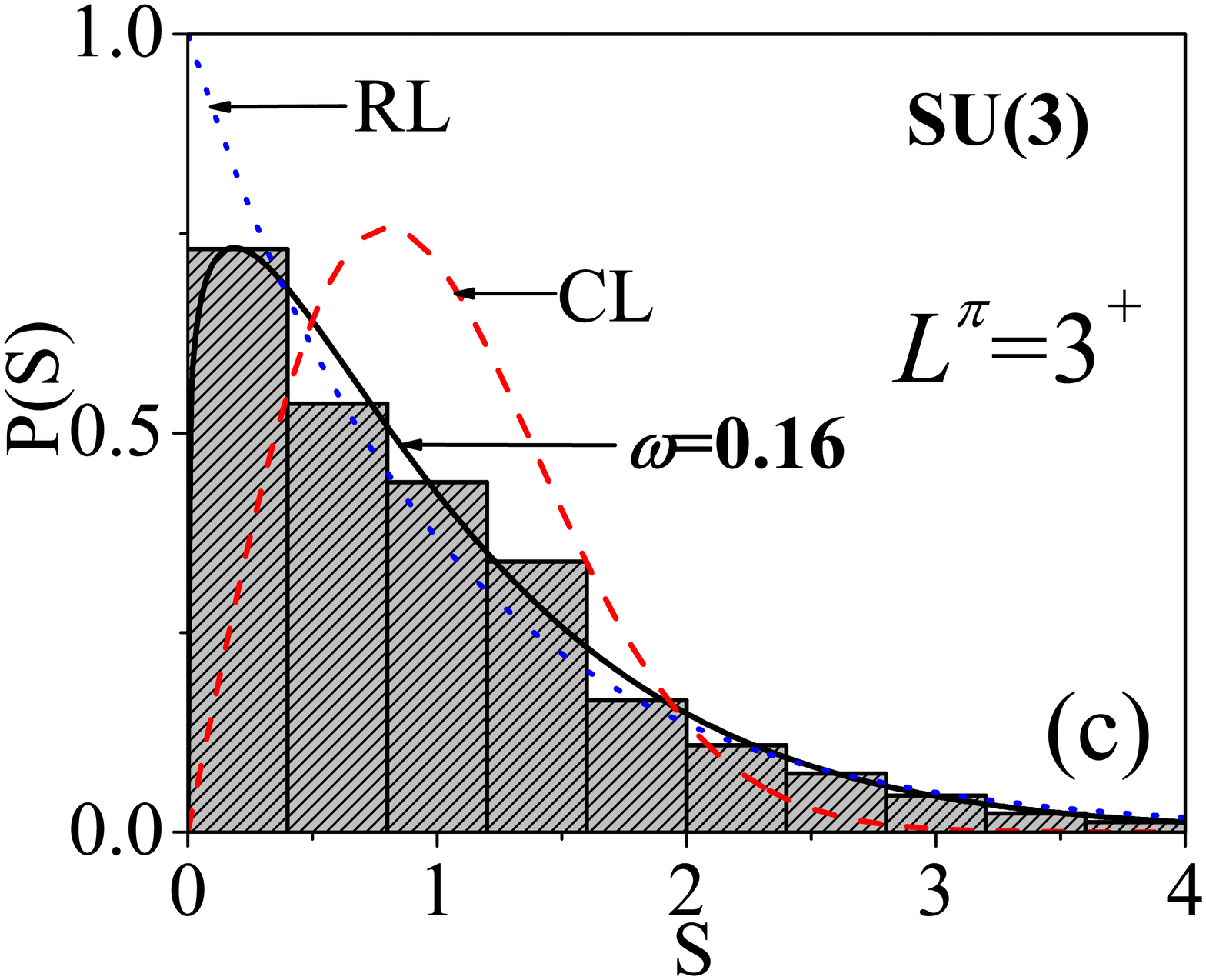}
\includegraphics[scale=0.165]{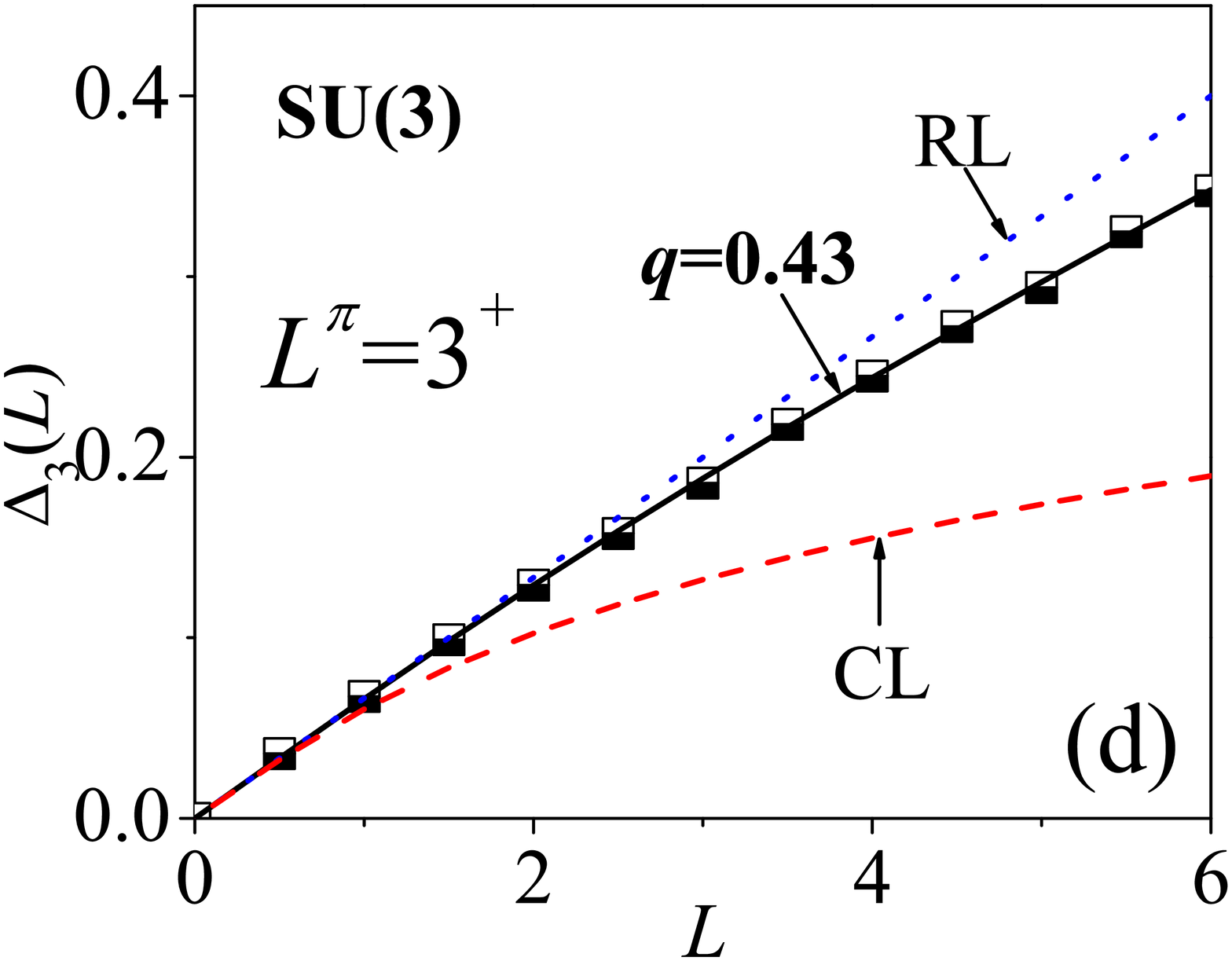}
\includegraphics[scale=0.165]{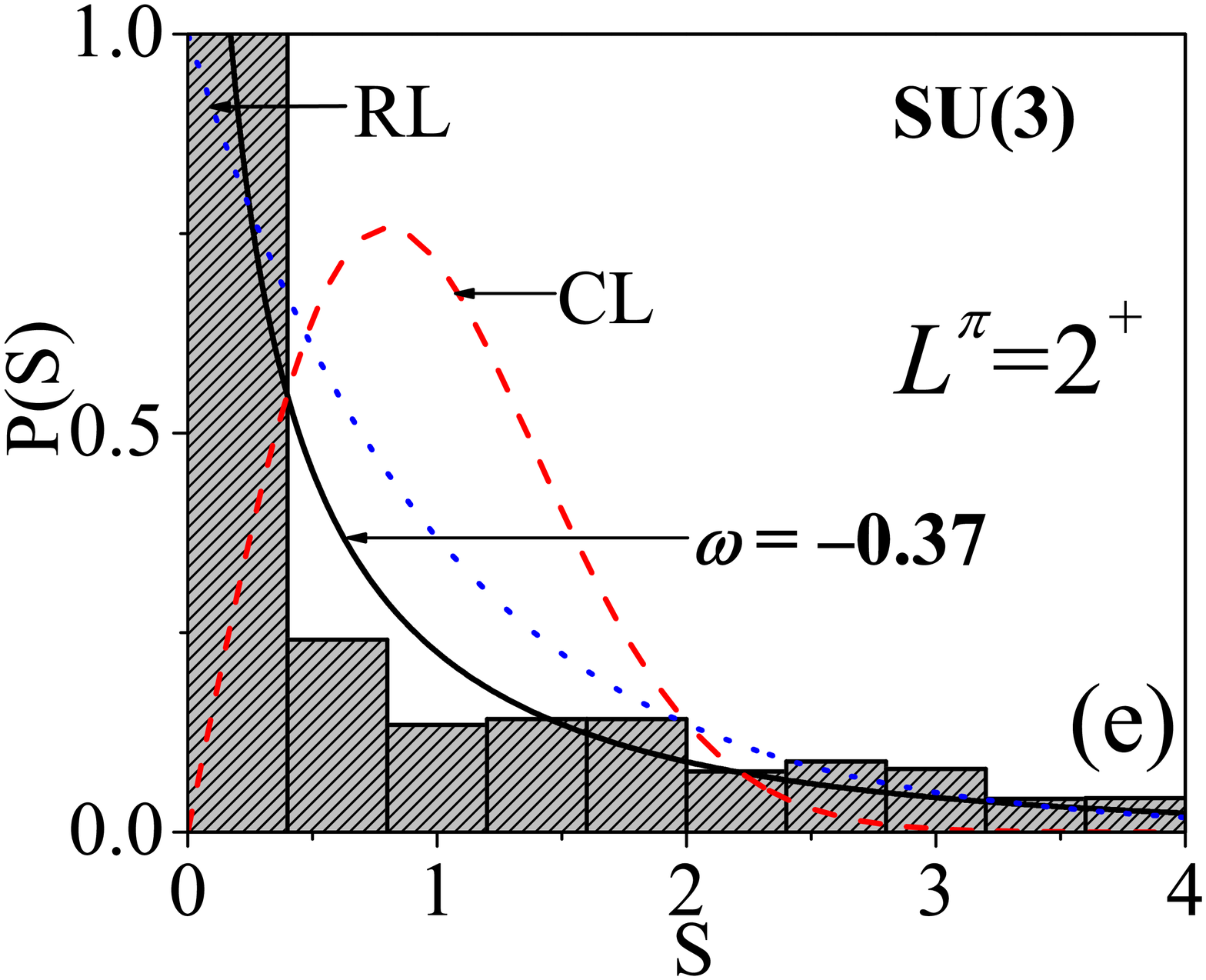}
\includegraphics[scale=0.165]{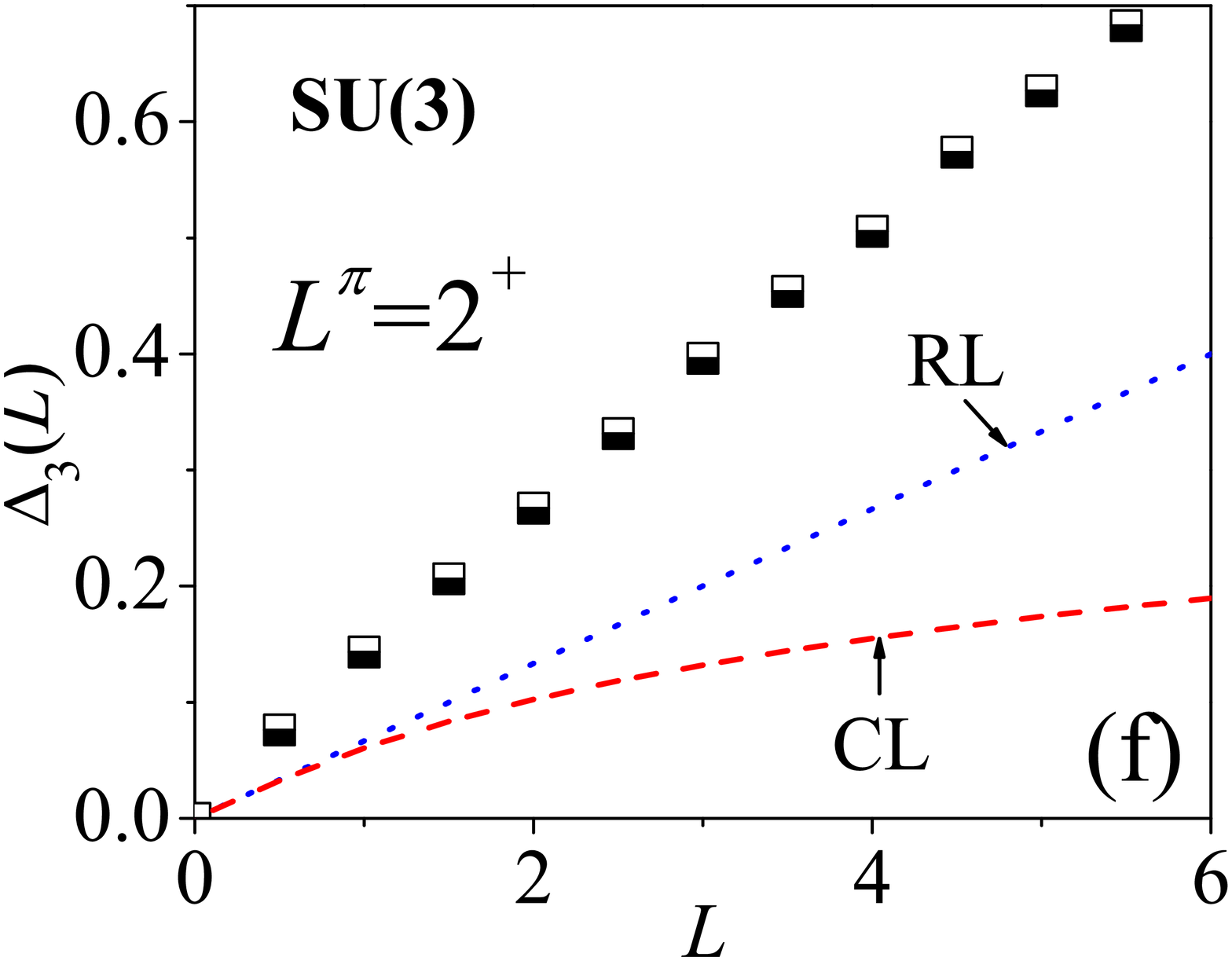}
\figcaption{(a) The $P(S)$ statistics for the states with
$L^\pi=0^+$ in the SU(3) limit with $\omega$ fitted from
Eq.~(\ref{brody}) are shown to compare the regular limit (RL) and
the chaotic limit (CL), which are denoted by the dotted and dashed
lines, respectively; (b) The same as in (a) but for the
$\Delta_3(L)$ statistics with $q$ fitted from Eq.~(\ref{delta}). The
panels (c)-(f) show the same statistics as in (a) or in (b) but for
those with $L^\pi=3+$ and $L^\pi=2^+$. The calculation is performed
for $N=200$.}\label{F2}
\end{center}

To exemplify the $P(S)$ and $\Delta_3$ statistics, the calculated
results for the spectra with $L^\pi=0^+,~2^+,~3^+$ in the SU(3)
limit are shown in Fig.~\ref{F2}. Since the SU(3) limit in the IBM
corresponds to a completely integrable situation, its dynamics are
expected to be close to the regular limit (the Poisson statistics).
As seen in Fig.~\ref{F2}, the results do agree with such a
expectation, giving $\omega\sim0.12$ ($q\sim0.40$) for the $0^+$
spectrum and $\omega\sim0.16$ ($q\sim0.43$) for the $3^+$ spectrum,
which thus provides an example of regular system for reference. As
for the $2^+$ spectrum in the SU(3) limit, it is shown that the
$P(S)$ statistics indicate a negative value of $\omega$ and the
$\Delta_3$ statistics present the values even larger than the
Poisson statistics (the regular limit). This is a result of
degeneracies and related to missing labels according to the analysis
given in \cite{Alhassid1992}. In the SU(3) limit, the missing label
is the quantum number $K$, which usually gives two possible values
$K=0,~2$ for the $2^+$ states in a given SU(3) irrep
$(\lambda,~\mu)$, but this point may not affect the statistics for
the $0^+$ and $3^+$ spectra as only one value is possible ($K=0$ or
$K=2$, respectively) for $L^\pi=0^+$ and
$L^\pi=3^+$~\cite{Alhassid1992}. The similar situations can also
occur in the other symmetry limits or even in the cases being close
to a dynamical symmetry~\cite{Alhassid1991I,Alhassid1992}, where the
degeneracies accordingly become the approximate ones. Since the $q$
values fitted from Eq.~(\ref{delta}) can not be negative and will be
thus set as $q=0$ in the following discussions if the $\Delta_3$
statistics is larger than the Poisson limit like those shown in
Fig.~\ref{F2}(f). Comparatively, $\omega$ seems to be always a
reliable indicator of the spectral fluctuations in different cases.

\section{Evolution of the spectral fluctuations in the IBM}

In this section, we will discuss the spectral fluctuations at the
selected parameter points and analyze their evolutional characters
with the excitation energy. It should be mentioned that the
energy-dependent quantum statistics for the large-$N$ $0^+$ spectra
in the vicinity of the AW arc was previously performed in
\cite{Karampagia2015} with the states being divided into equal
number from low energy to high energy. Here, we focus on more
general situations and the states are separated according to the
energy cutoff instead of equal number to do the statistical
calculations. In particular, both zero and nonzero spins will be
considered in the study.

As shown above, the spectral fluctuations can be measured from the
fitted $\omega$ and $q$ values with $\omega\in[0,~1]$ and
$q\in[0,~1]$. In the following, both $\omega$ and $q$ will be given
as a function of the excitation energy with each value representing
the result being solved from the $P(S)$ or $\Delta_3(L)$
calculations for all the states below a given energy cutoff $E/N$.
Clearly, the higher the excitation energy, the more the states
involved in the statistics. It means that any changes in the
statistical results with the energy cutoff just reflect the
evolution of the spectral chaos with the excitation energy. To
obtain a reasonable statistical result, the level numbers in the
calculations at the lowest energy cutoff are required to be larger
than 200. To complete the statistical analysis, we also perform the
$P(S)$ and $\Delta_3(L)$ statistics from high energy to low energy
with each value of $\omega$ or $q$ being solved from the statistics
for all the states above a given energy cutoff. Similarly, the level
numbers in the calculations at the highest energy cutoff are also
required to be larger than 200. In short, the energy dependence of
the spectral fluctuations will be tested from two directions, from
low energy to high energy and from high energy to low energy. In the
calculations, the total boson number is taken as $N=200$, which
means that there are totally 3434 $0^+$ states, 6767 $2^+$ states
and 3333 $3^+$ states involved in the statistics calculations for
each parameter point. For higher angular momentum, the number of
states in the case of $N=200$ may increase rather rapidly, which
makes the production of the states much more difficult. In
consideration of the computing time, the present discussions are
confined to $L^\pi=0^+,~2^+,~3^+$. In the following, the selected
parameter points are divided into three groups: the points A, B and
C describe the U(5)-SU(3) transition, C and D characterize the
situation inside the triangle but lying on and off the AW arc,
respectively, and the point D represents a typical case in the
SU(3)-O(6) transition.
\begin{center}
\includegraphics[scale=0.25]{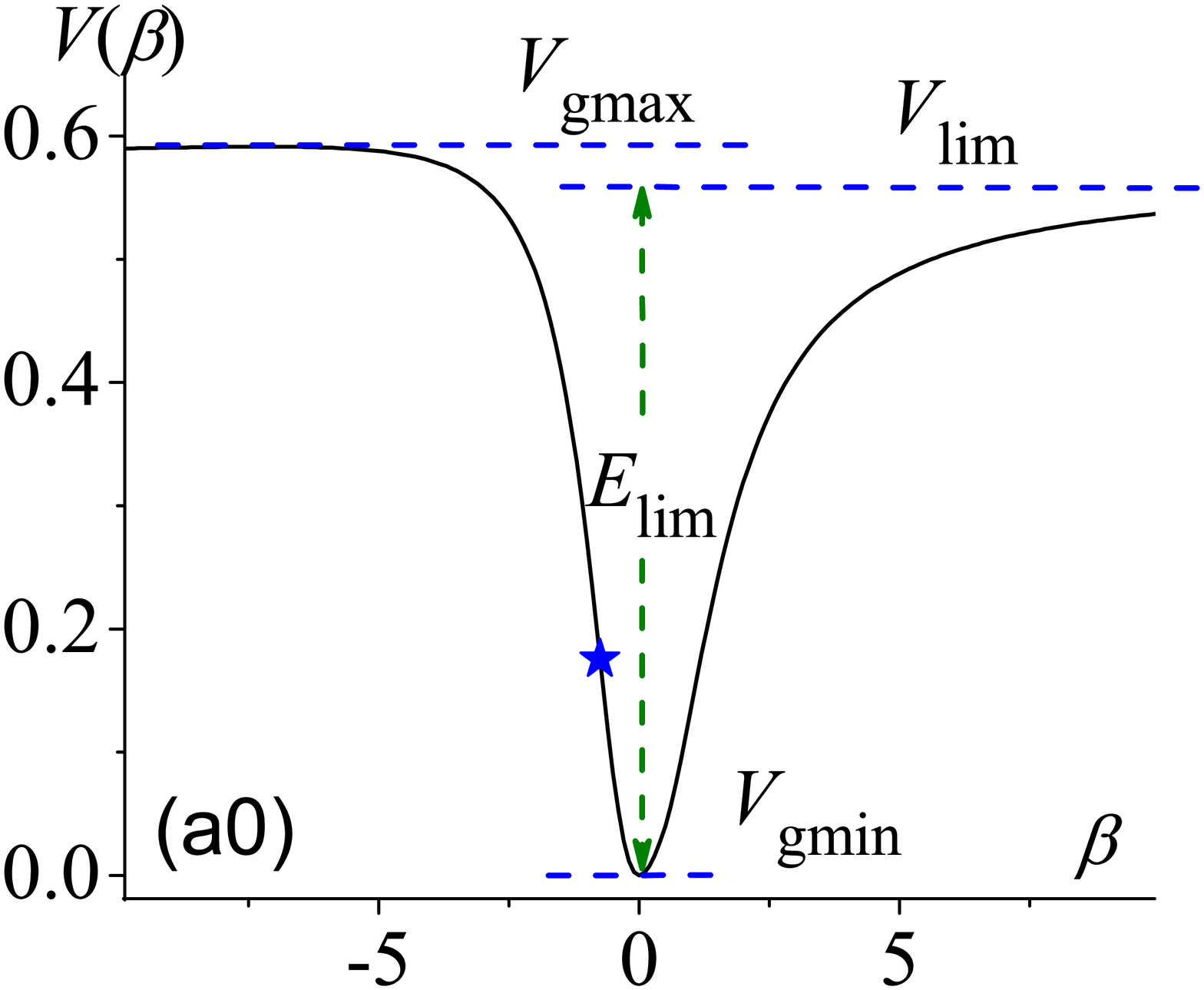}
\includegraphics[scale=0.16]{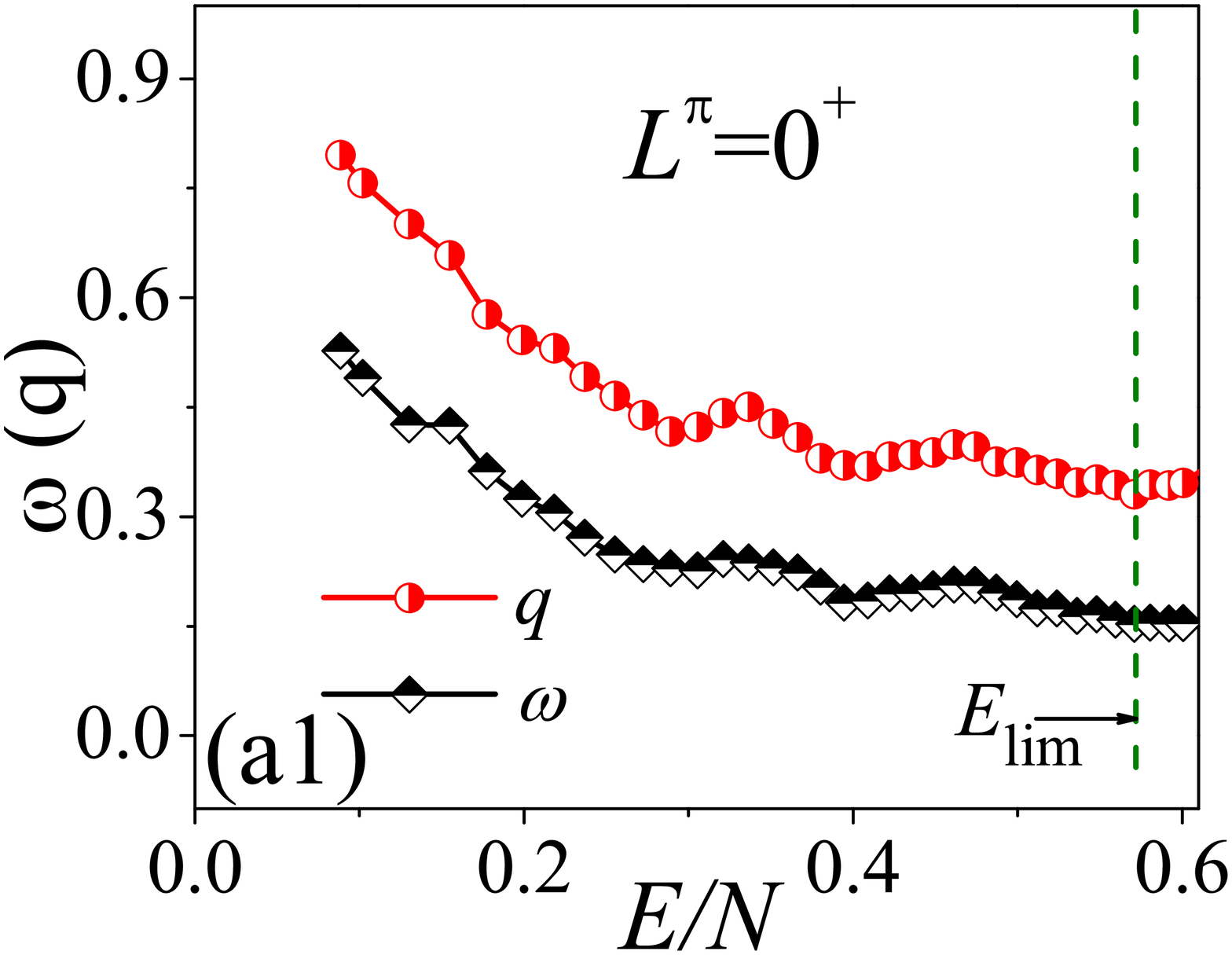}
\includegraphics[scale=0.16]{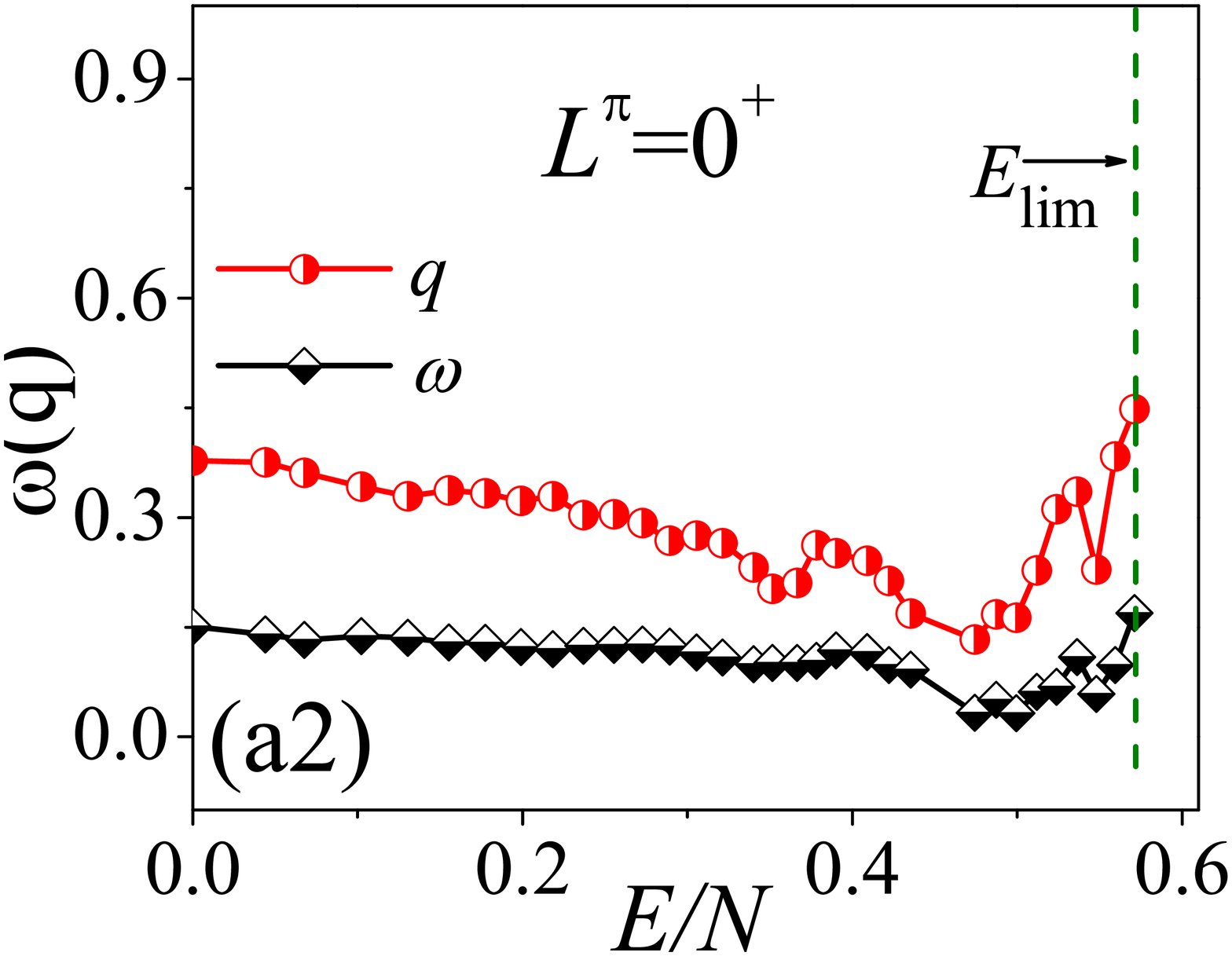}
\includegraphics[scale=0.16]{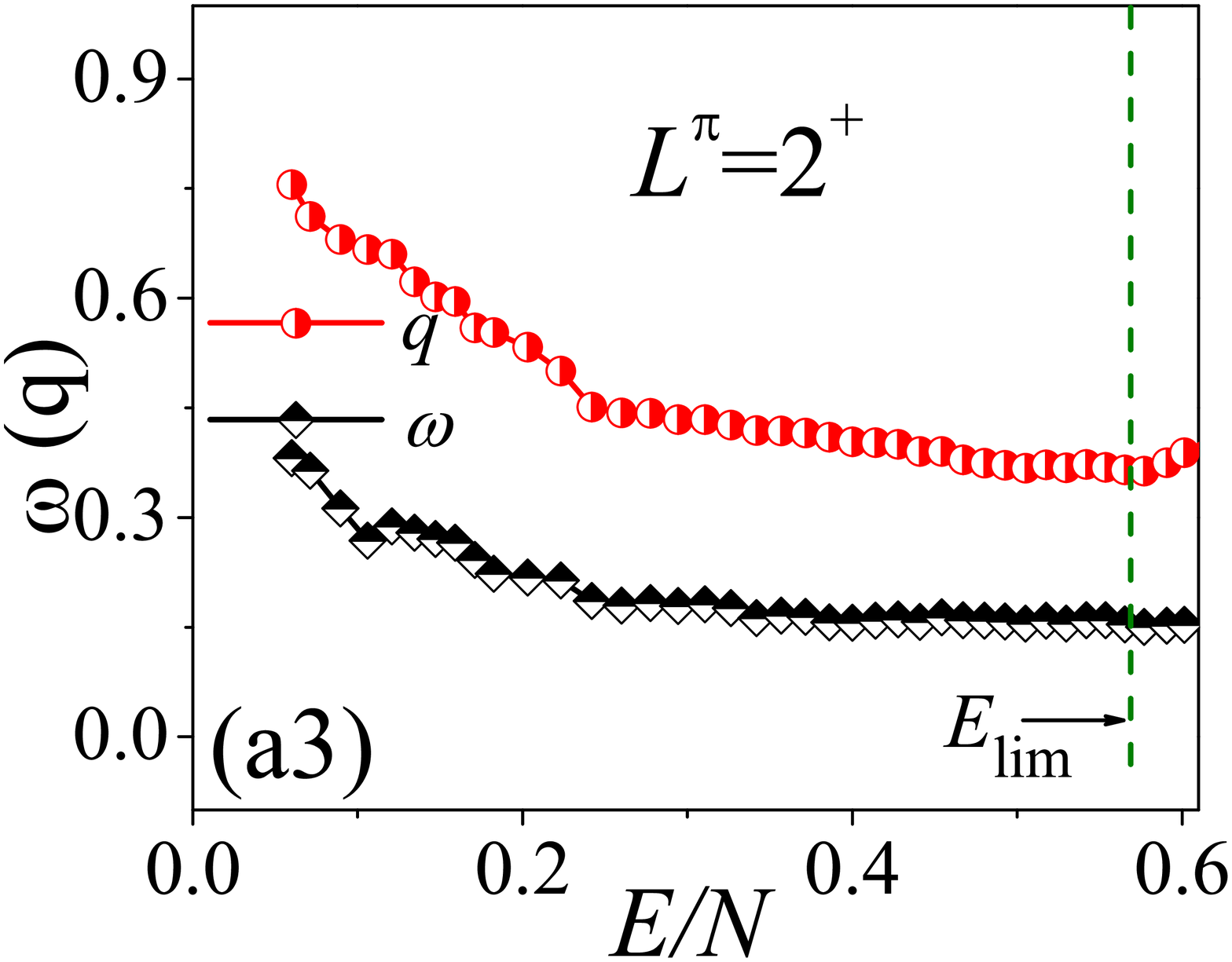}
\includegraphics[scale=0.16]{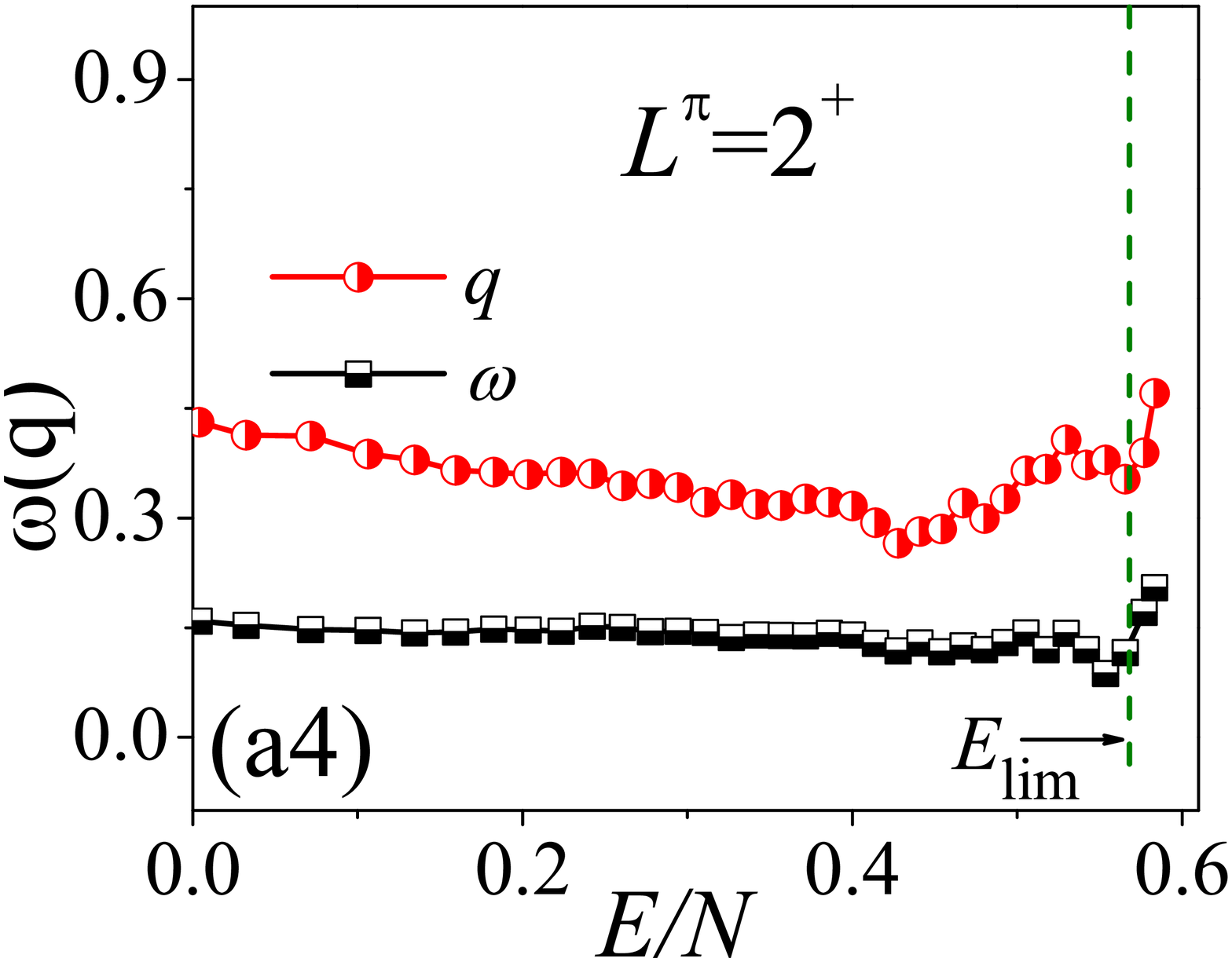}
\includegraphics[scale=0.16]{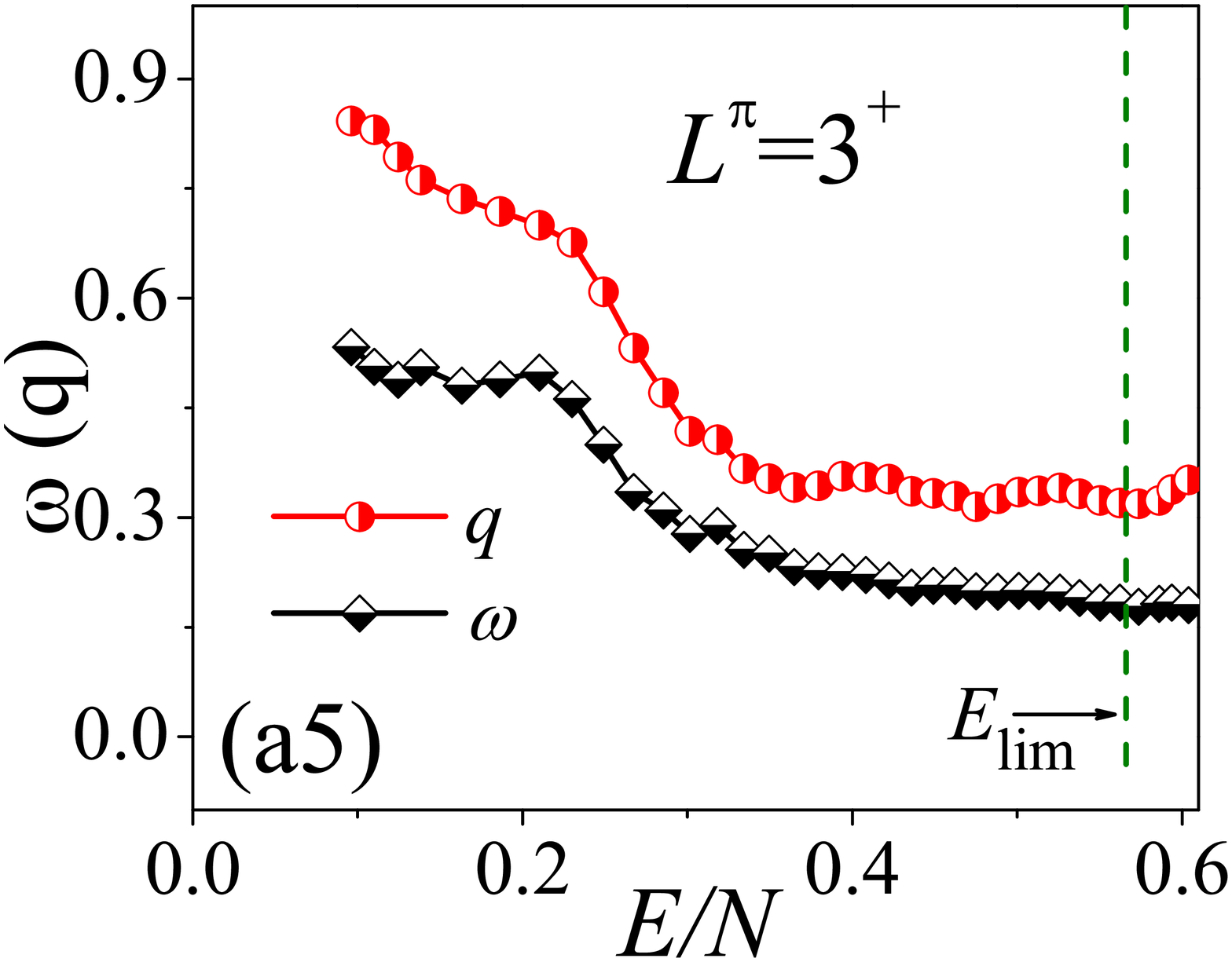}
\includegraphics[scale=0.16]{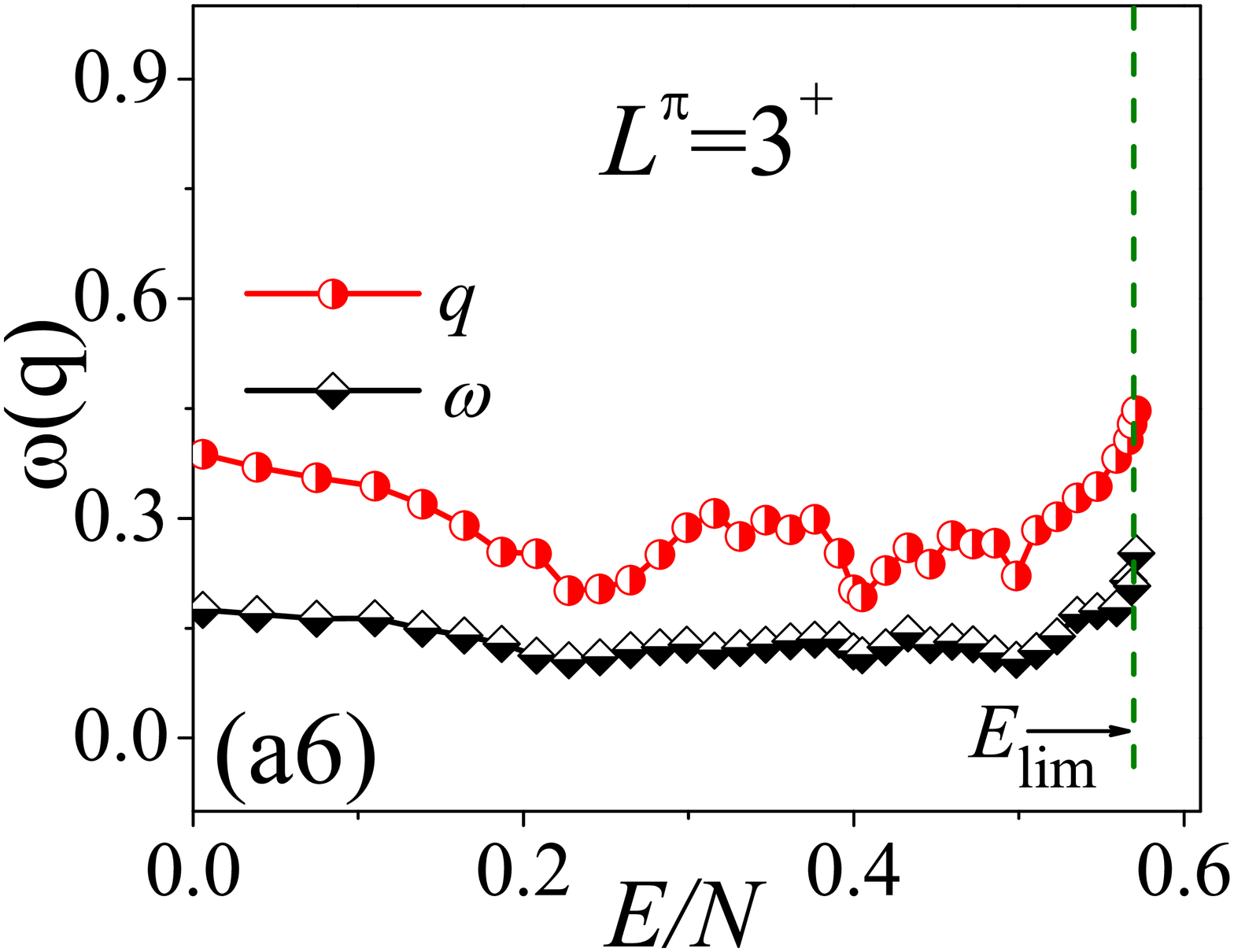}
\figcaption{The parameter point A: (a0) The potential curve
$V(\beta)$ with the blue star signifying
$\frac{\partial^2V(\beta)}{\partial\beta^2}=0$; (a1) The fitted
$\omega$ and $q$ values evolve as functions of the per-boson energy
cutoff $E/N$ (any unit) with each $w(q)$ value being obtained from
the $P(S)$ ($\Delta_3(L)$) statistics on all the $0^+$ levels below
given $E/N$; (a2) The same as in (a1) but with the statistics on the
levels above given $E/N$. The panels (a3) and (a5) show the same
results as in (a1) but for $L^\pi=2^+$ and $L^\pi=3^+$, while (a4)
and (a6) give the same results as in (a2) but for $L^\pi=2^+$ and
$L^\pi=3^+$.}\label{F2a}
\end{center}
\subsection{The U(5)-SU(3) transition}

The U(5)-SU(3) transition at $\eta_c\simeq0.47$ is proven to be a
1st-order GSQPT. The results solved from the parameter point A, B
and C in this transition are shown in Fig.~\ref{F2a}-\ref{F2c},
respectively. As seen in Fig.~\ref{F2a}, the potential curve at the
point A is rather simple with only one stationary point
$V_{\mathrm{lim}}$ lying between the global minimum and maximum. It
is further shown that the entire spectrum for a given spin could be
rather regular if taking all the levels into the statistical
calculation, which gives, for example, $\omega\approx0.15$ and
$q\approx0.38$ for $L^\pi=0^+$. On the other hand, the results
suggest that the spectral fluctuations are not uniform in energy but
the fitted $\omega$ and $q$ values as functions of the excitation
energy are shown to be consistent with each other during the entire
evolutional process. Specifically, $\omega$ and $q$ for a given
$L^\pi$ may slowly decrease as functions of the energy as seen in
the panels (a1), (a3) and (a5). If observing the statistics from
high energy to low energy, one can find from the panels (a2), (a4)
and (a6) that the $\omega$ and $q$ values present the nearly
constant evolutions except for some small fluctuations appearing
near $E_{\mathrm{lim}}$. Anyway, it seems that the evolution of the
spectral fluctuations in the spherical case with the simple
potential configuration is relatively simple.

A similar picture can be also found in the case at the critical
point (the point B) as seen from Fig.~\ref{F2b}. It is shown that
the global configuration of the potential curve at this point is as
simple as the one at the point A except that two degenerate minima
appear on the bottom of the potential indicating the 1st-order
GSQPT. As expected, the statistical results indicate that the entire
spectrum at the critical point becomes less regular compared to the
spherical case. For example, $\omega\approx0.45$ is given in the
former but $\omega\approx0.15$ in the latter if involving all the
levels in the $P(S)$ statistics. Similar to the spherical case, the
spectral fluctuations at the critical point are not uniform in
energy. One can find from (b1), (b3) and (b5) that $\omega$ and $q$
exhibit the consistent evolutions with the excitation energy and may
reach their maximal values around $E/N\sim0.2$, giving
$\omega_{\mathrm{max}}>0.84$ and $q_{\mathrm{max}}>0.9$,
respectively. However, this feature is not easy to be illustrated
from the mean-field structure as no stationary point appear nearby
except the fastest changing point in $V(\beta)$ shown around
$E/N\sim0.15$ as seen from the panel (b0). In addition, the
consistence between the $\omega$ and $q$ evolutions seems to be
broken near $E_{\mathrm{lim}}$ as shown in the panels (b2) and (b6).
This can be roughly explained by that the calculated $\Delta_3(L)$
results in the related cases are already larger than the Poisson
limit and thus set with $q=0$. Usually, such situations only occur
at $\omega\sim0$.

\begin{center}
\includegraphics[scale=0.25]{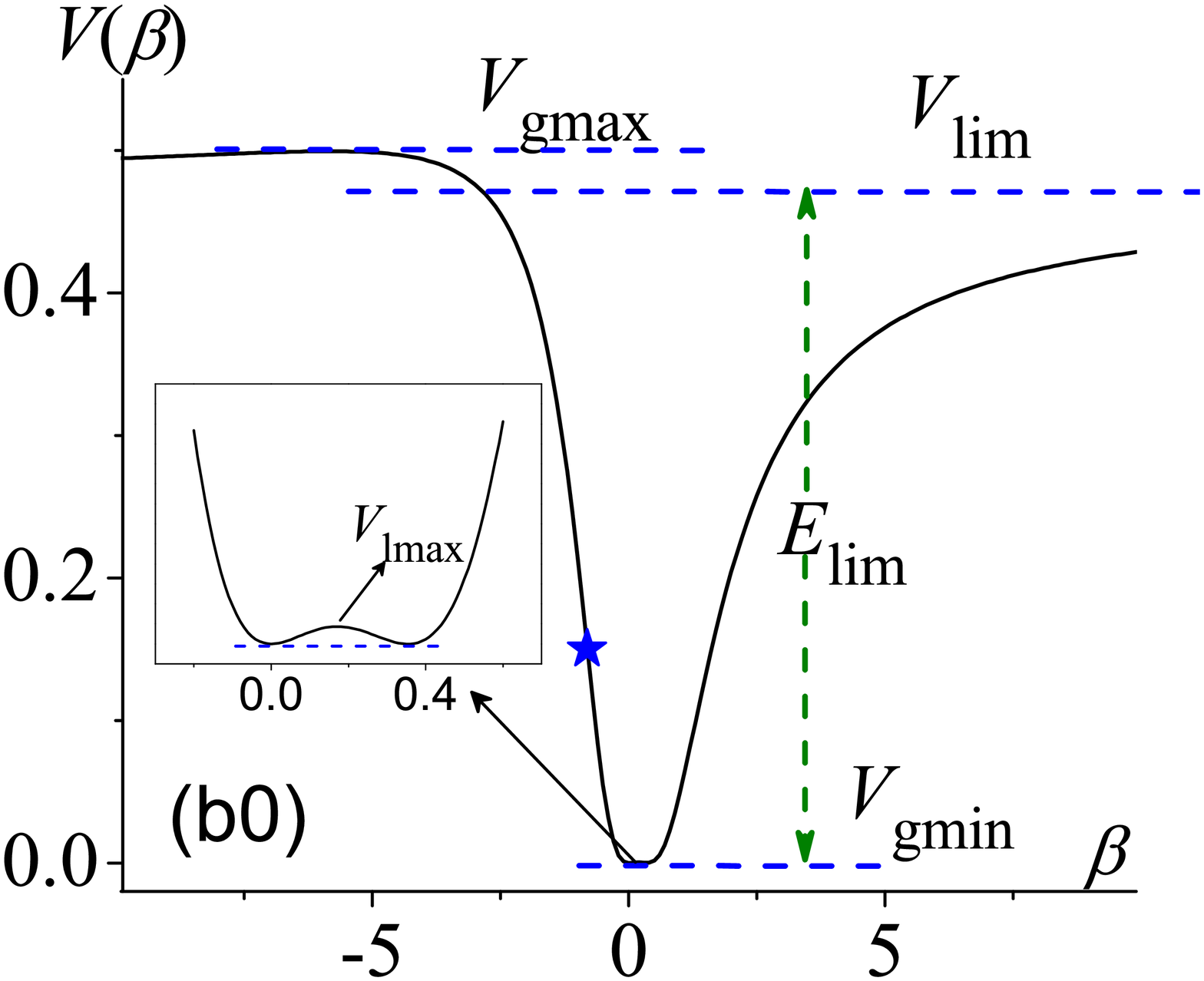}
\includegraphics[scale=0.16]{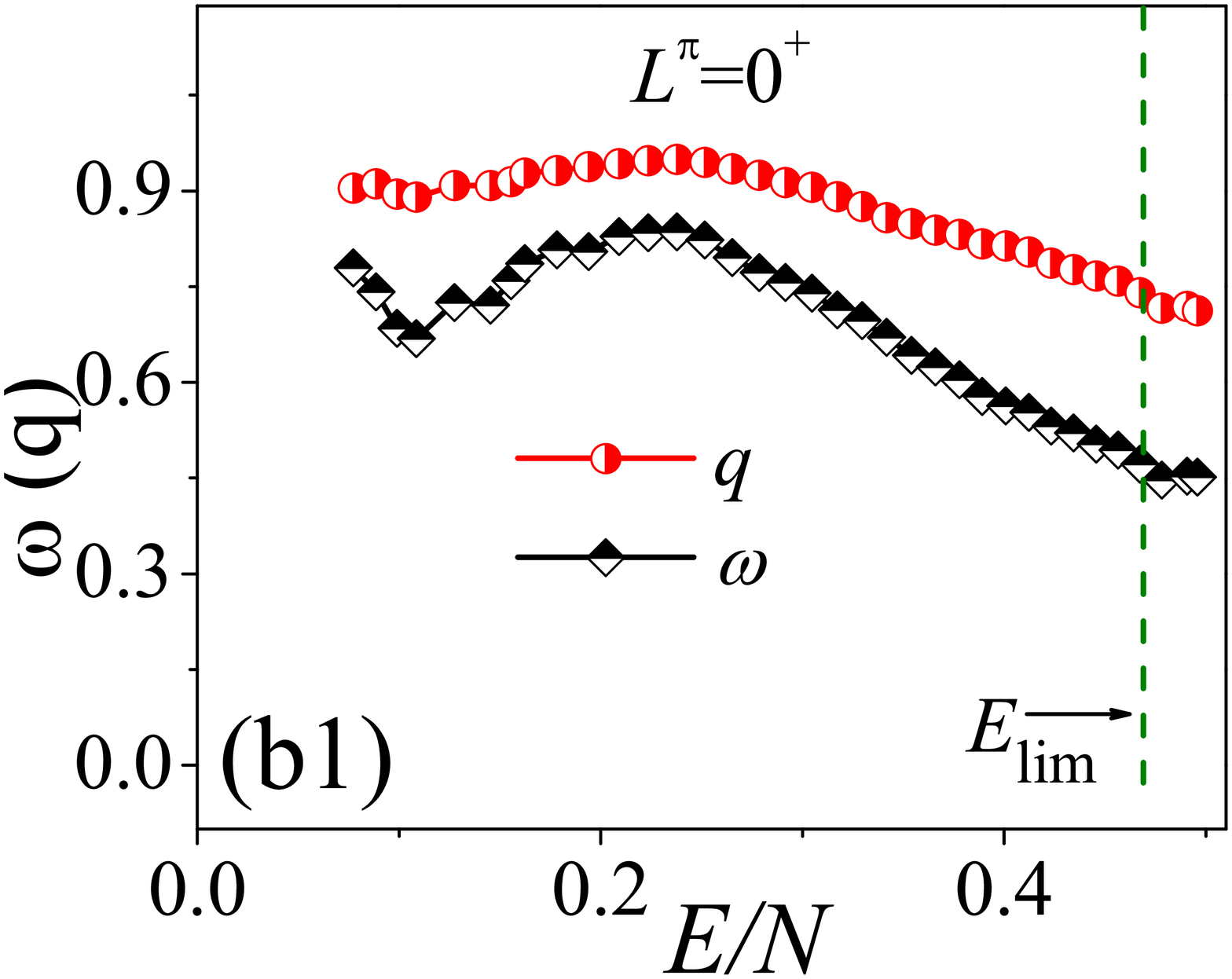}
\includegraphics[scale=0.16]{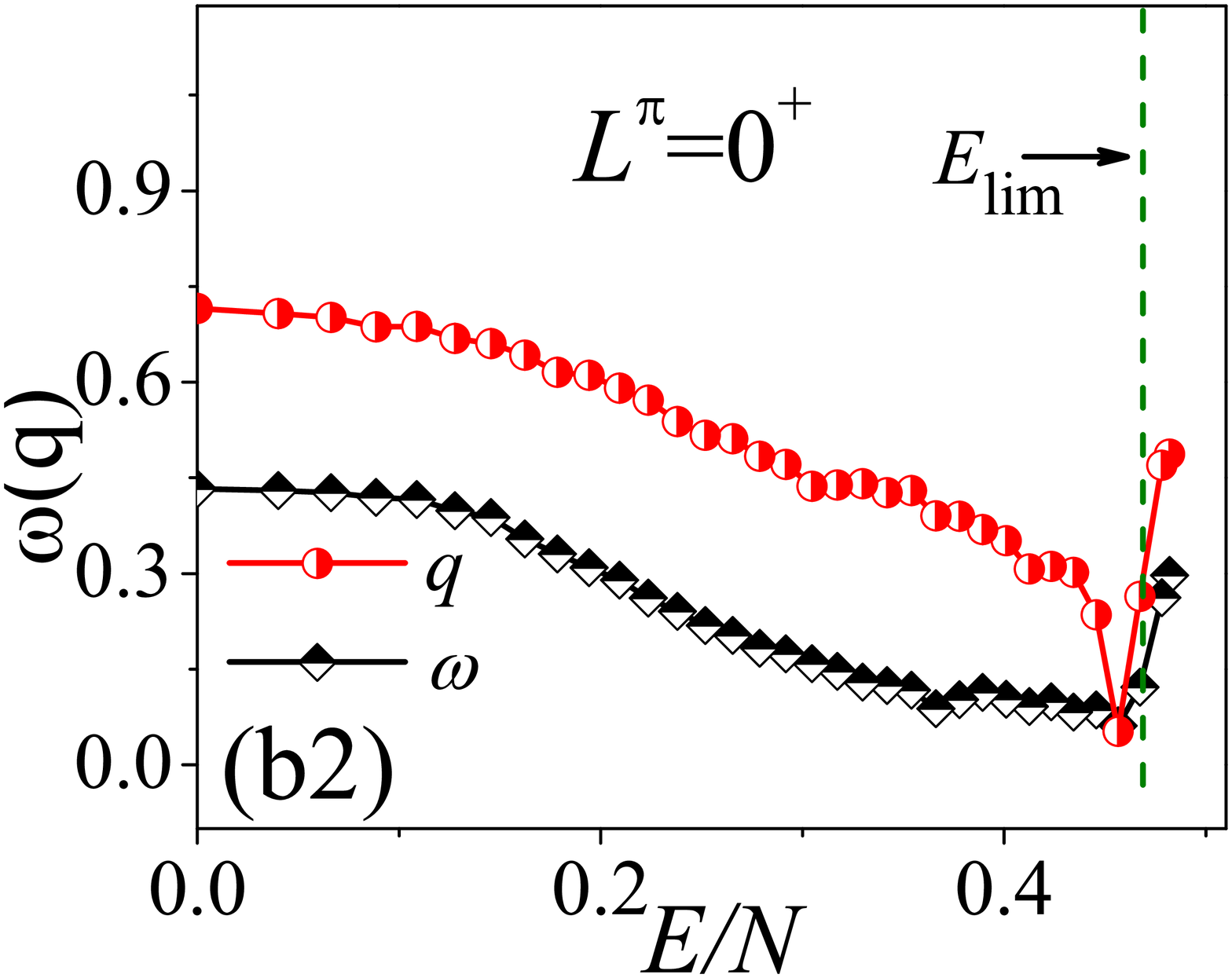}
\includegraphics[scale=0.16]{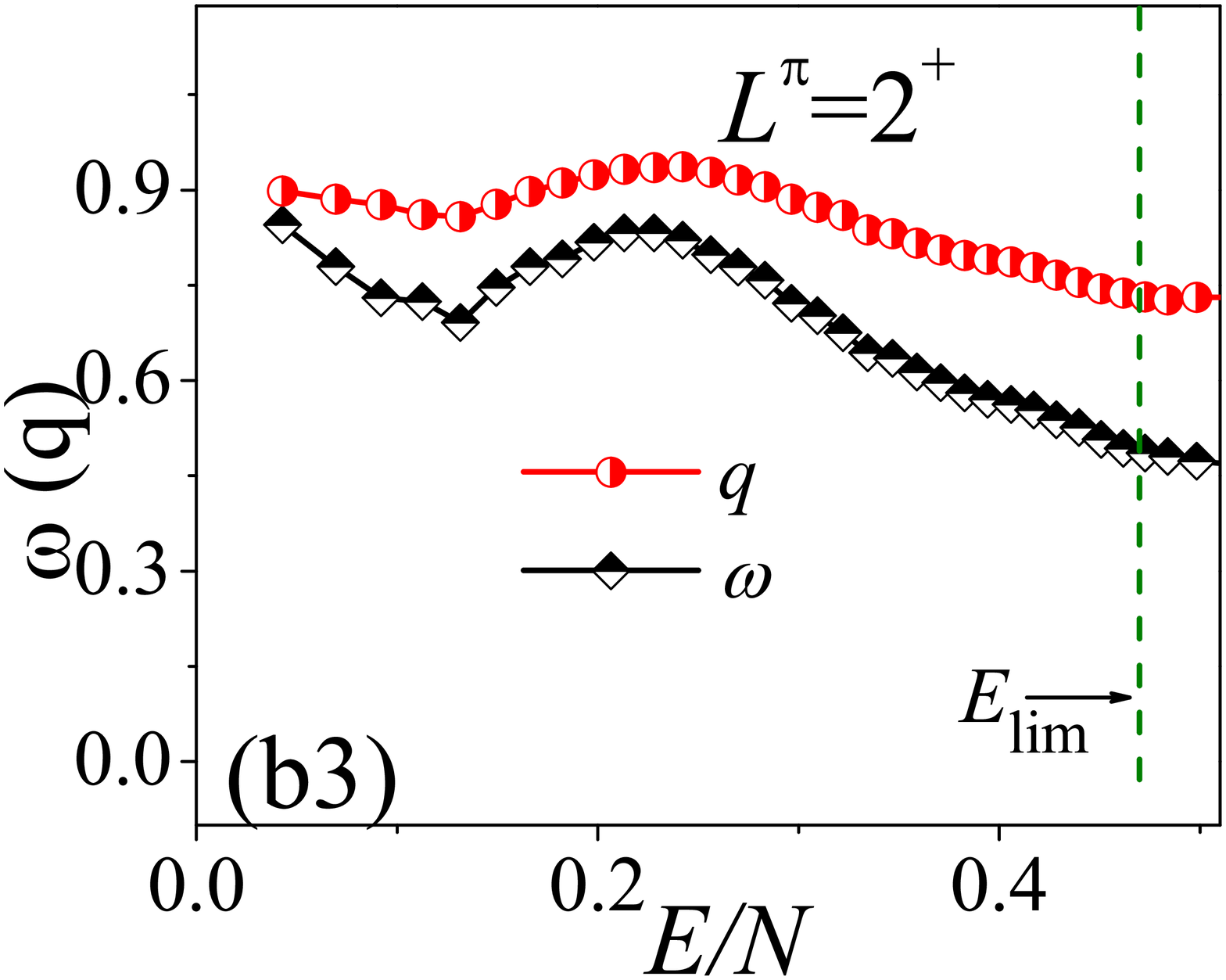}
\includegraphics[scale=0.16]{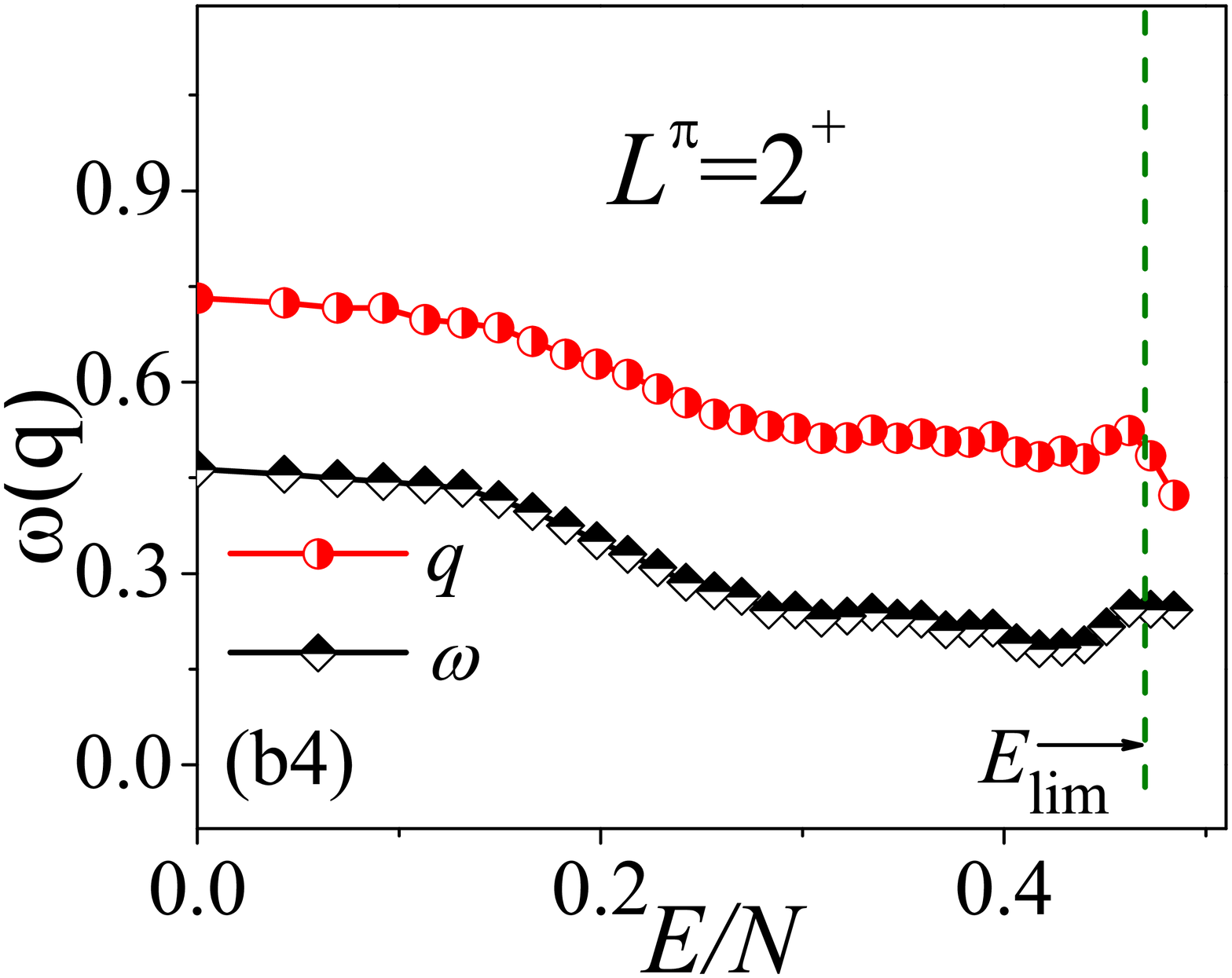}
\includegraphics[scale=0.16]{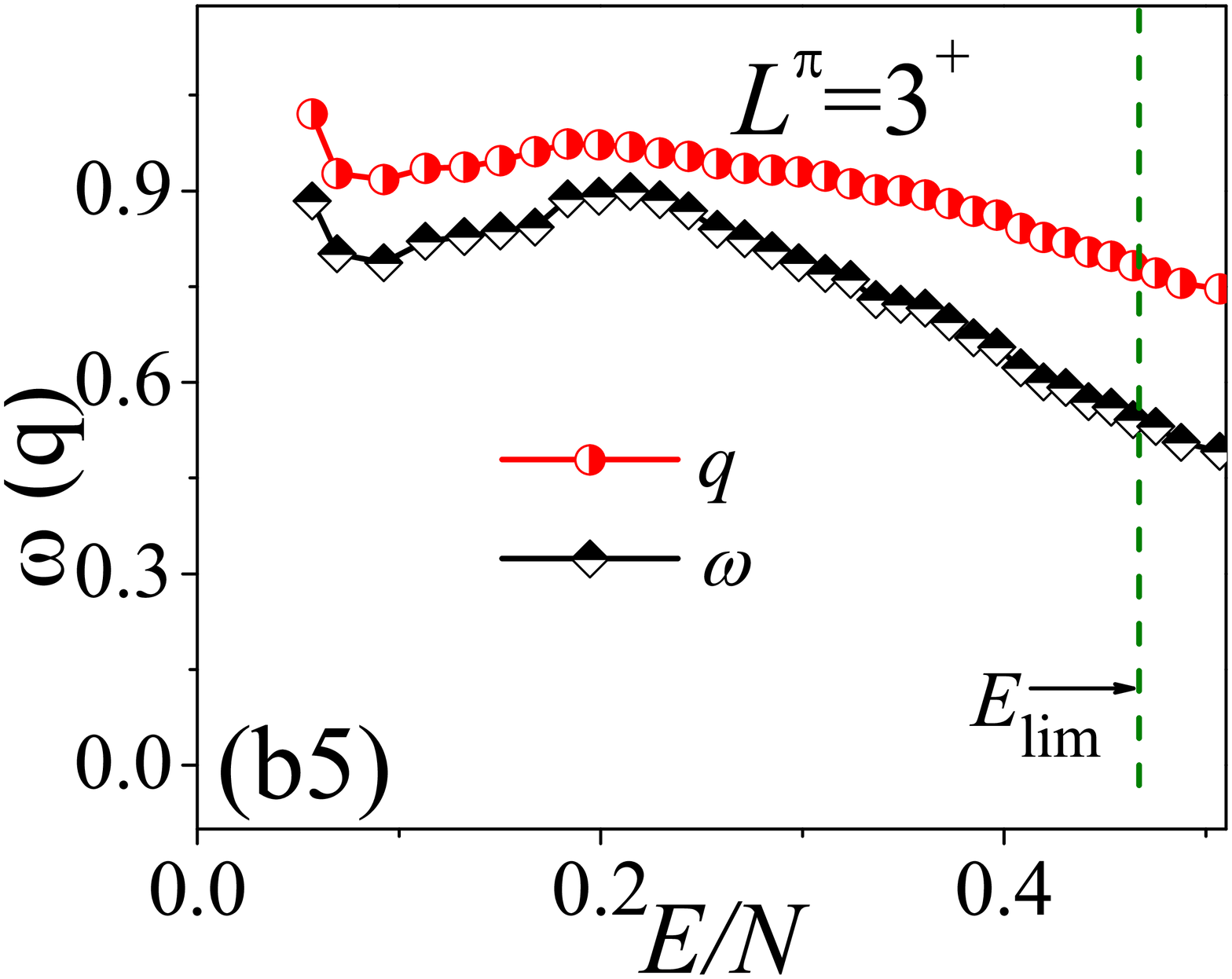}
\includegraphics[scale=0.16]{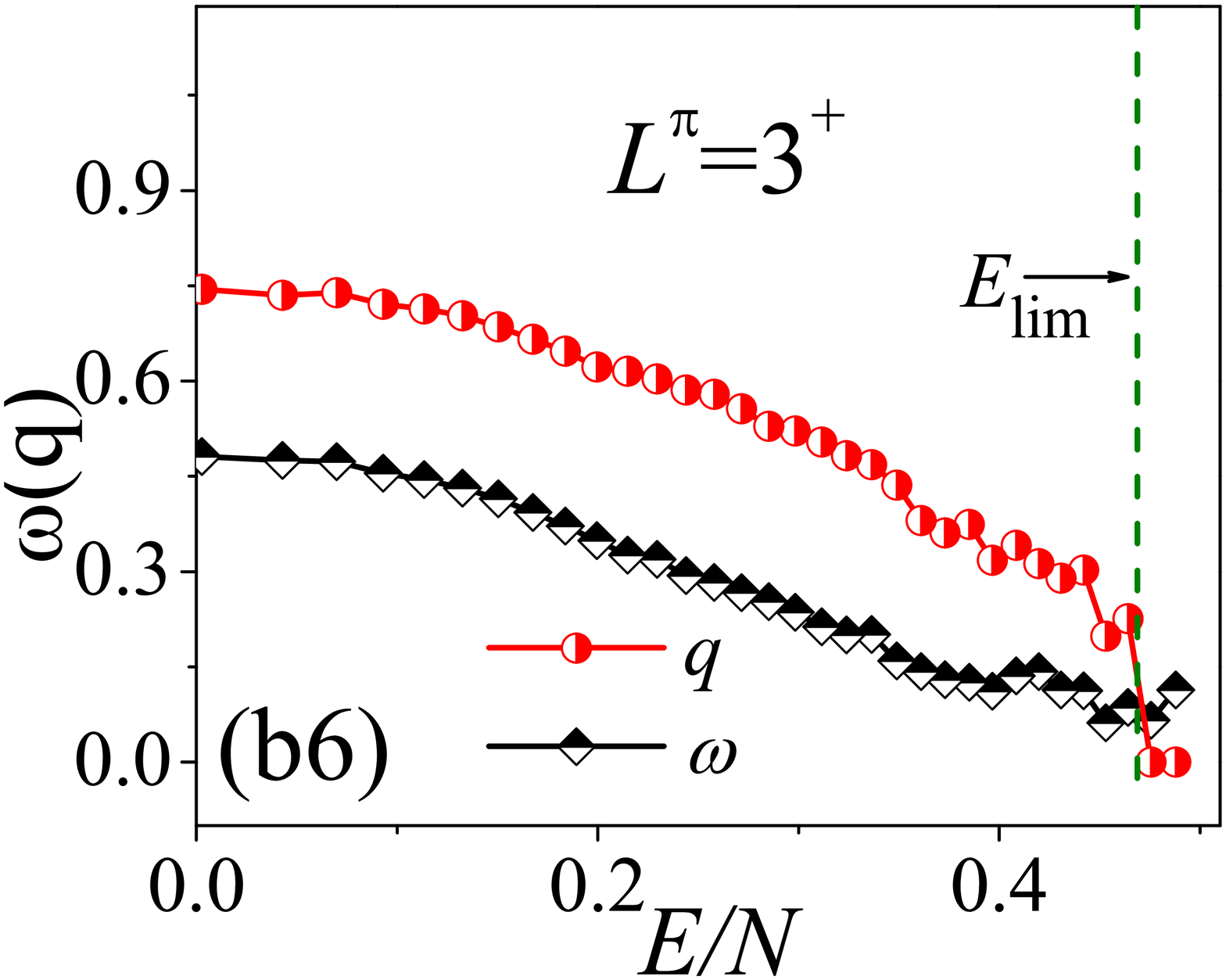}
\figcaption{The same as in Fig.~\ref{F2a} but for those
corresponding to the parameter point B. The inset in (b0) shows an
amplified picture of the potential bottom with two degenerated
minimal points indicating the 1st-order GSQPT at this parameter
point.}\label{F2b}
\end{center}

\begin{center}
\includegraphics[scale=0.25]{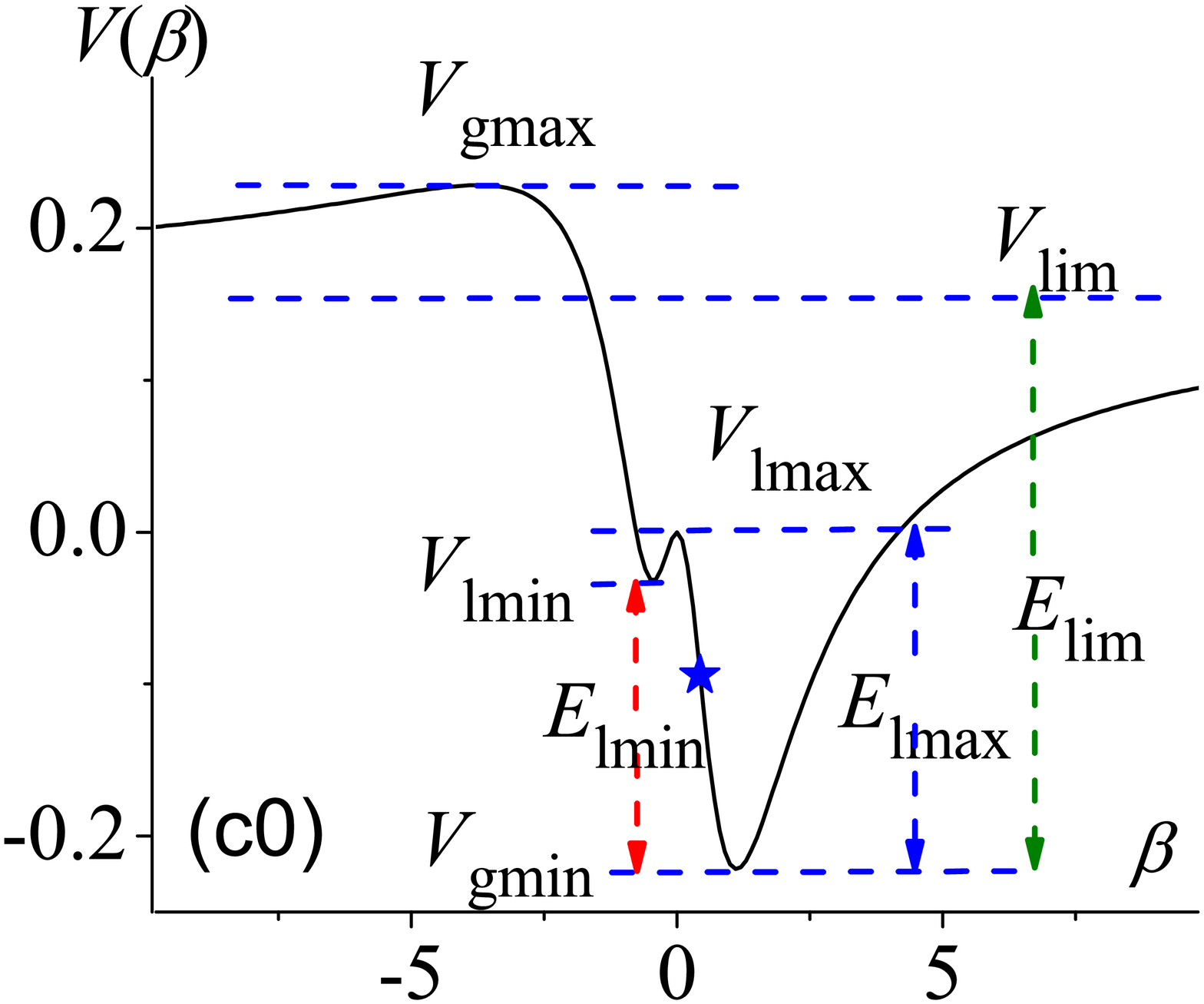}
\includegraphics[scale=0.16]{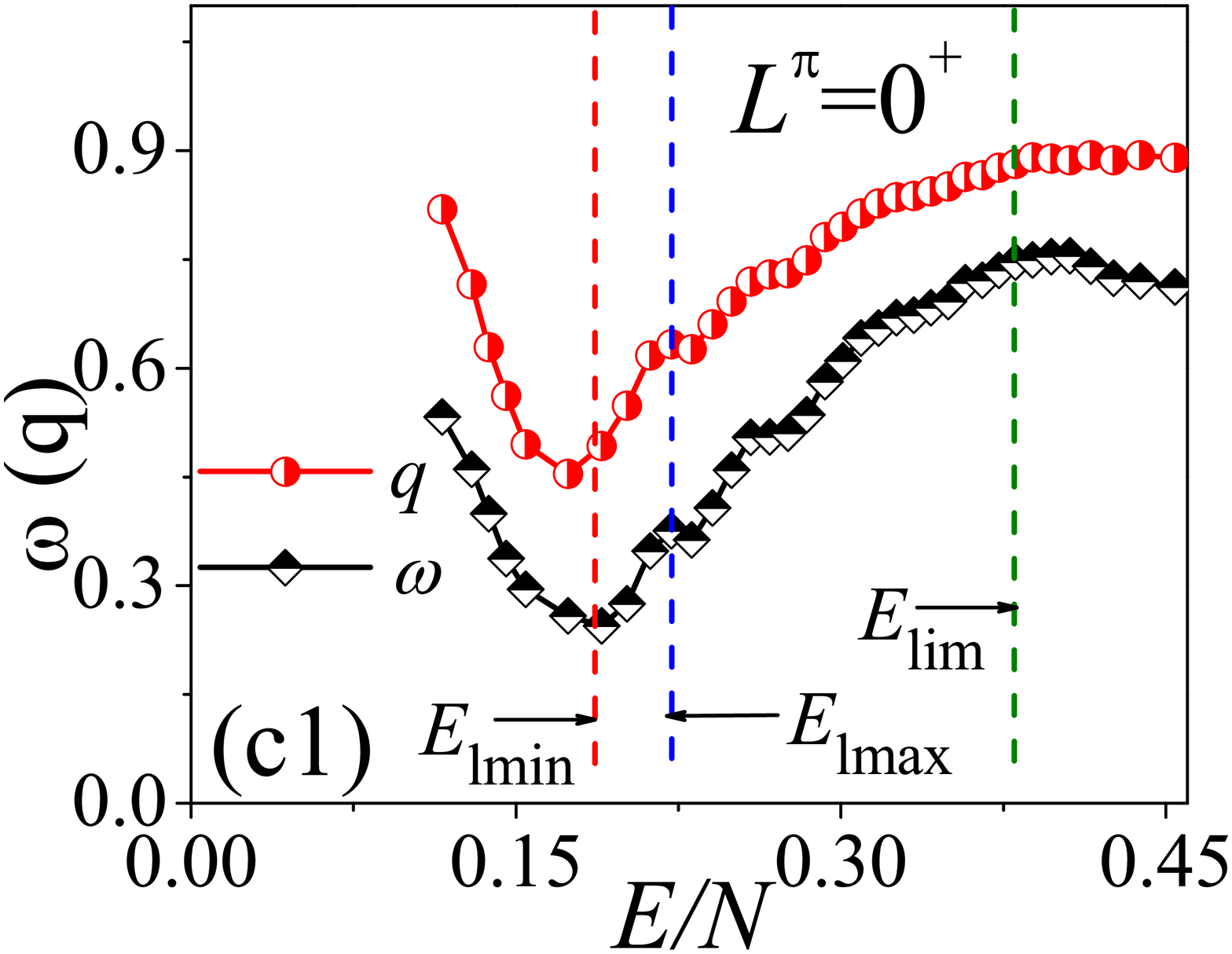}
\includegraphics[scale=0.16]{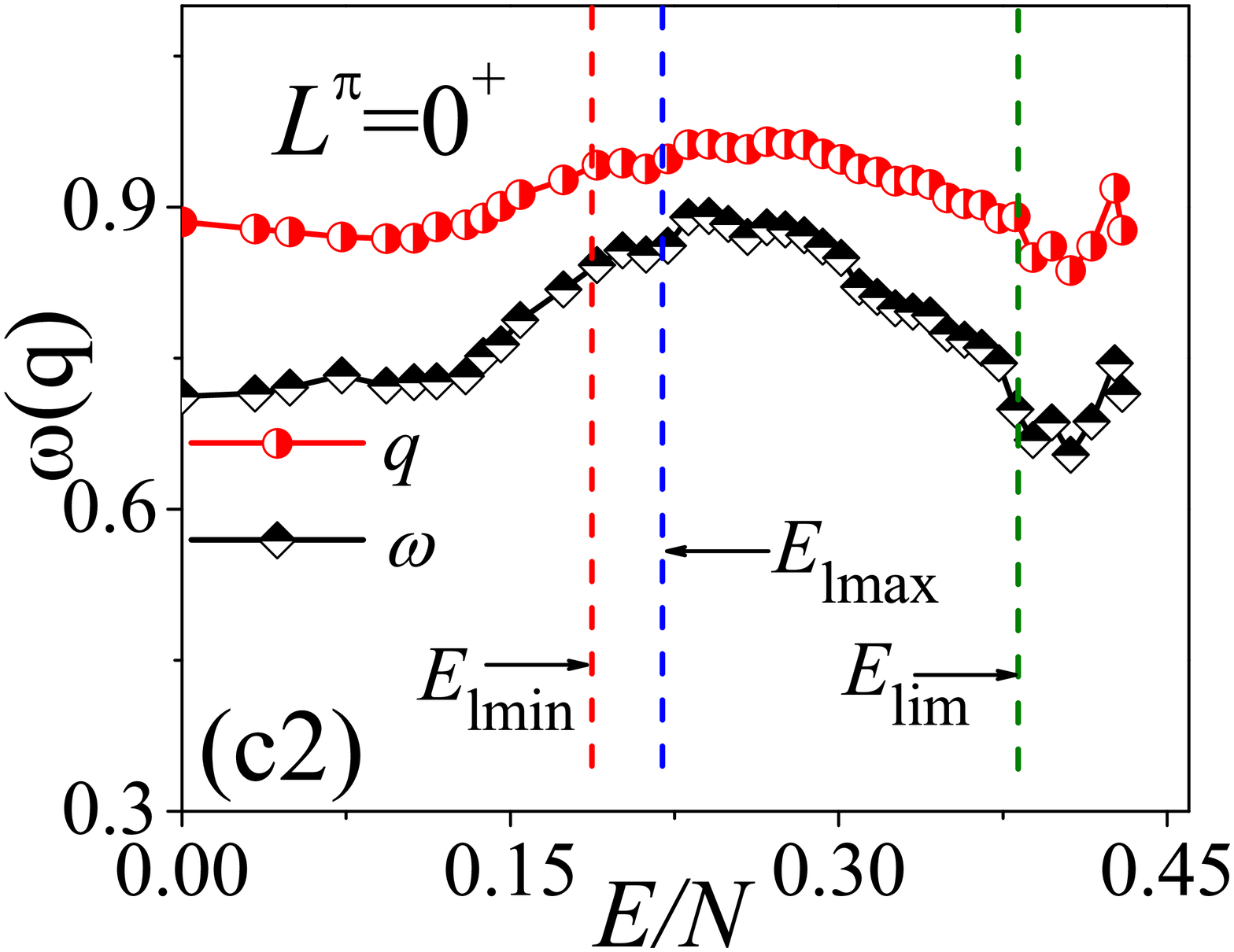}
\includegraphics[scale=0.16]{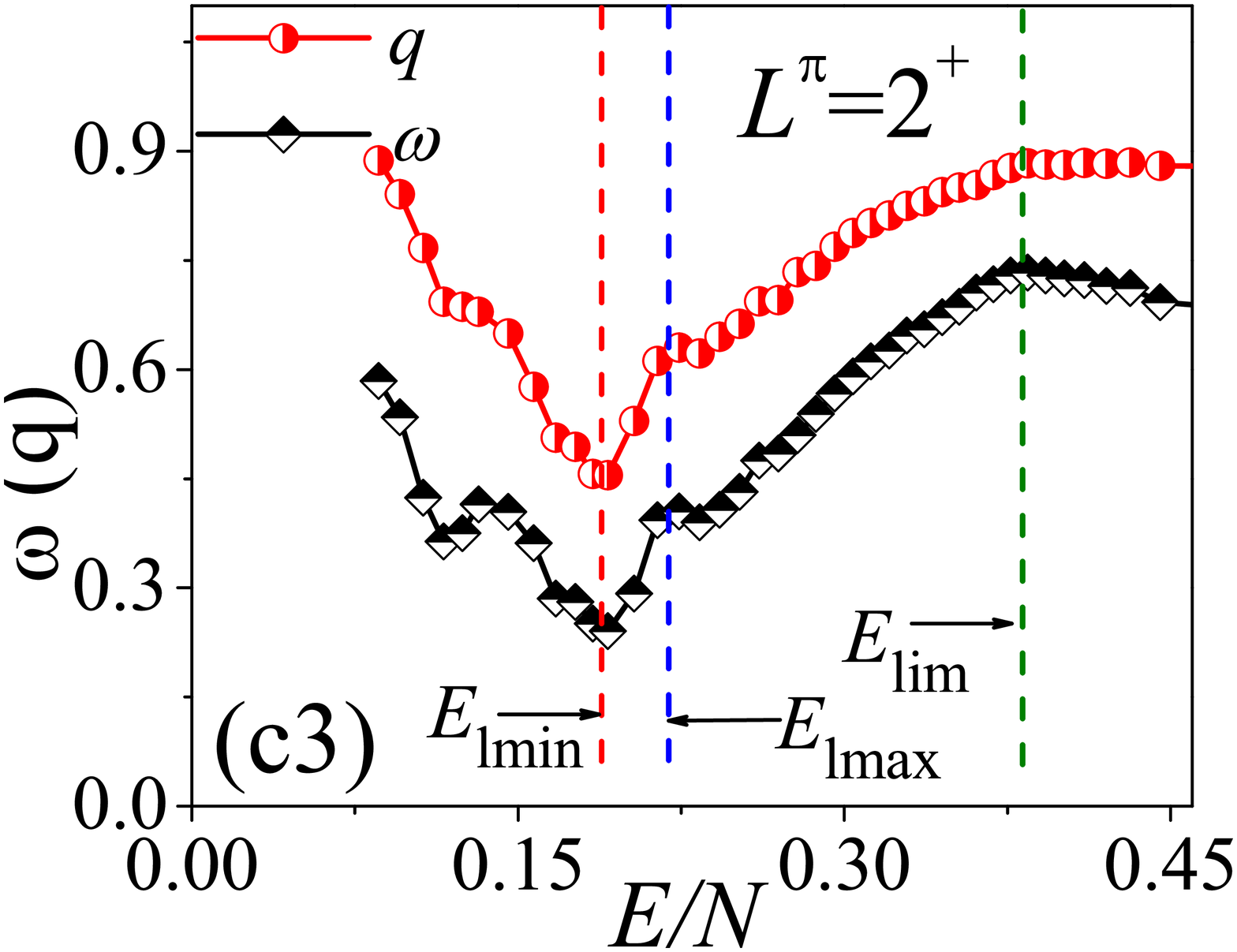}
\includegraphics[scale=0.16]{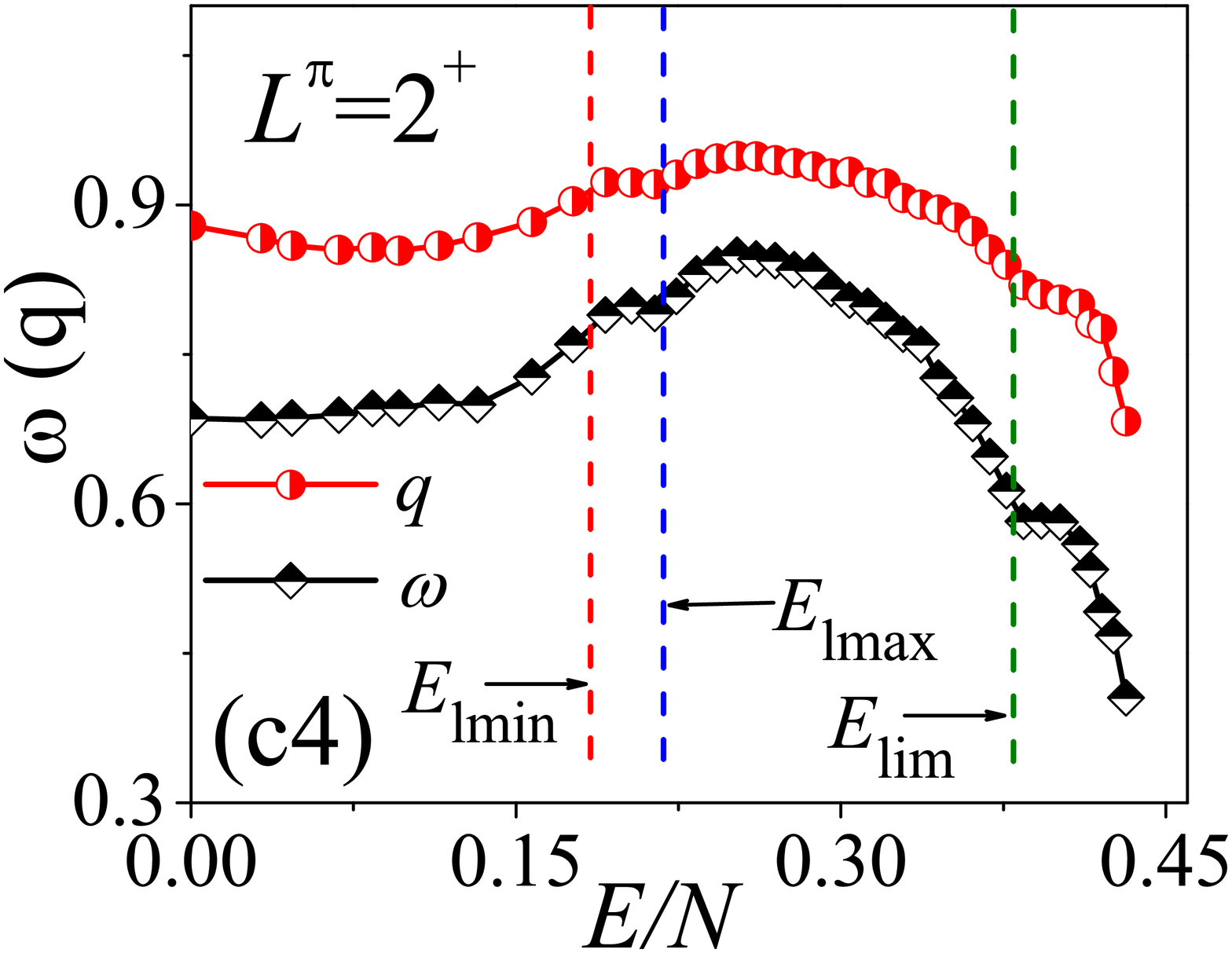}
\includegraphics[scale=0.16]{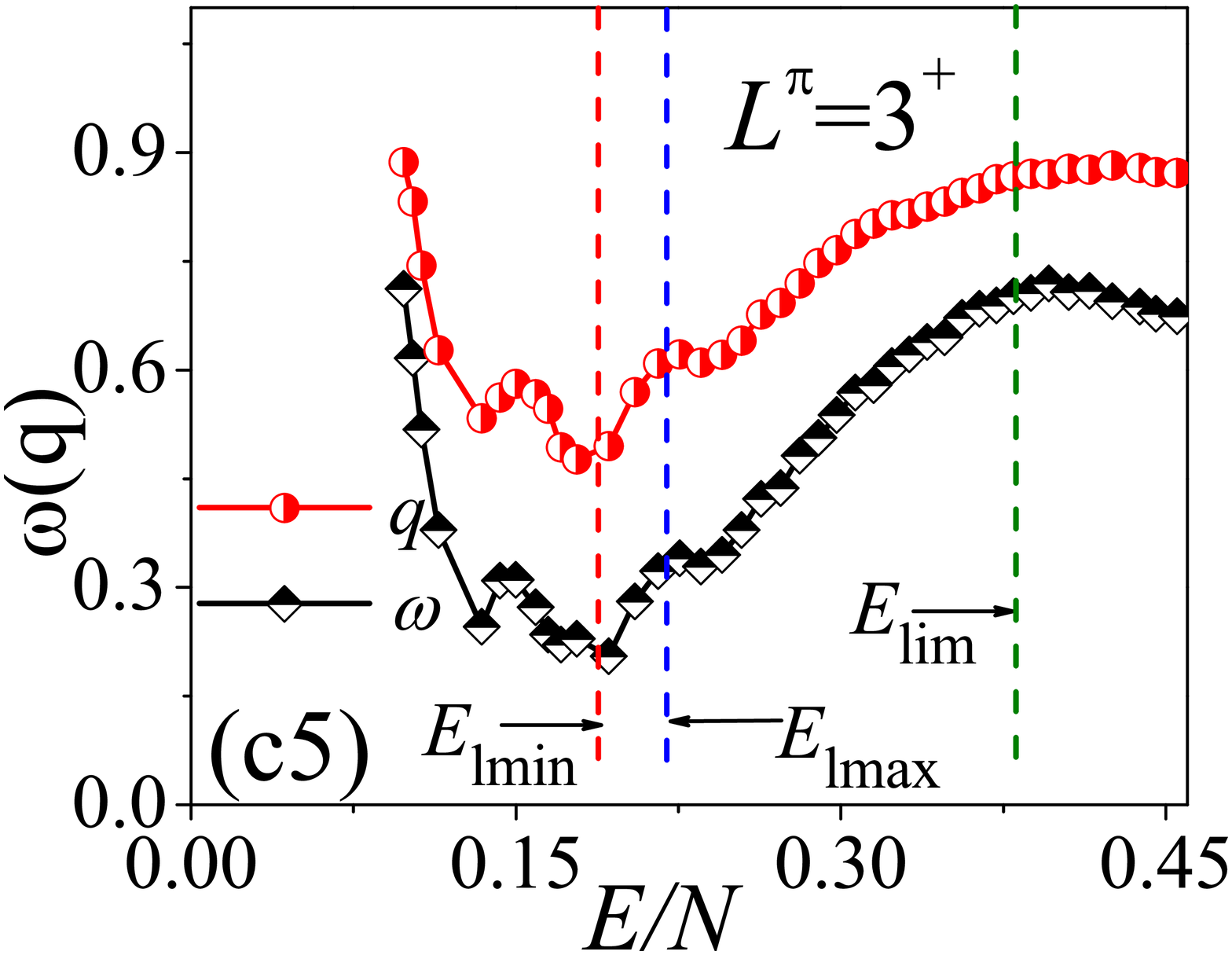}
\includegraphics[scale=0.16]{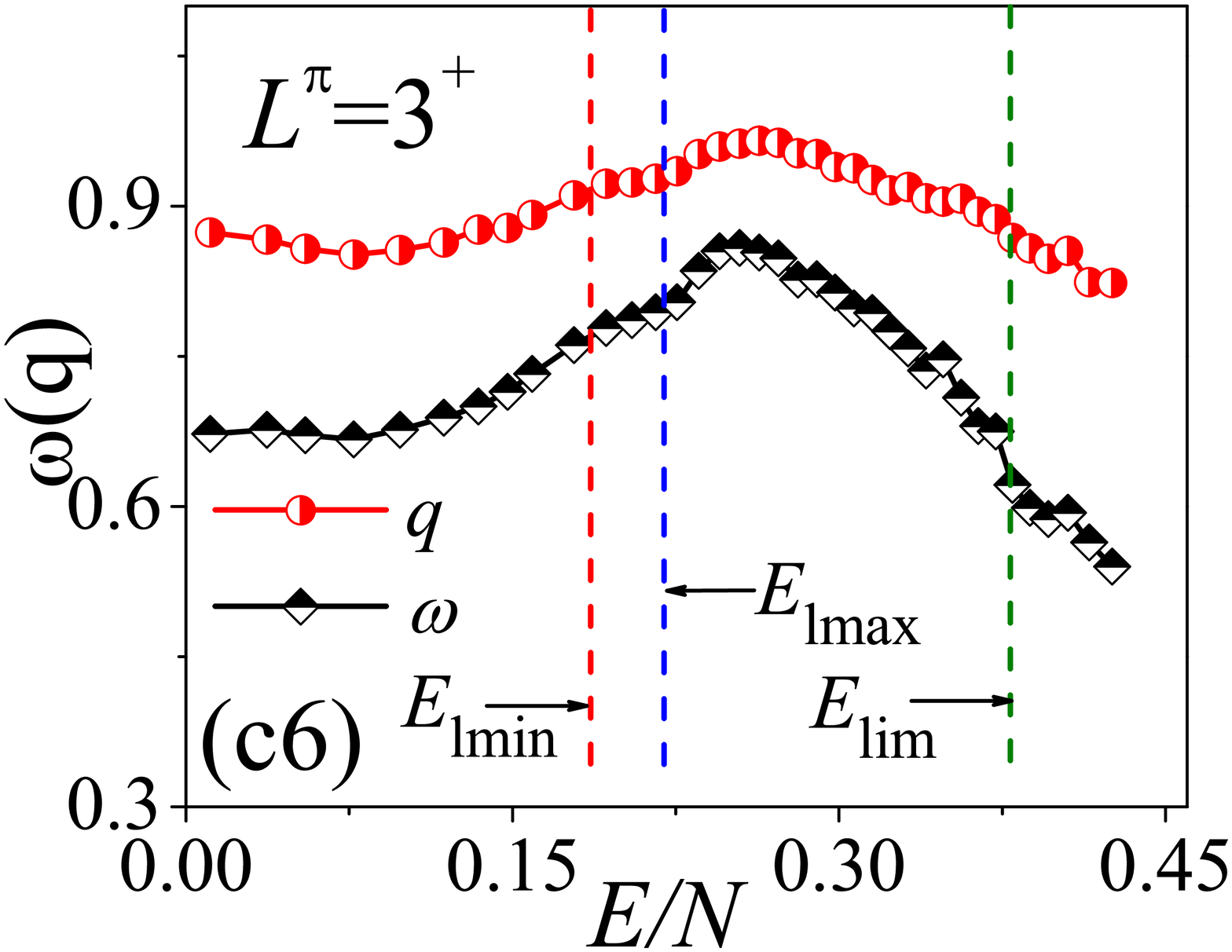}
\figcaption{The same as in Fig.~\ref{F2a} but for those
corresponding to the parameter point C.}\label{F2c}
\end{center}
The more interesting case should be the one at the parameter point C
because the potential $V(\beta)$ in a deformed system can hold a
richer stationary point structure. As seen from Fig.~\ref{F2c}(c0),
three stationary points $V_\mathrm{lmax}$, $V_\mathrm{lmin}$ and
$V_\mathrm{lim}$ locate in between $V_\mathrm{gmax}$ and
$V_\mathrm{gmin}$. More importantly, the effects of these stationary
points can be clearly observed in the evolutions of the statistical
results as a function of the excitation energy. As shown in
Fig.~\ref{F2c}(c1), the $\omega(q)$ values for $L^\pi=0^+$ present
the non-monotonic evolutions with the minimal values appearing
exactly at $E_\mathrm{lmin}$, above which the influence of the
stationary point can be also clearly observed near
$E_\mathrm{lmax}$. Apart from the similar evolutional features
appearing around the stationary points, the $\omega(q)$ values for
$L^\pi=2^+,~3^+$ meanwhile exhibit small fluctuations near
$E/N\sim0.15$ as seen in the panels(c3) and (c5). This point to some
extent reflects the effects of nonzero spins on the spectral chaos.
As further seen from the panels (c2), (c4) and (c6), the influences
of $V_\mathrm{lmin}$ and $V_\mathrm{lmax}$ on the spectral
fluctuations have not been implicitly exhibited in the statistics
from high energy to low energy but the influence of $V_\mathrm{lim}$
can be still observed. This point is not difficult to be understood
since the energy levels bound above $V_\mathrm{lmax}$ in this case
may occupy more than 75 percent of the total number for a given
$L^\pi$. As a result, the influences of $V_\mathrm{lmin}$ and
$V_\mathrm{lmax}$ on the spectral fluctuations may more or less be
screened by the high-energy levels far from the two stationary
points.

In the analysis given in \cite{Zhang2021}, the stationary points
$V_\mathrm{lmax}$ and $V_\mathrm{lmin}$ were taken as the phase
boundaries to do the statistical analysis of the excited state
quantum phase transitions (ESQPTs). Clearly, the present results
further justify the reasonability of doing so. In addition, the
analysis of the quantum optical models given in \cite{Fernandez2011}
indicate that the precursors of ESQPT phenomenon in a nonintegrable
case may be accompanied by a abrupt emergence of spectral chaos. The
similar situation is suggested to occur here too. Specifically, the
results in Fig.~\ref{F2c}(c1) indicate that the degree of chaoticity
characterized by both $\omega$ and $q$ will quickly increase after
$E_\mathrm{lmin}$, which is alternatively defined as the critical
energy of the ESQPT in the U(5)-SU(3) transition~\cite{Zhang2016}.
It is thus confirmed that the ESQPT in the nonintegrable U(5)-SU(3)
cases will be also accompanied by a sudden emergence of spectral
chaos.

\begin{center}
\includegraphics[scale=0.165]{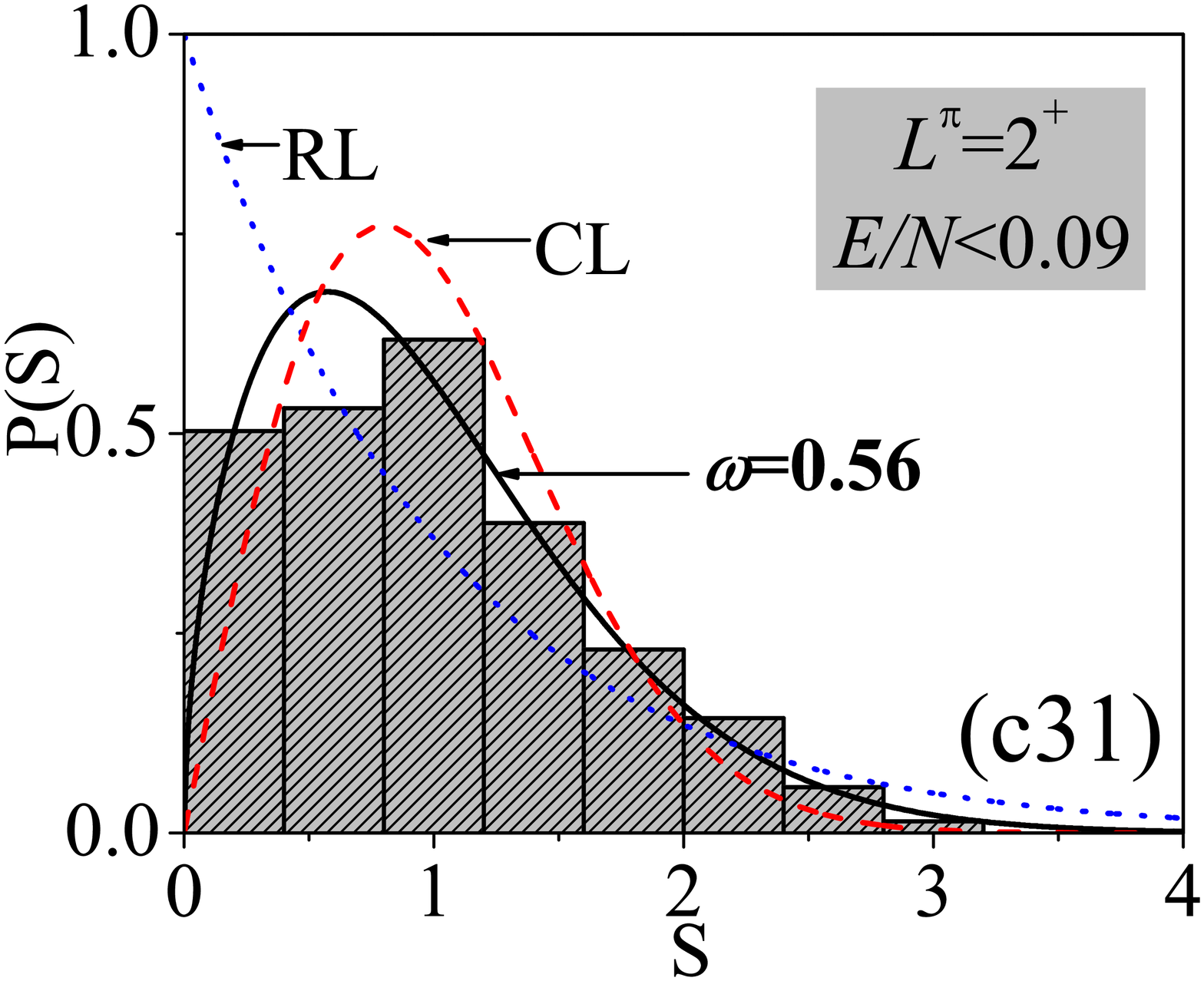}
\includegraphics[scale=0.165]{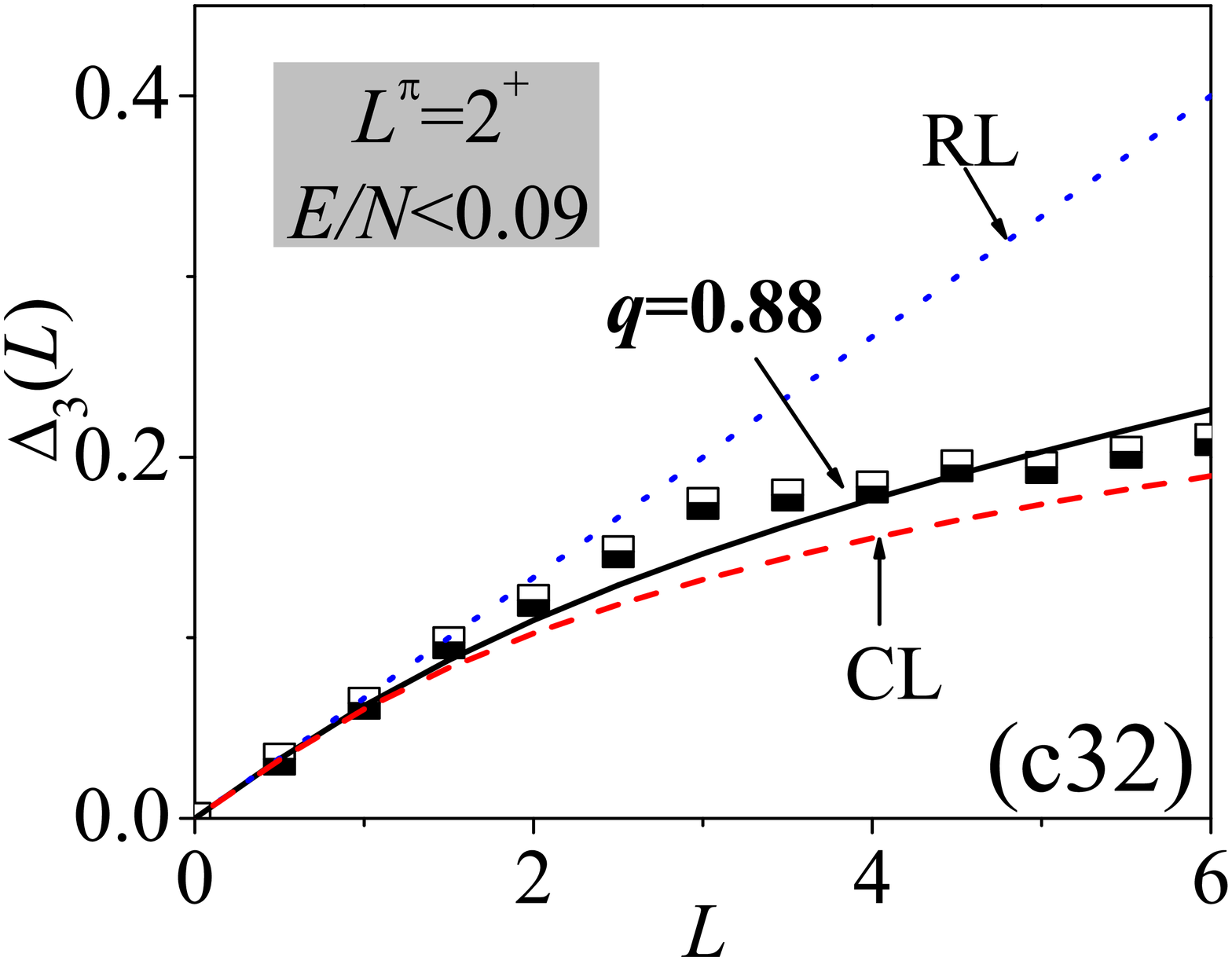}
\includegraphics[scale=0.165]{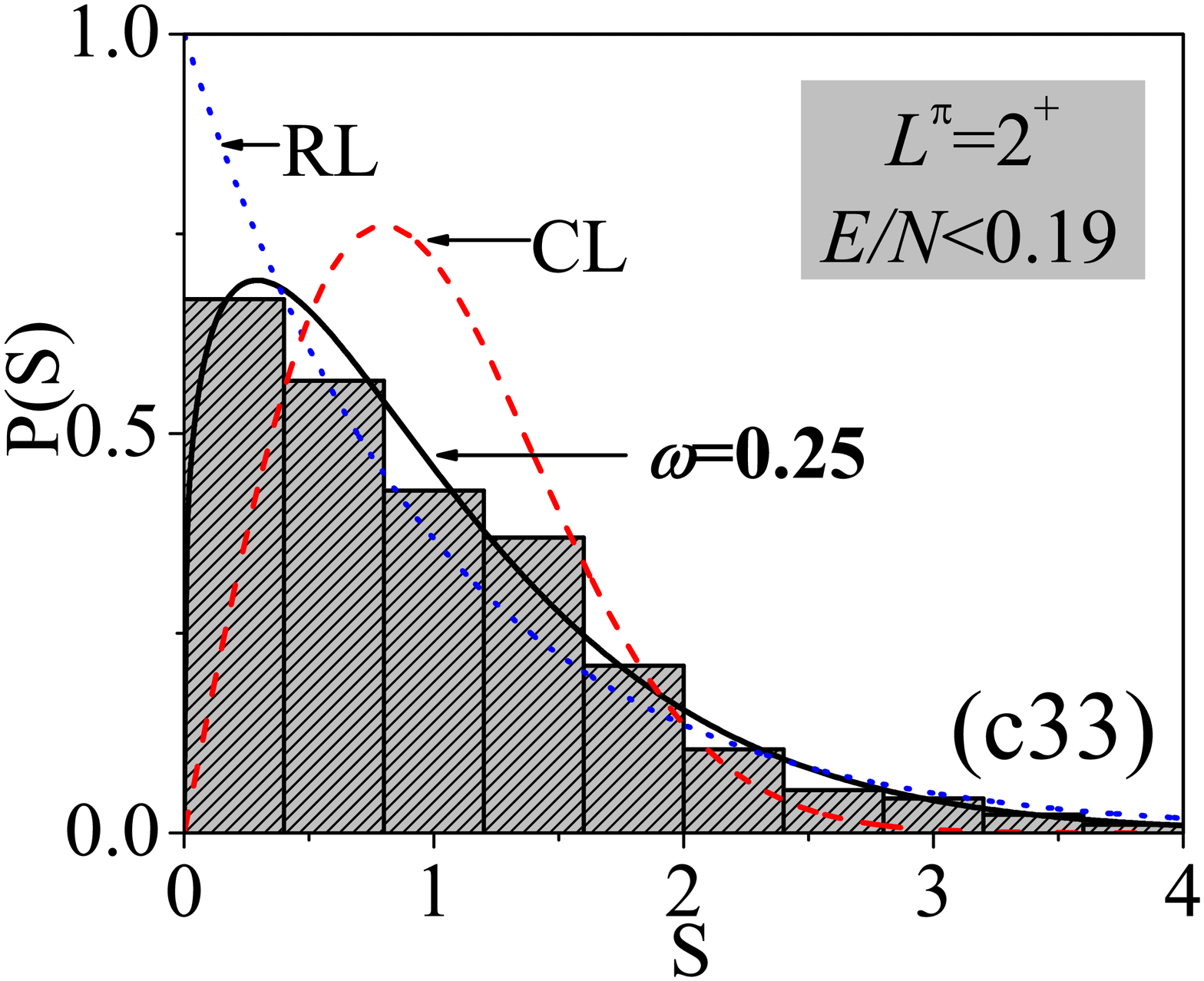}
\includegraphics[scale=0.165]{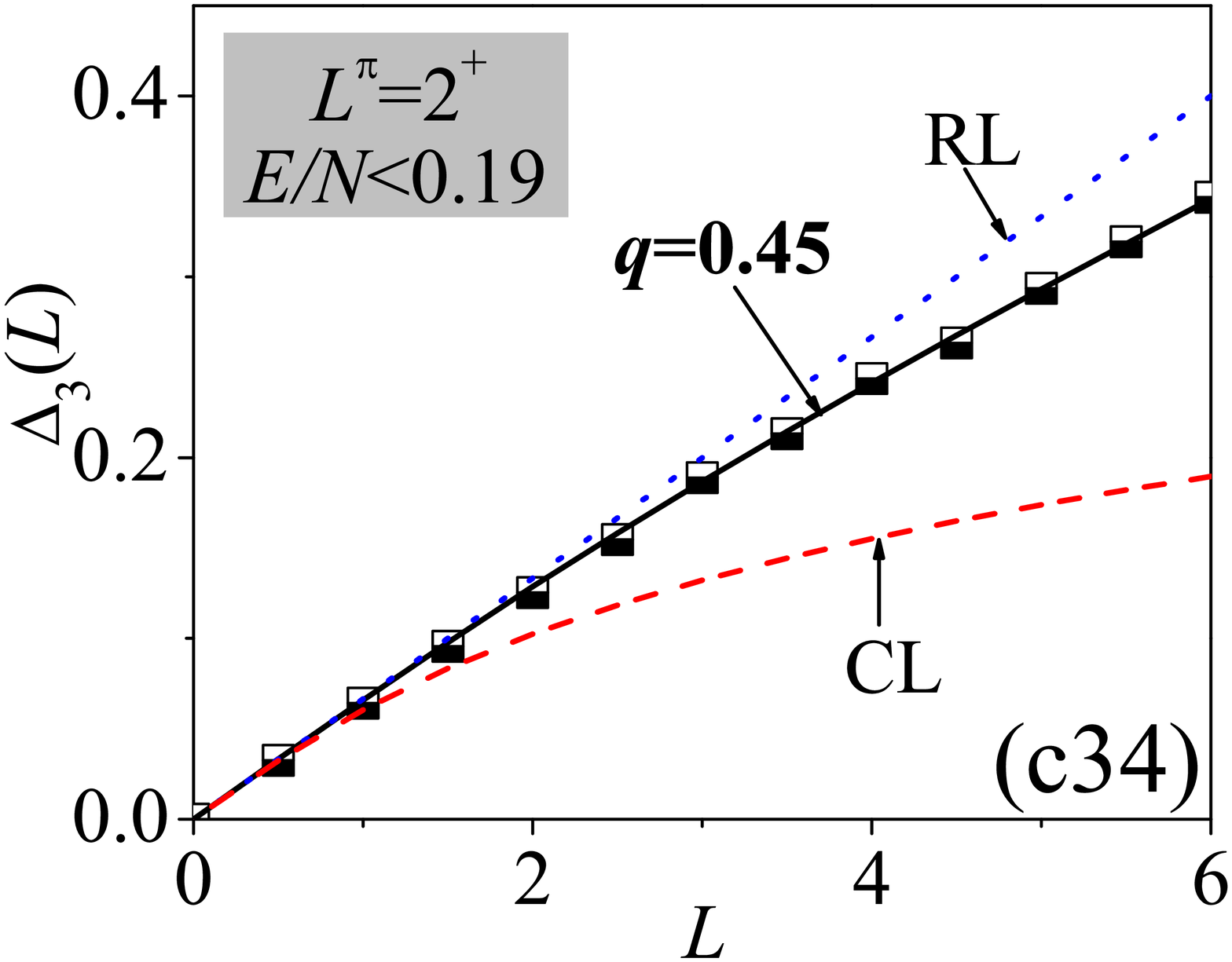}
\includegraphics[scale=0.165]{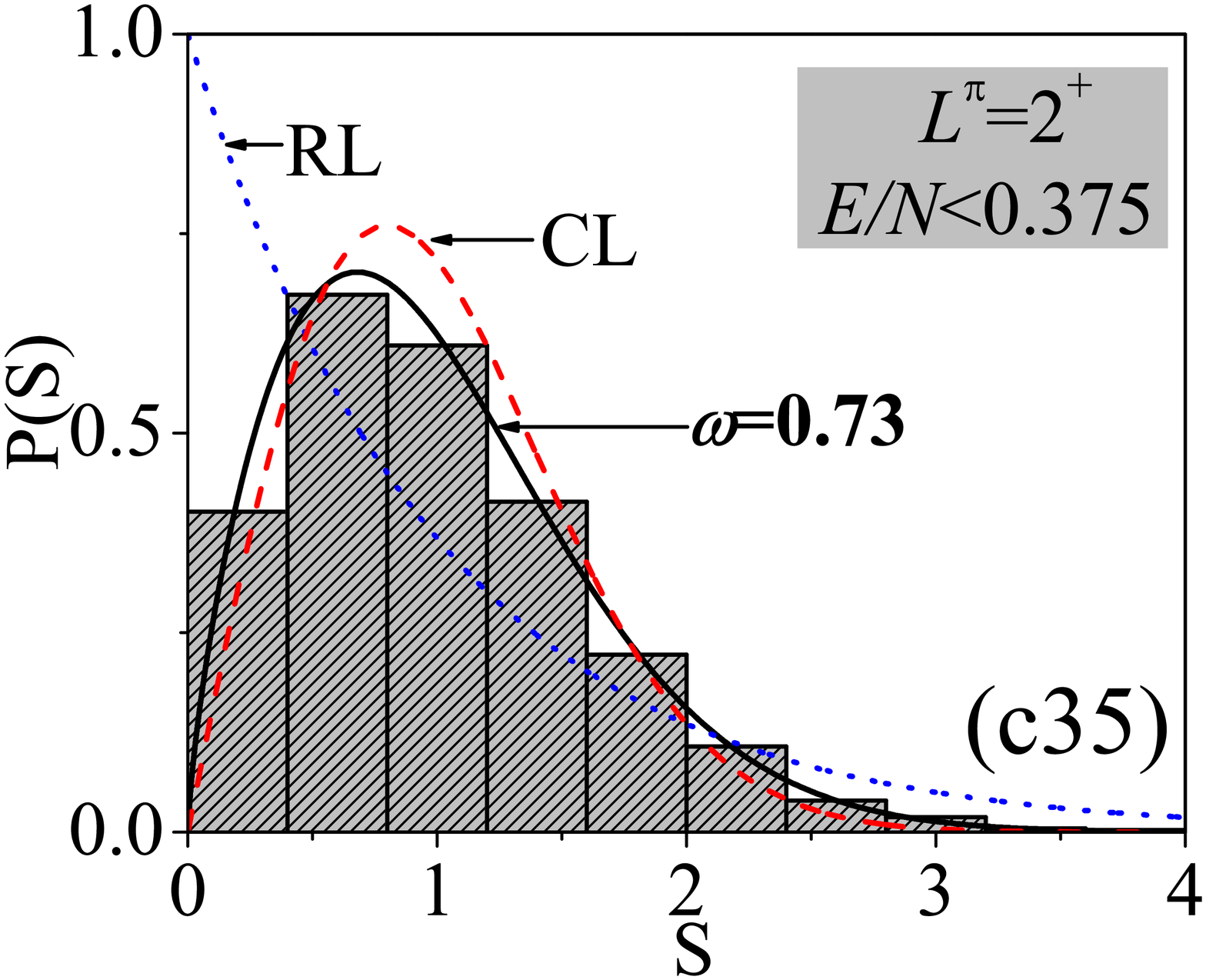}
\includegraphics[scale=0.165]{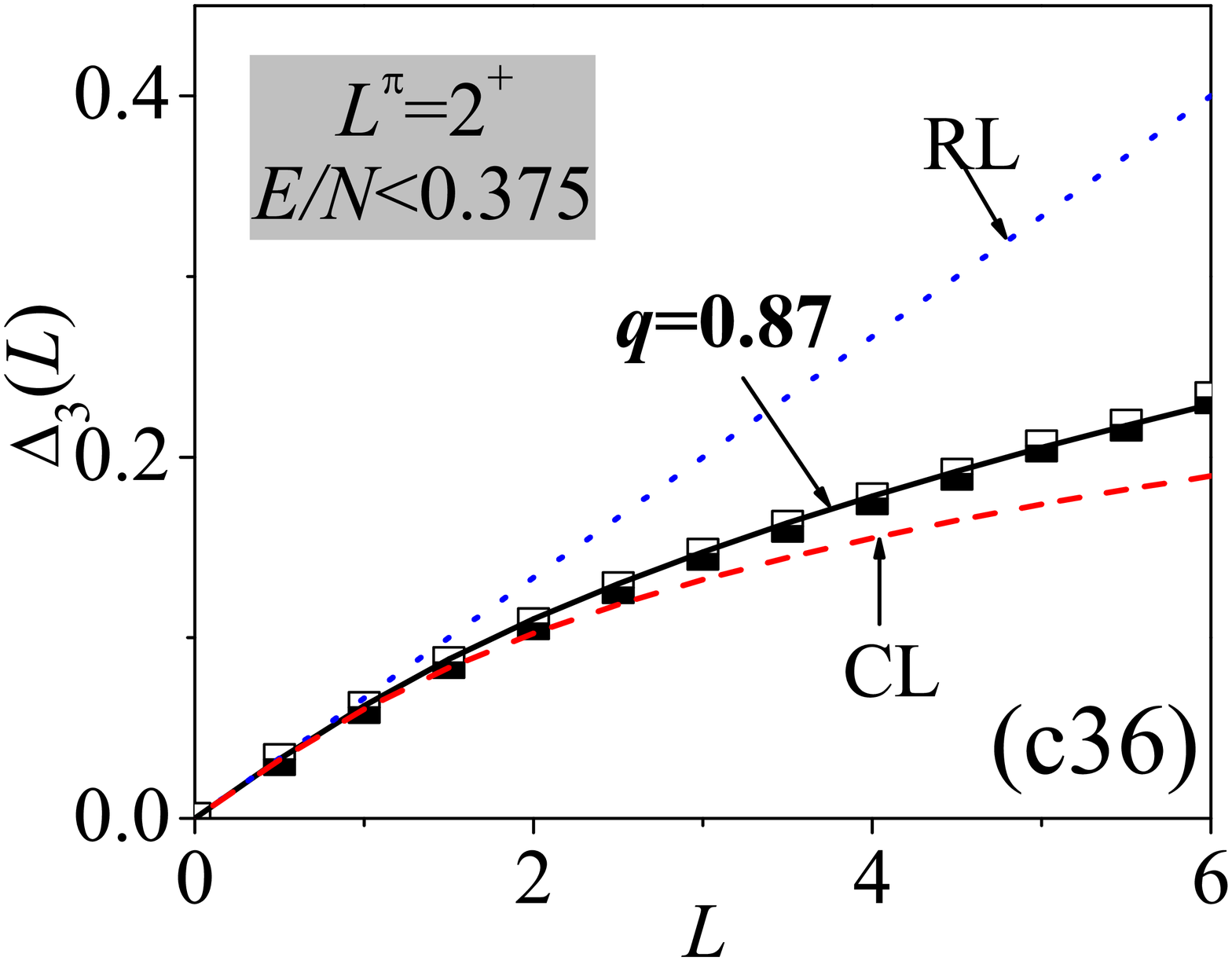}
\figcaption{The statistics for the parameter point C are shown with
the statistical sample chosen as the $2^+$ levels bound below
different energy cutoffs.}\label{F2c1}
\end{center}

\begin{center}
\includegraphics[scale=0.165]{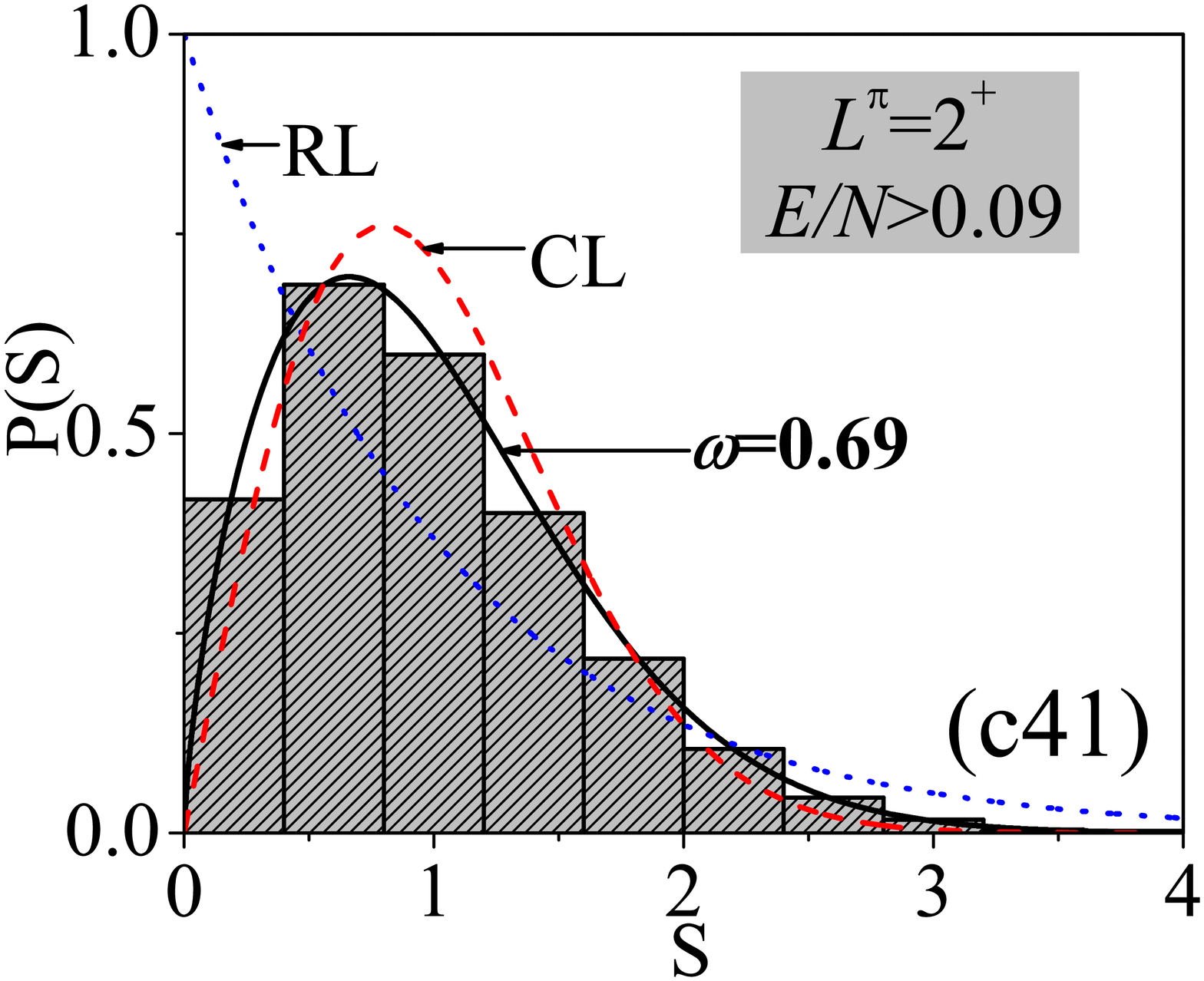}
\includegraphics[scale=0.165]{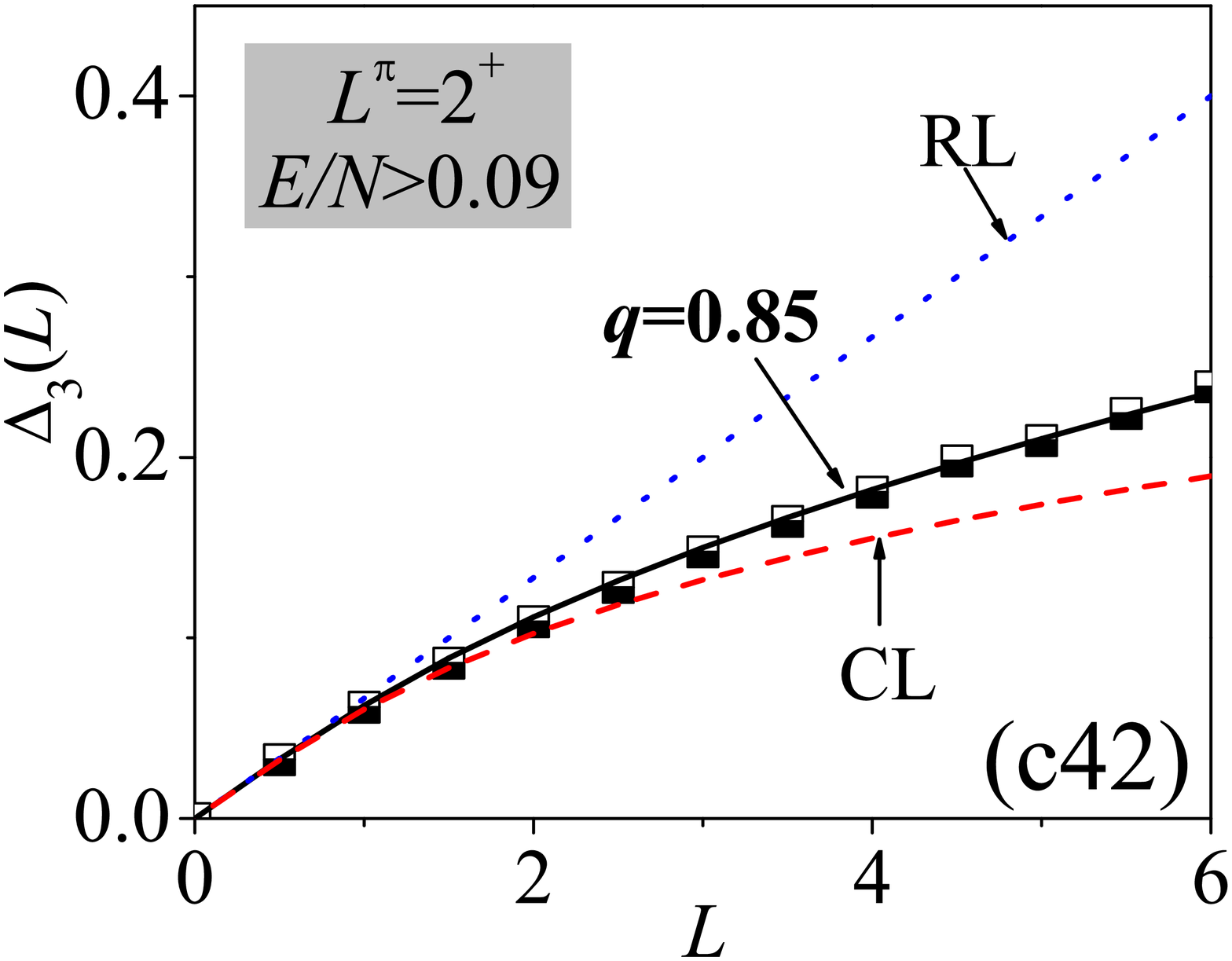}
\includegraphics[scale=0.165]{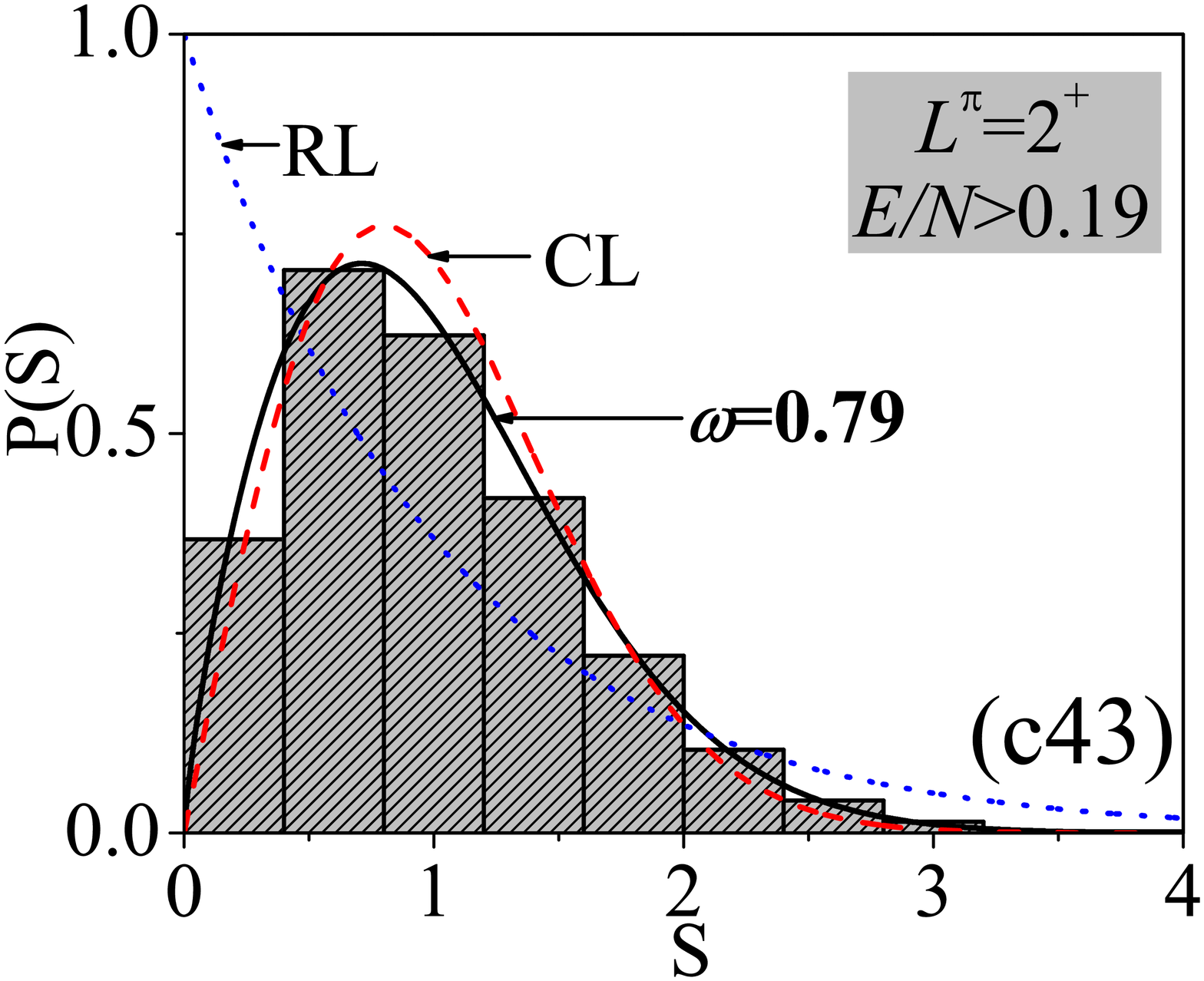}
\includegraphics[scale=0.165]{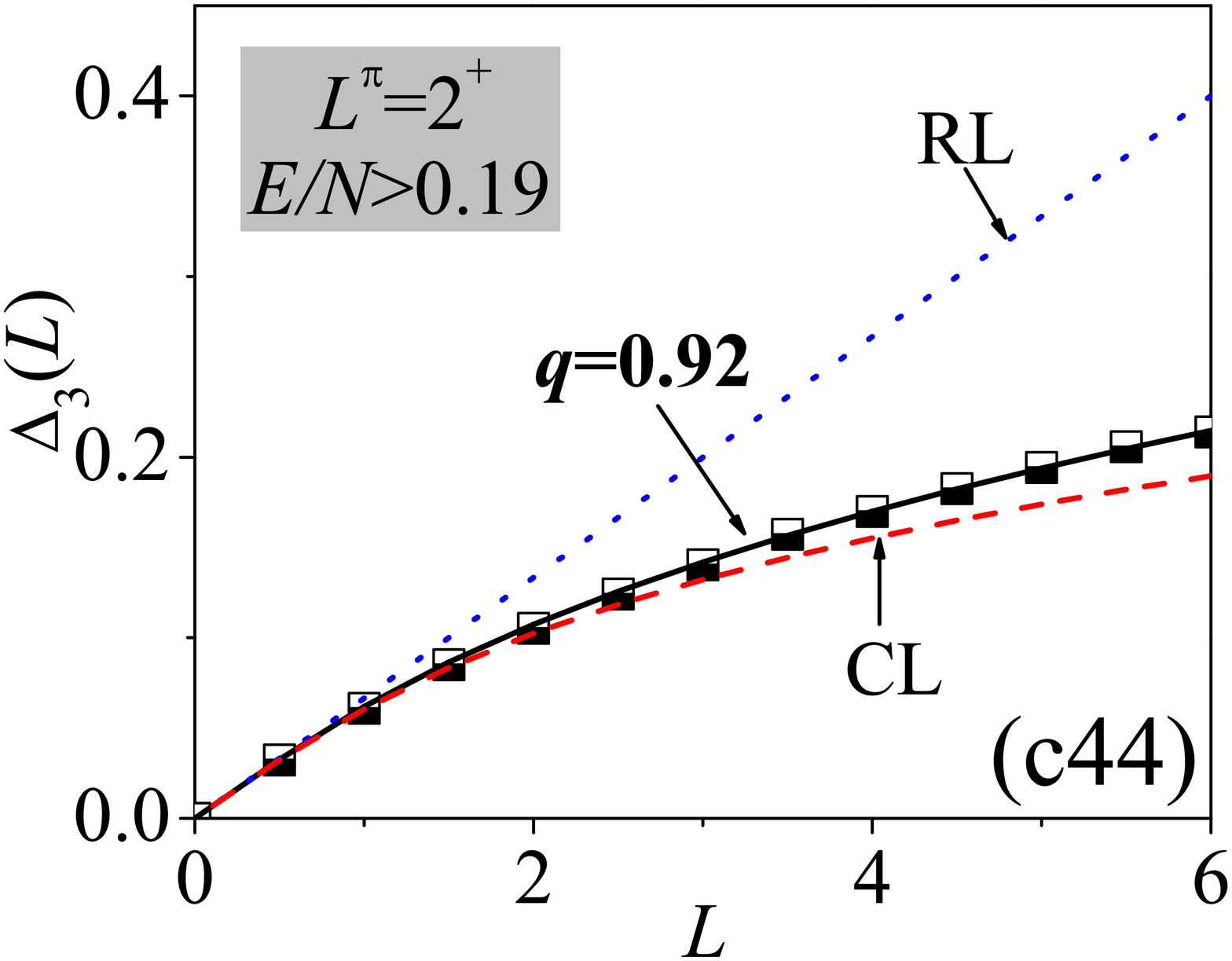}
\includegraphics[scale=0.165]{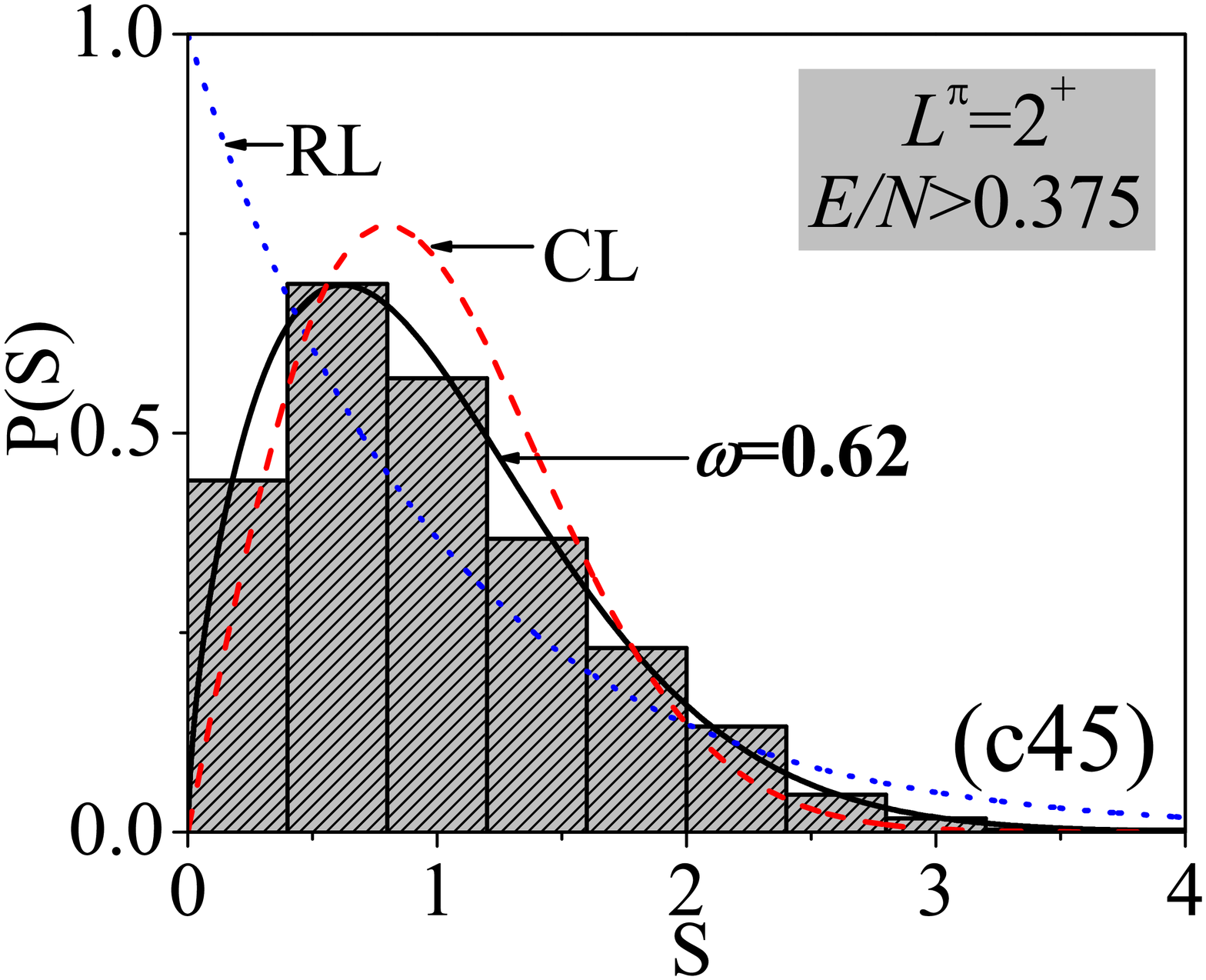}
\includegraphics[scale=0.165]{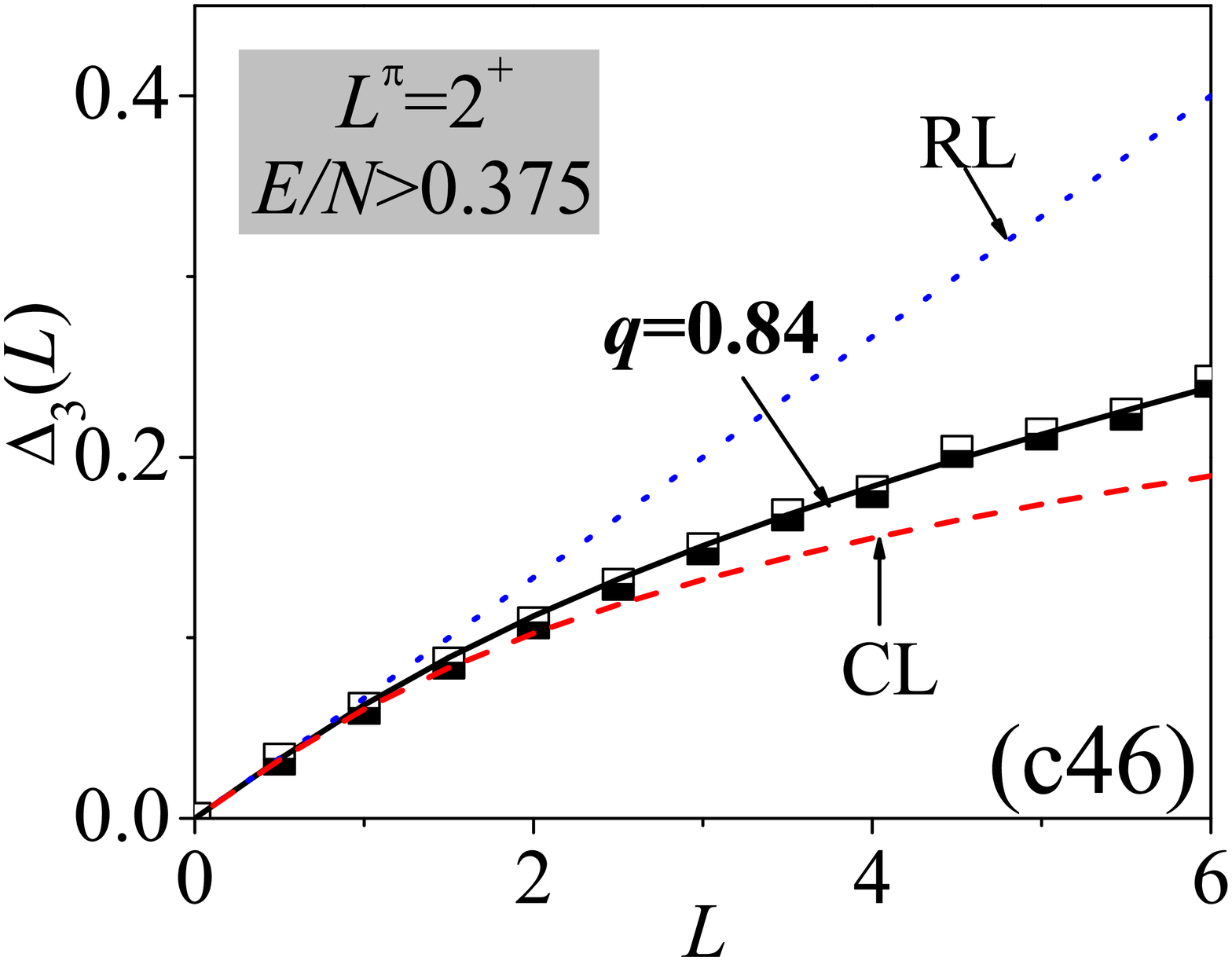}
\figcaption{The same as in Fig.~\ref{F2c1} but taking those bound
above different energy cutoffs.}\label{F2c2}
\end{center}
To have a close look at the energy dependence of the $P(S)$ and
$\Delta_3(L)$ statistics, we select the cases at three different
energy cutoffs from the panels (c2) and (c4) of Fig~\ref{F2c},
respectively. One can find from Fig~\ref{F2c1} and Fig~\ref{F2c2}
that the statistical results for $L^\pi=2^+$ could be very different
at different cutoffs. It is given by $\omega=0.25$ ($q=0.45$) for
the spectrum below $E/N=0.19$ but by $\omega=0.79$ ($q=0.92$) for
that above $E/N=0.19$. Since $E/N=0.19$ is very close to the
critical energy $E_\mathrm{lmin}$, the results confirm again that
the spectral fluctuations could be significantly different by taking
the stationary point $V_\mathrm{lmin}$ as a boundary as did in
\cite{Zhang2021}. In short, the spectral fluctuations in the
deformed phase of the U(5)-SU(3) GSQPT are shown to be more energy
dependent due to the richer stationary point structure.

\subsection{Inside the triangle}

Inside the triangle phase diagram, the parameter points D and E
describe the situations on and off the AW arc, respectively. The
systems on the AW arc are always expected to give regular spectra
for zero~\cite{Karampagia2015} and nonzero spins~\cite{Whelan1993}.
As shown in Fig.~\ref{F2d}, the results indicate that the entire
spectrum on the AW arc is indeed very regular if involving all the
levels in statistical calculations for a given $L^\pi$. This is
highlighted by the results for $L^\pi=2^+$ as seen in the panels
(d3) or (d4), where one can find that the statistical calculations
for the entire $2^+$ spectrum give $\omega\sim0$ and $q\sim0$.
Nonetheless, the energy dependence of the spectral fluctuations on
the AW arc can be still clearly seen~\cite{Karampagia2015}. In
particular, it is shown that the $\omega$ and $q$ values as a
function of the excitation energy may change evidently near
$E_{\mathrm{lmax}}$. The results in the panel (d1) indicate that the
spectral fluctuations in the $0^+$ states will decrease until
$E_{\mathrm{lmax}}$ and then slightly increase followed by another
decrement near $E_{\mathrm{lim}}$. It means that the highest-lying
states are always regular but the states near or below the
stationary point $V_{\mathrm{lmax}}$ are also relatively regular.
This picture actually agree with the classical analysis of the AW
arc given in \cite{Macek2007}, where the results for the regular
fraction $f_{\mathrm{reg}}$ (see Fig.~4 in \cite{Macek2007})
indicate that the relative regularity of the AW arc can be to some
extent enhanced near the absolute energy $E_{\mathrm{cl}}=0$
corresponding to the stationary point $E_{\mathrm{lmax}}$ discussed
here.

On the other hand, the ground state and the adjacent lowest states
for the AW arc should be also regular according to the analysis
given in \cite{Macek2007}. However, this point cannot be directly
reflected from the $P(S)$ and $\Delta_3(L)$ calculations present
here since the statistics require enough number of levels rather
than a single ground state or a few number of lowest-lying states.
As a result, the relatively larger $\omega(q)$ values at low energy
shown in the panel (d1) only imply that some relatively chaotic
states may occupy a certain proportion below the corresponding
energy cutoff. In fact, the results for the regular fraction given
in \cite{Macek2007} also indicate that the degree of regularity on
the AW arc can be largely reduced in between the ground state energy
and the absolute energy $E_{\mathrm{cl}}=0$. While, such a "chaotic"
character at low energy will be quickly smoothed out when involving
more higher energy states in the statistical calculations. For
example, if the energy cutoff is taken as the saddle point energy
$E_{\mathrm{lmax}}$, the degree of chaos for the $0^+$ spectrum may
decrease to $\omega\sim0.3$ as seen from the panel (d1).

\begin{center}
\includegraphics[scale=0.25]{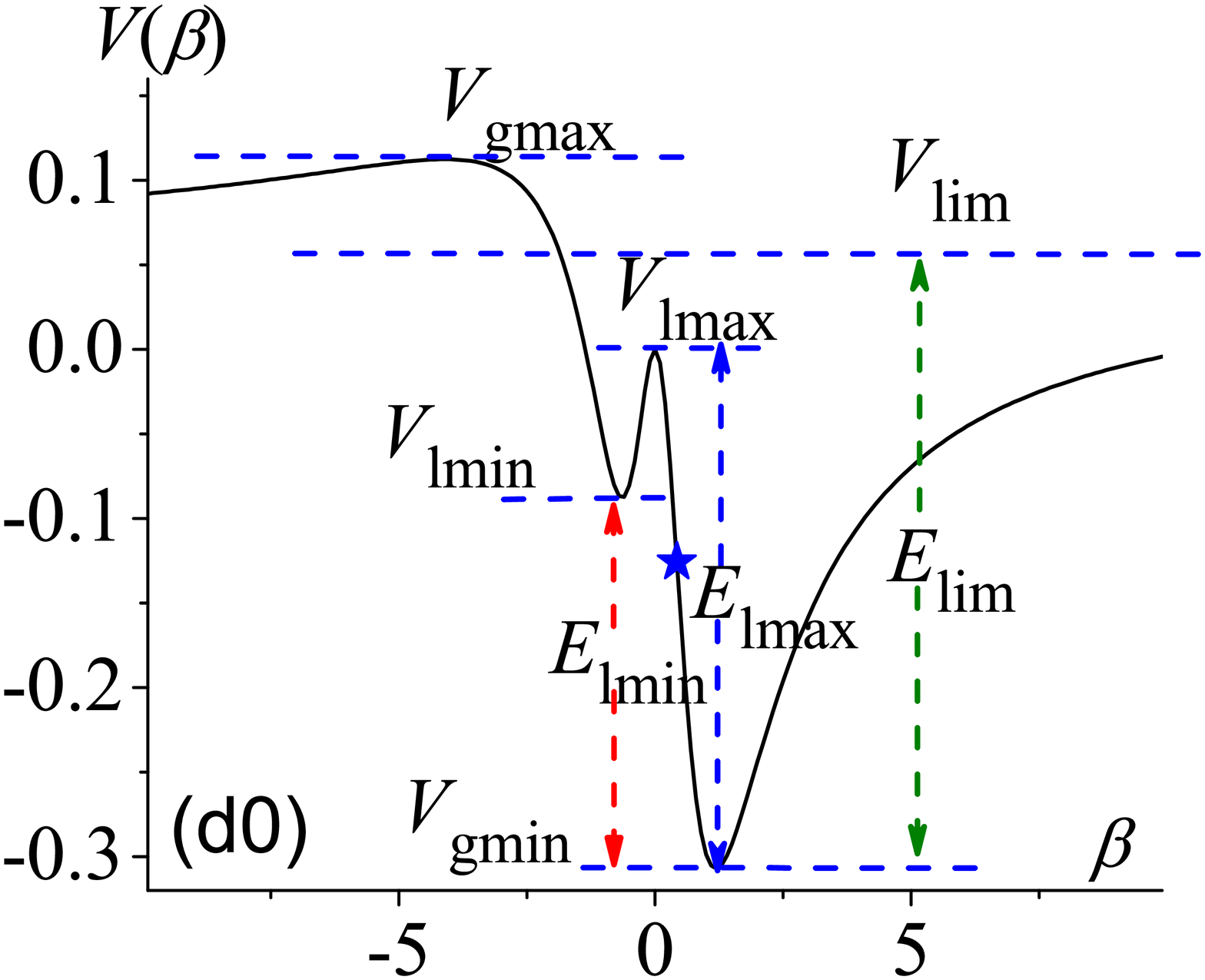}
\includegraphics[scale=0.16]{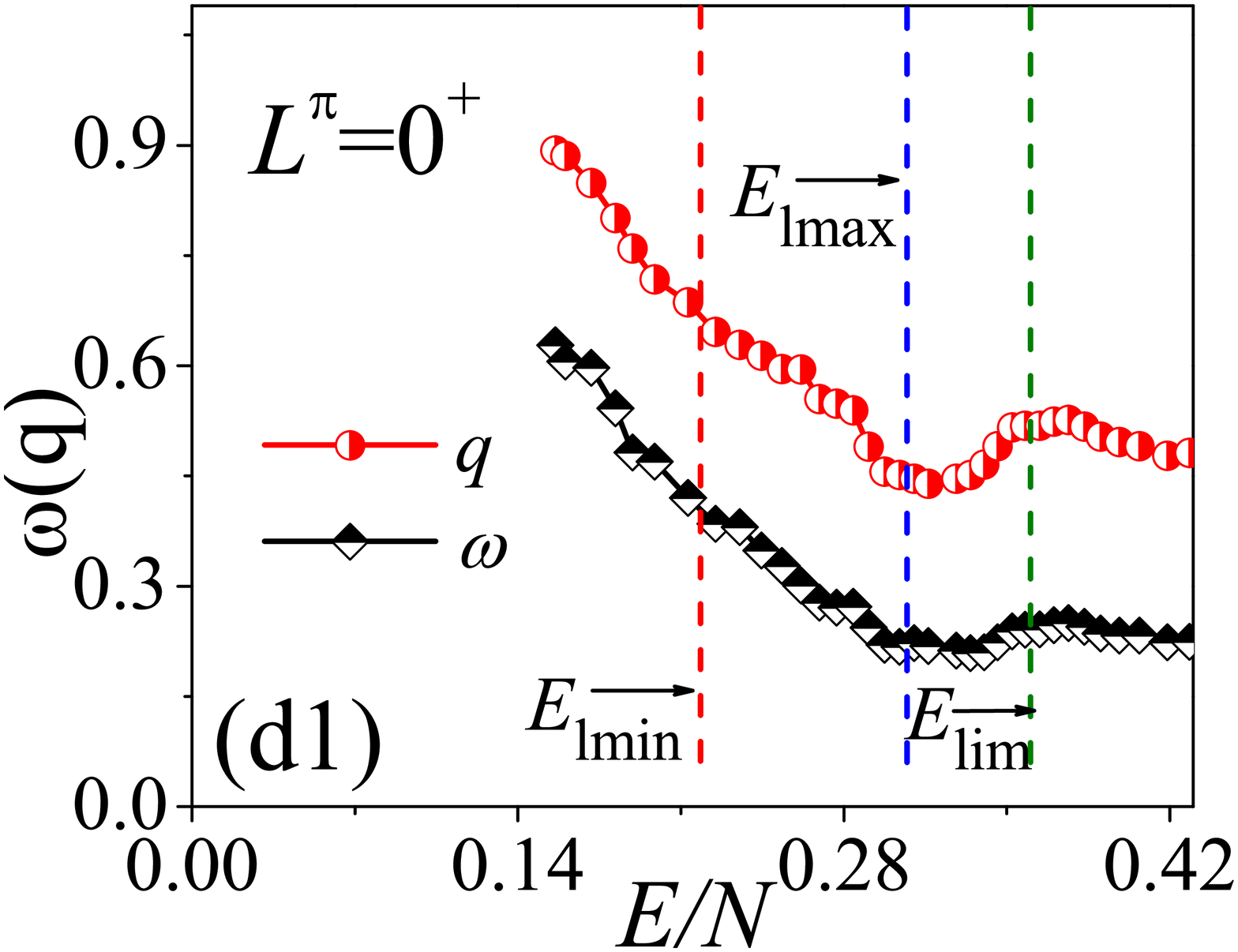}
\includegraphics[scale=0.16]{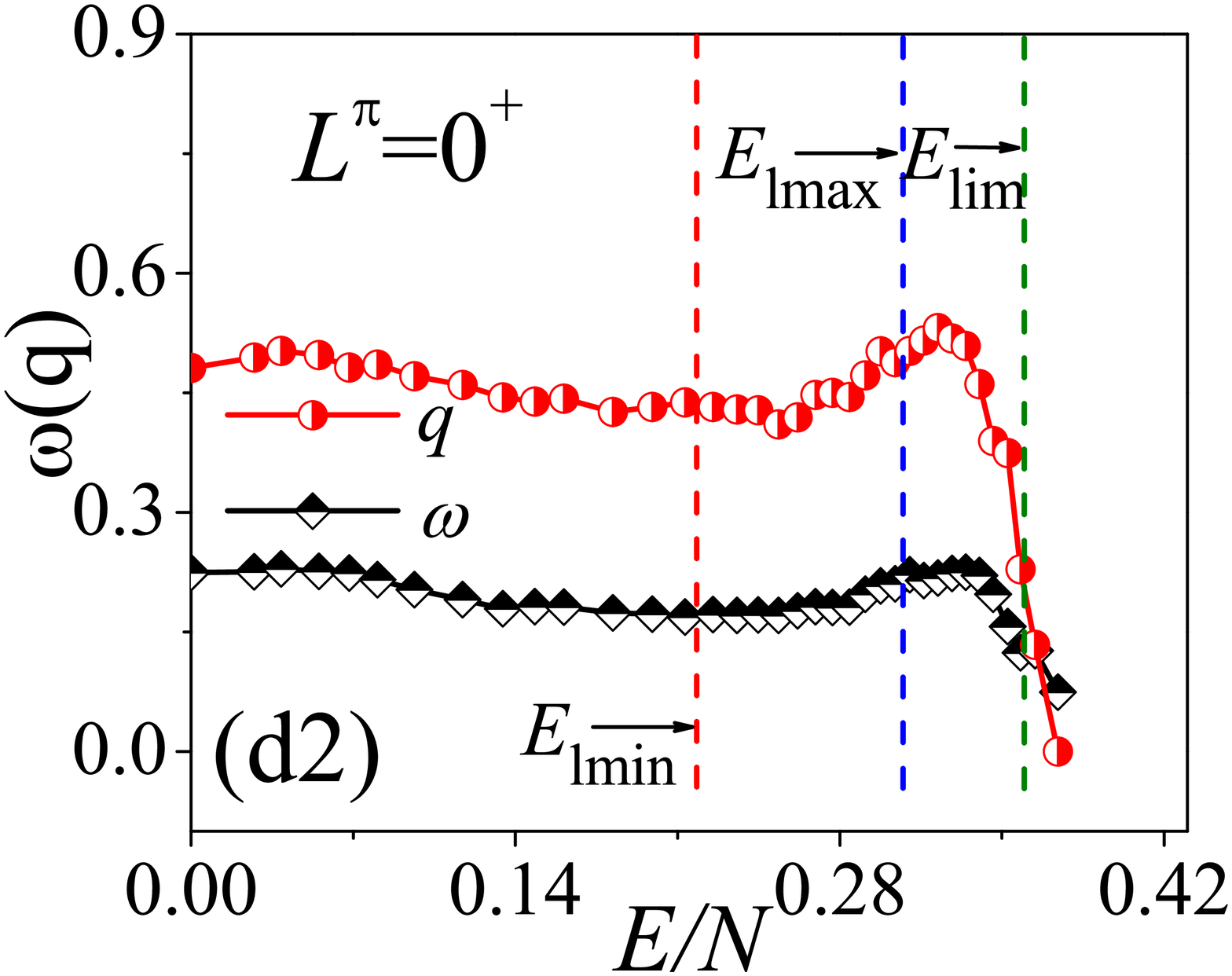}
\includegraphics[scale=0.16]{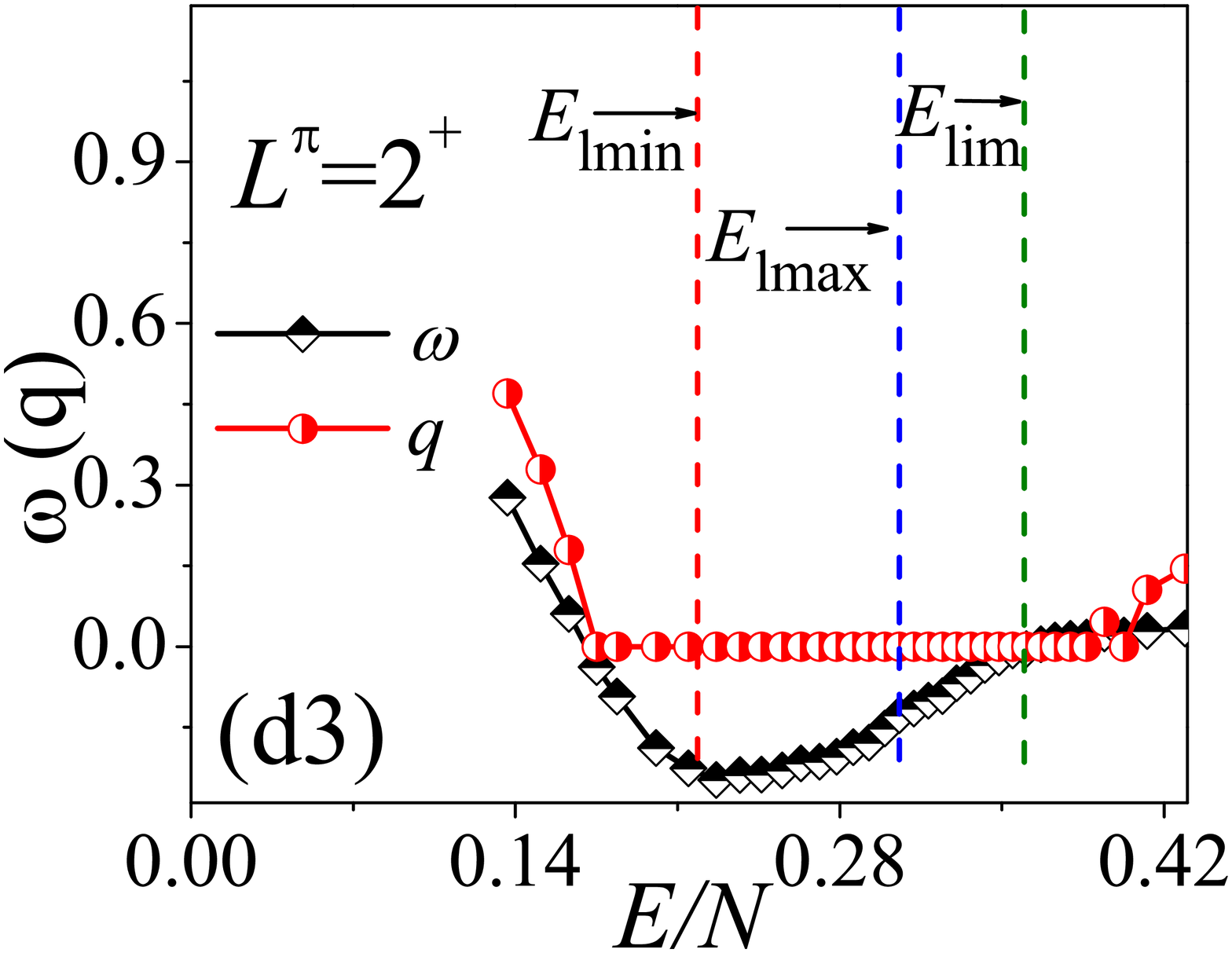}
\includegraphics[scale=0.16]{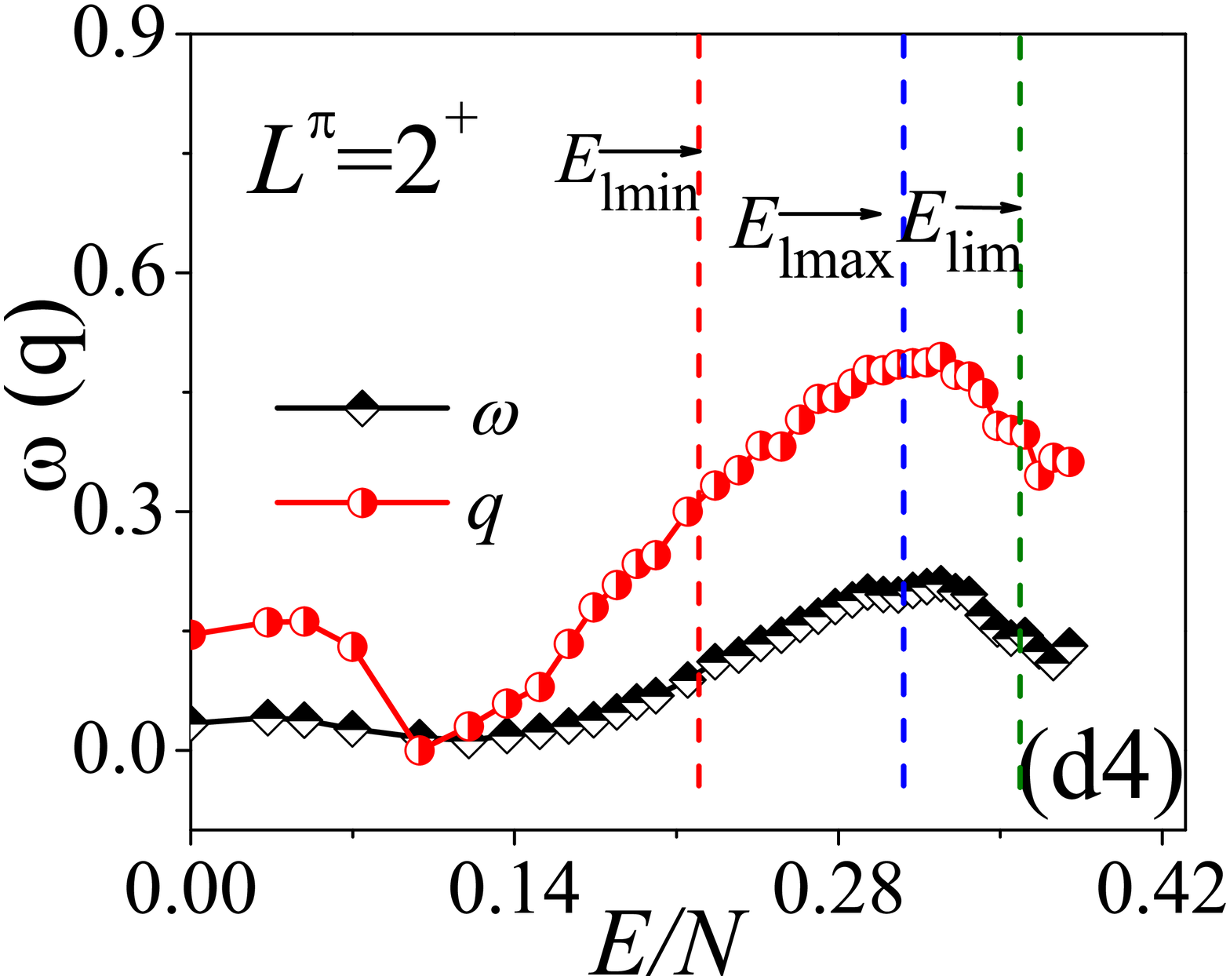}
\includegraphics[scale=0.16]{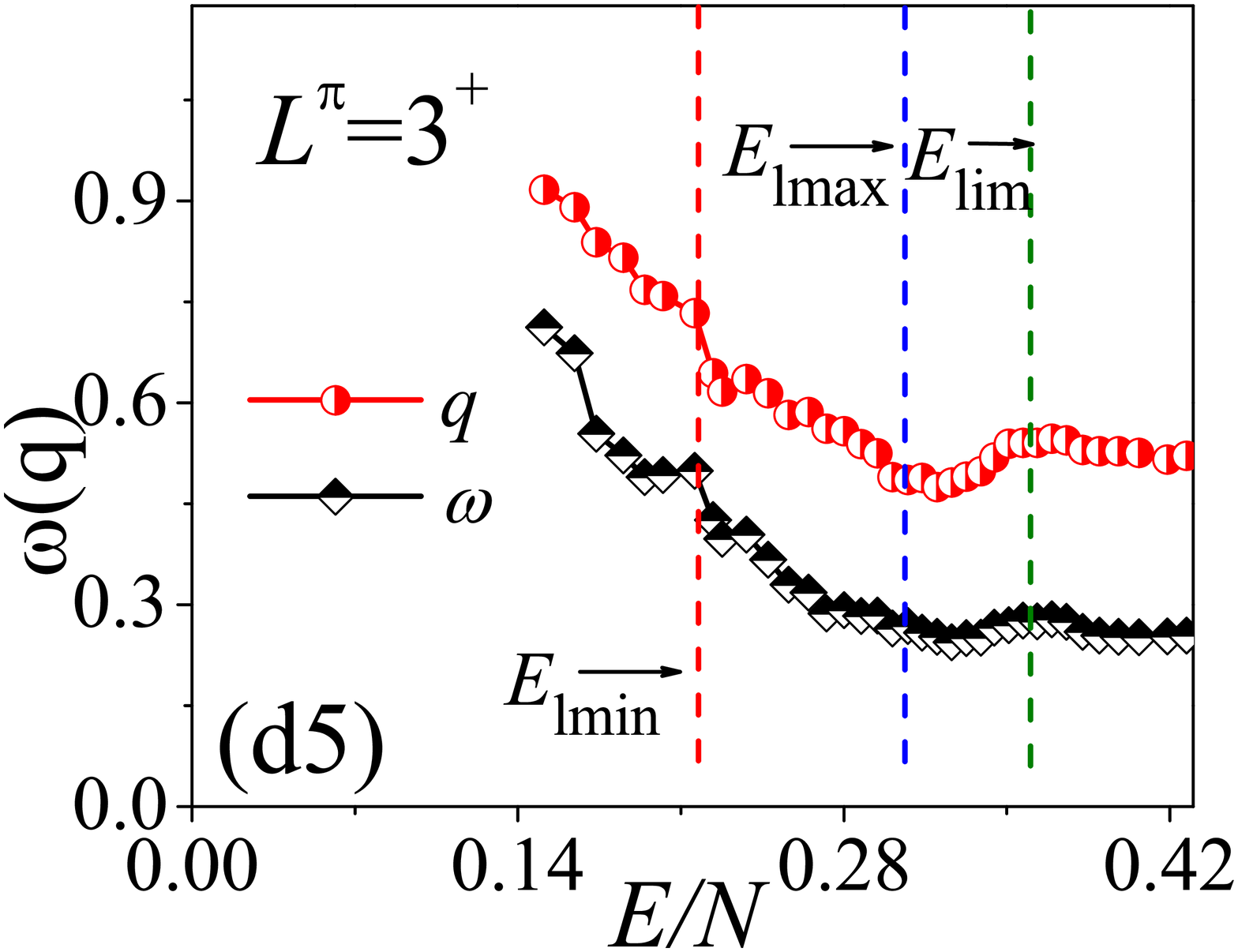}
\includegraphics[scale=0.16]{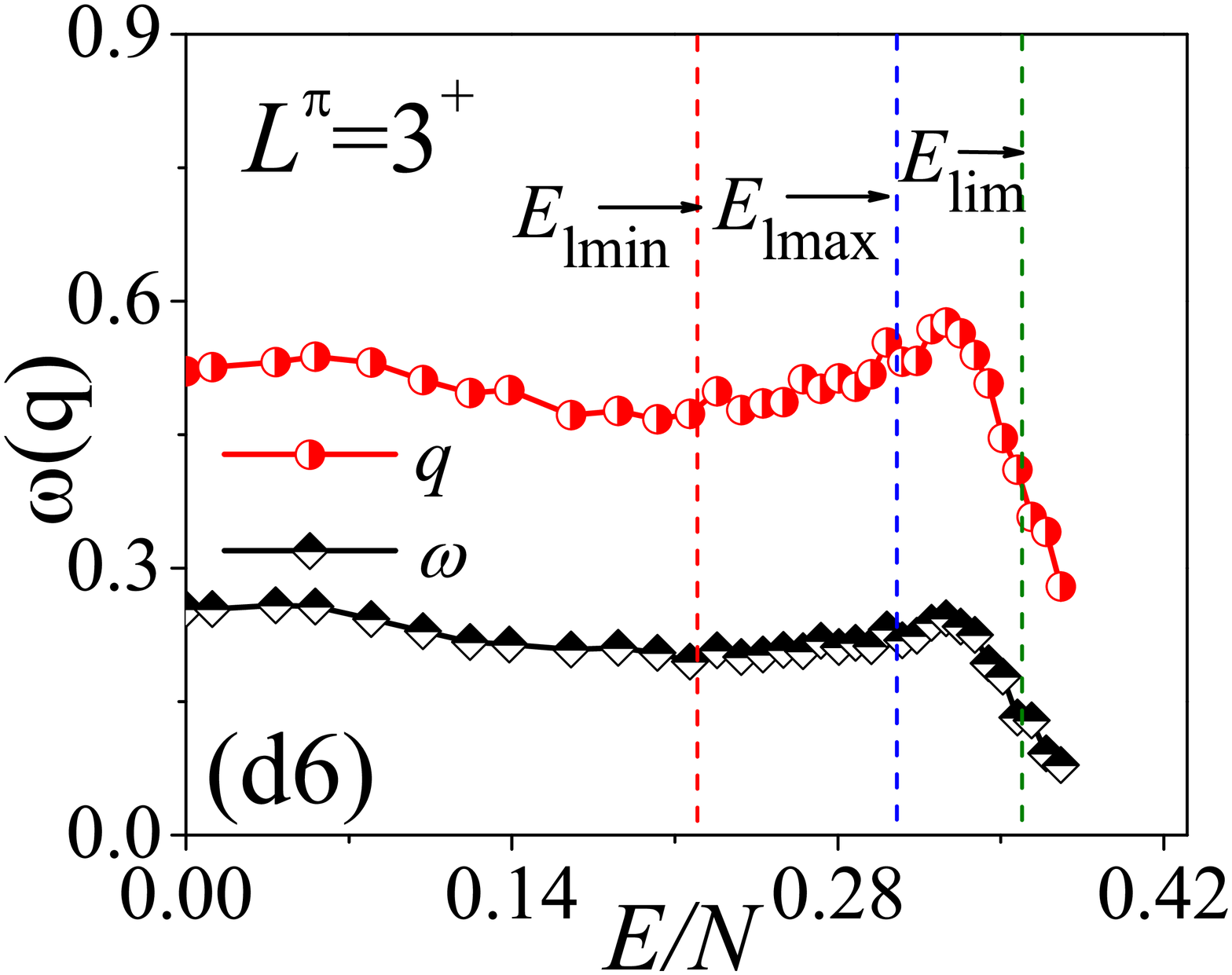}
\figcaption{The same as in Fig.~\ref{F2a} but those corresponding to
the parameter point D (on the AW arc).}\label{F2d}
\end{center}

One can find from the panels (d5) and (d6) that the energy
dependence of the spectral fluctuation for $L^\pi=3^+$ are very
similar to the one for $L^\pi=0^+$. In contrast, the spectral
fluctuations for $L^\pi=2^+$ exhibit some special features. An
impressive point is that $\omega$ will become negative in the energy
interval $[E_\mathrm{lmin},~E_\mathrm{lim}]$ with $q=0$ as shown in
the panel (d3). When doing the statistical calculations from high
energy to low energy, the $q=0$ results may also appear at very low
energy $E/N\sim0.1$ as shown in the panel (d4). According to the
convention, the $q$ values will be set by zero once the
$\Delta_3(L)$ results are larger than the Poisson limit. Similar to
the case in the SU(3) symmetry discussed in Sec.IIB, the results
$\omega<0$ and $q=0$ shown here may be also a result of
degeneracies. It is then deduced that the degeneracies (approximate)
in the $2^+$ spectrum may be associated with the SU(3)
quasidynamical symmetry~\cite{Bonatsos2010,Bonatsos2011} since its
parameter trajectory characterized by $E(2_\beta^+)=E(2_\gamma^+)$
was found to be very close to the AW arc therefore suggesting a
symmetry-based interpretation of the AW arc. The present results
indicate that such a symmetry-based explanation of the AW arc can be
tested from the point of statistics on the $2^+$ states. However, it
is not easy to point out which $2^+$ states are approximately
degenerate by directly observing their level energies in a
symmetry-broken case. In addition, the validity of the underlying
SU(3) quasidynamical symmetry inside was suggested to be limited to
low-lying states in the large-$N$ limit~\cite{Bonatsos2010}. So,
whether or not the negative $\omega$ values and $q=0$ can be fully
explored by the approximate degeneracies needs to be further
studied, which may be discussed elsewhere.

\begin{center}
\includegraphics[scale=0.25]{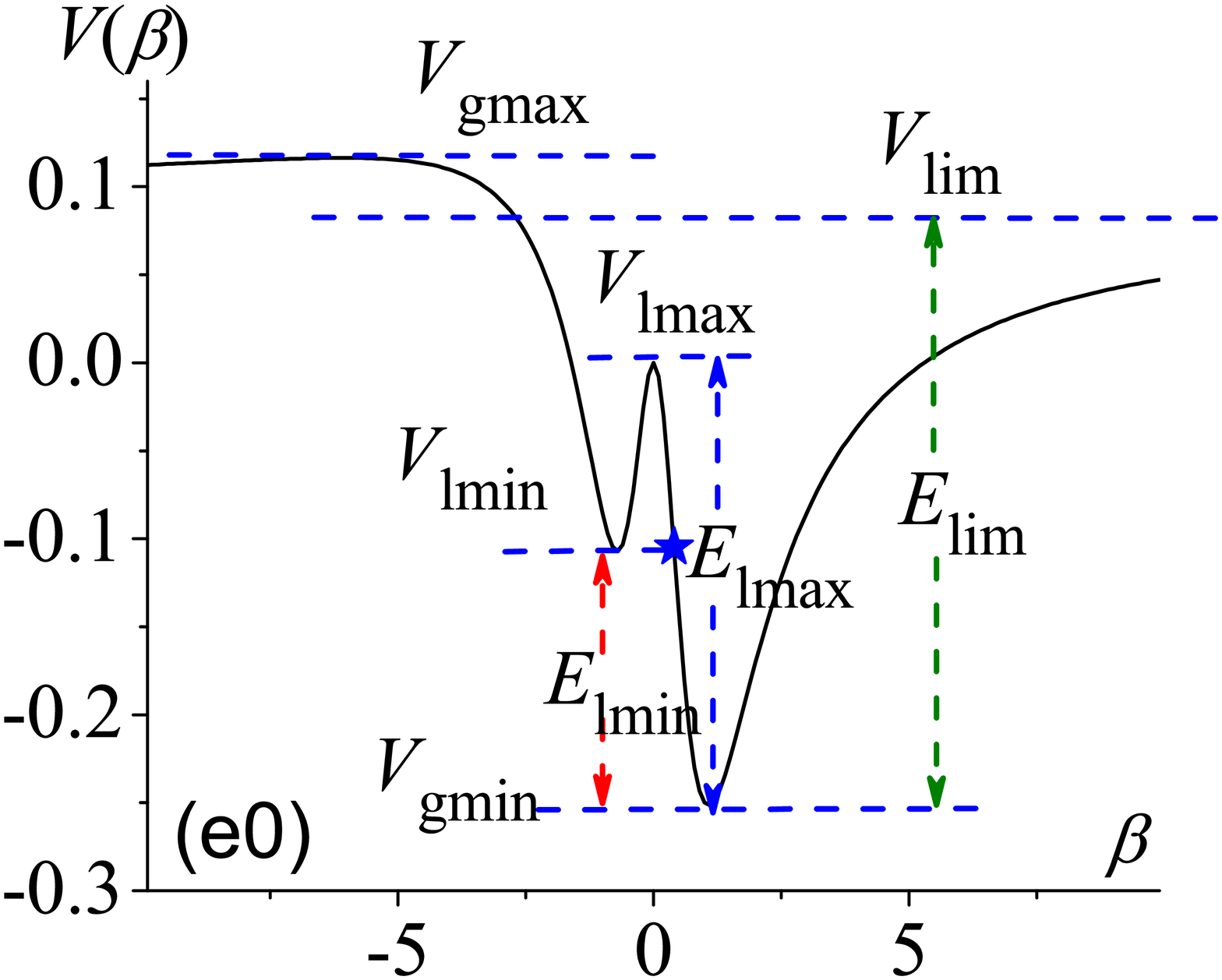}
\includegraphics[scale=0.16]{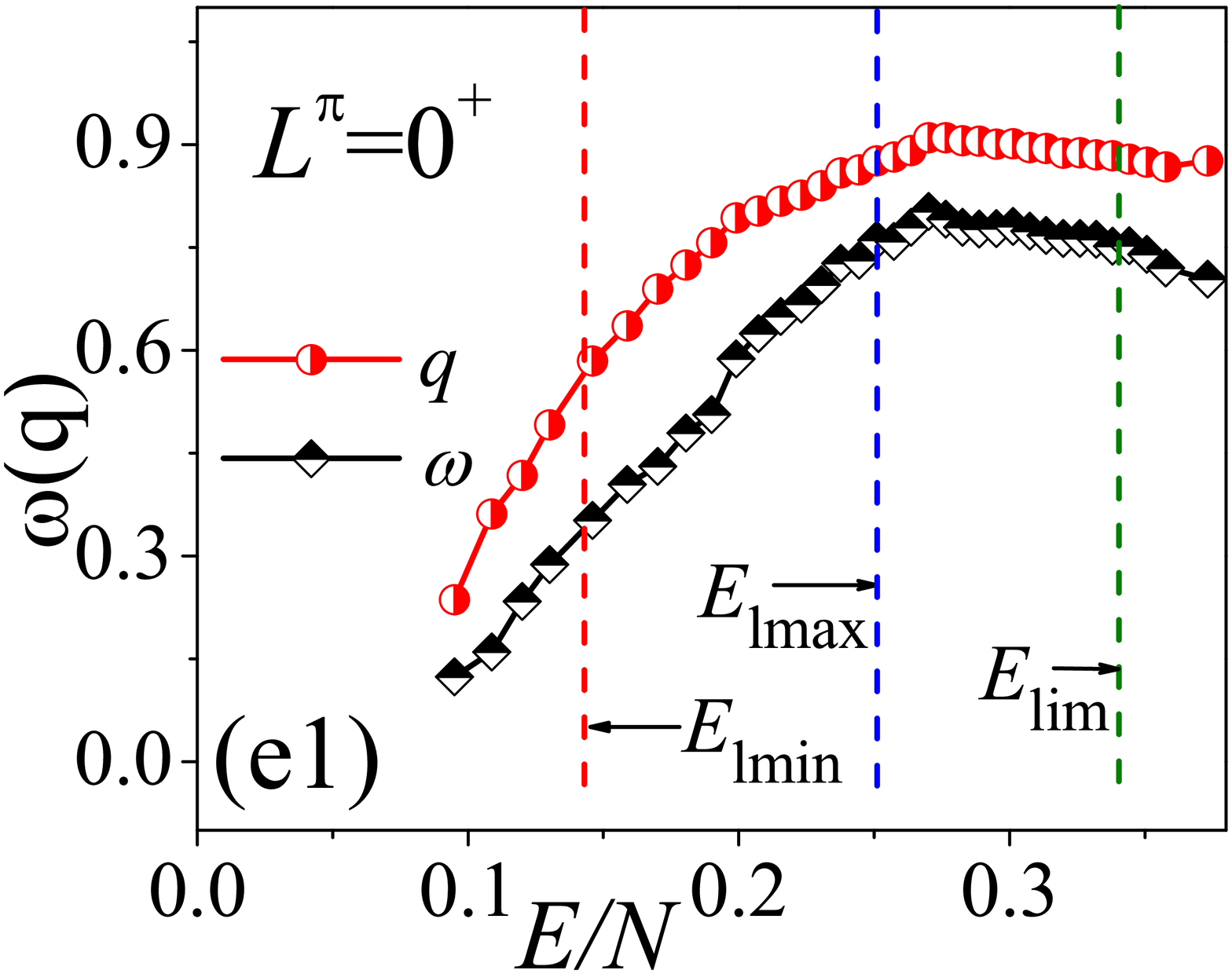}
\includegraphics[scale=0.16]{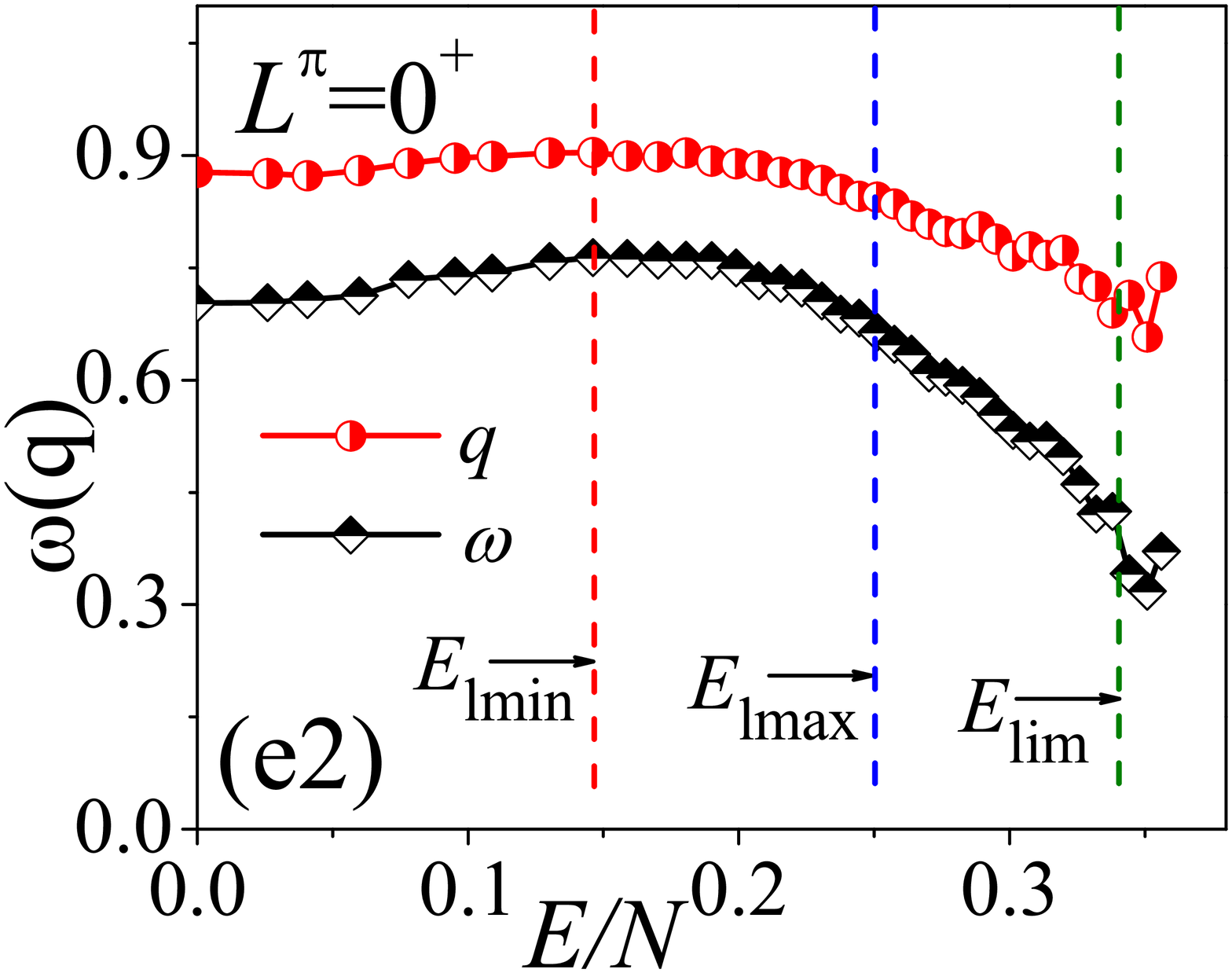}
\includegraphics[scale=0.16]{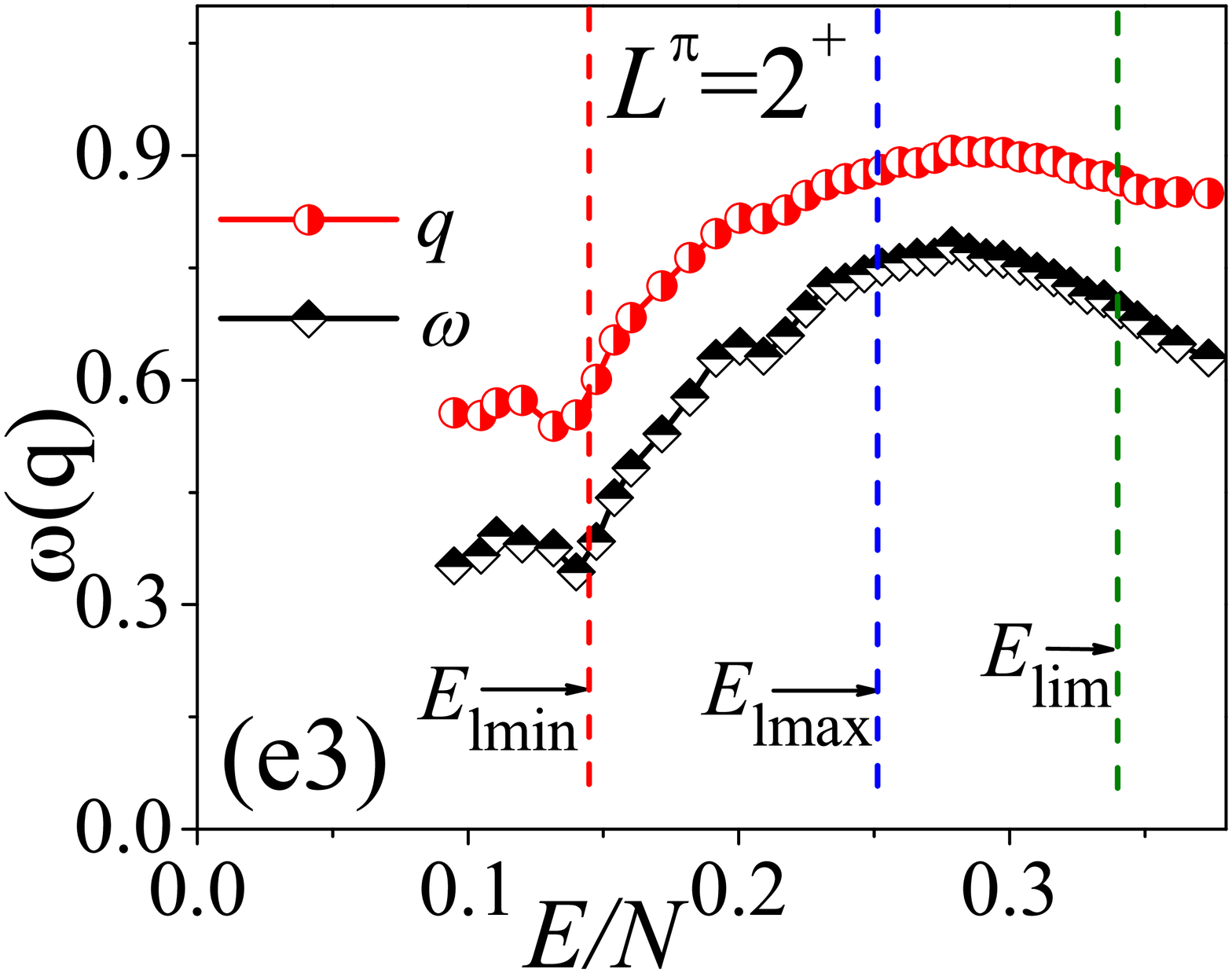}
\includegraphics[scale=0.16]{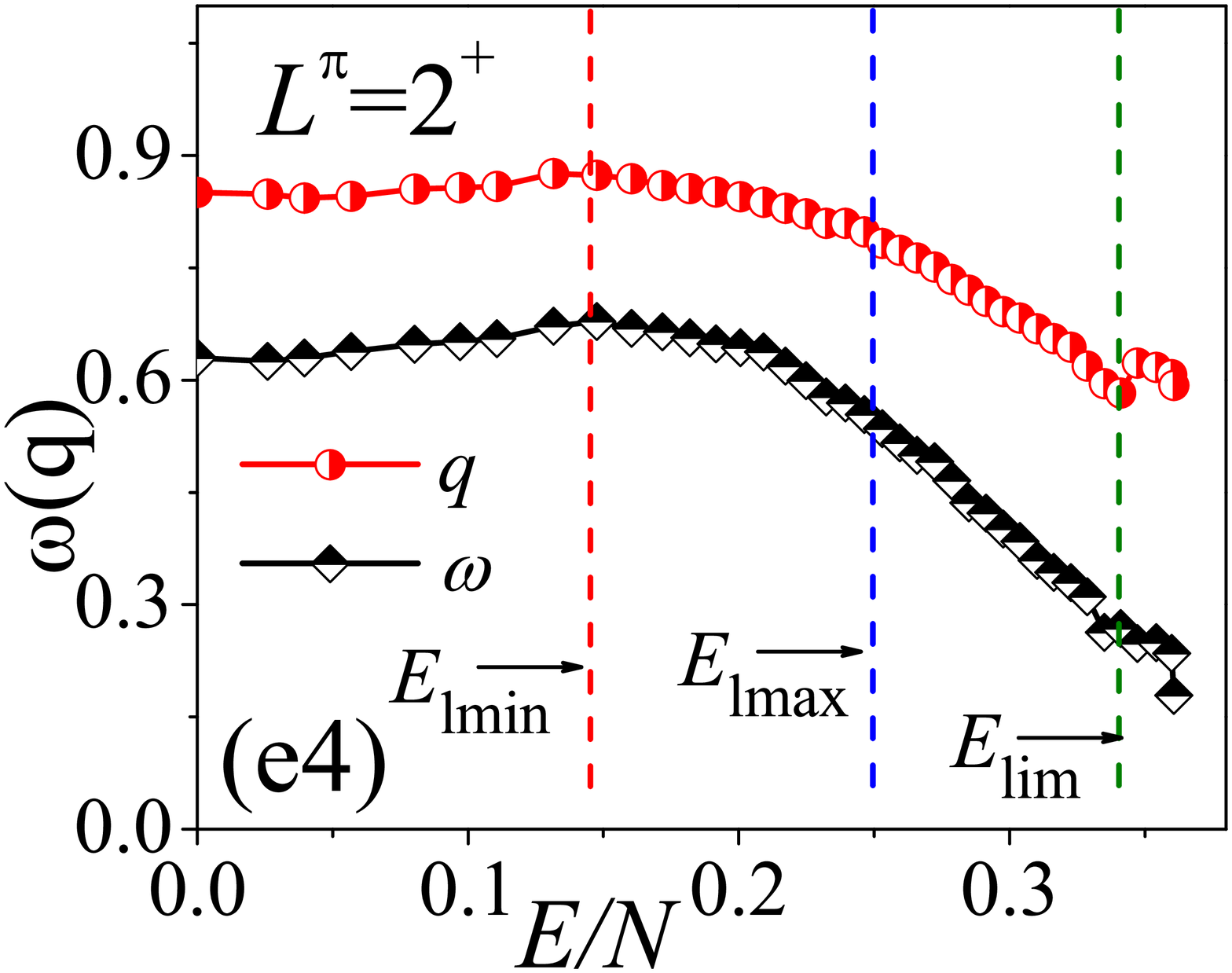}
\includegraphics[scale=0.16]{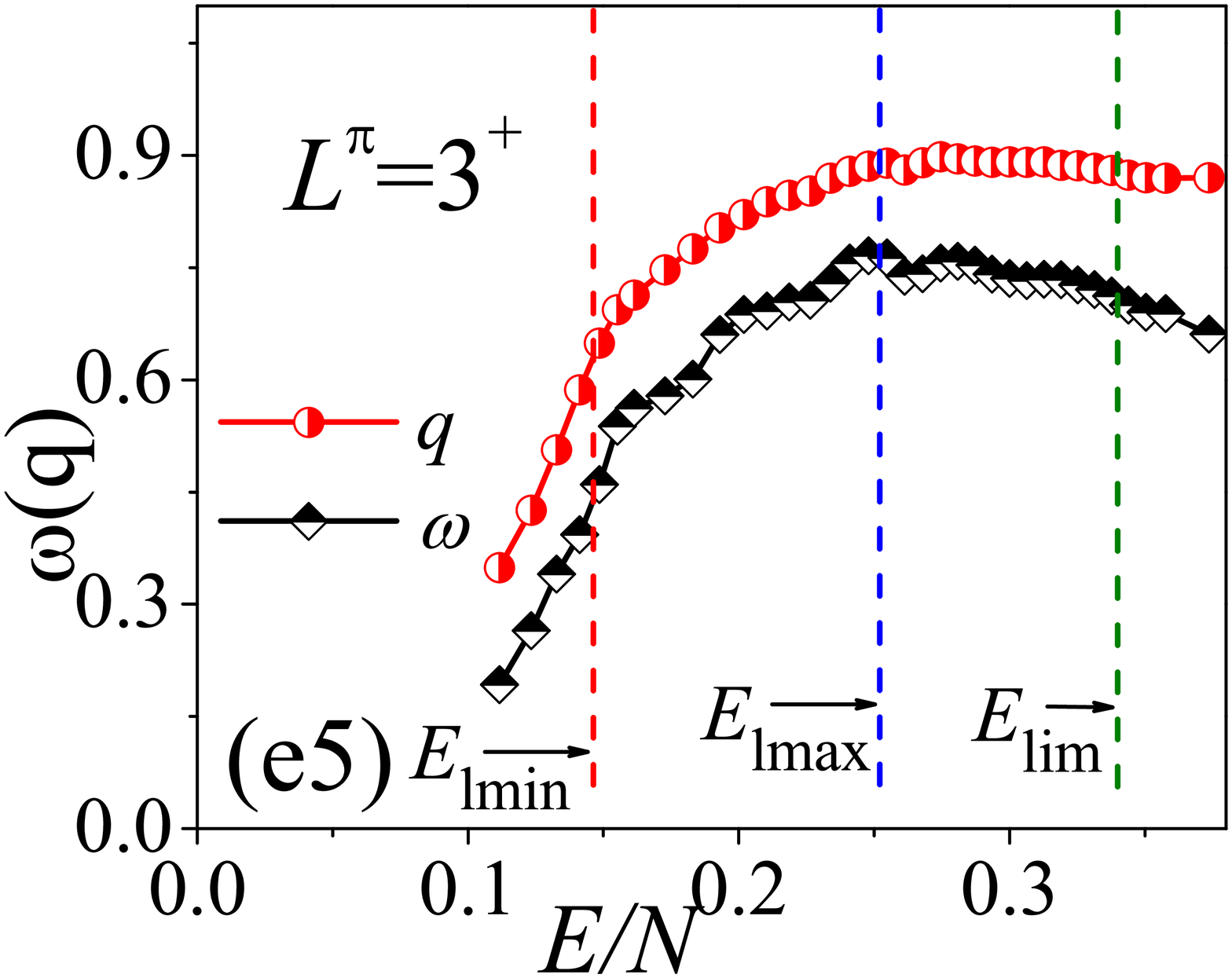}
\includegraphics[scale=0.16]{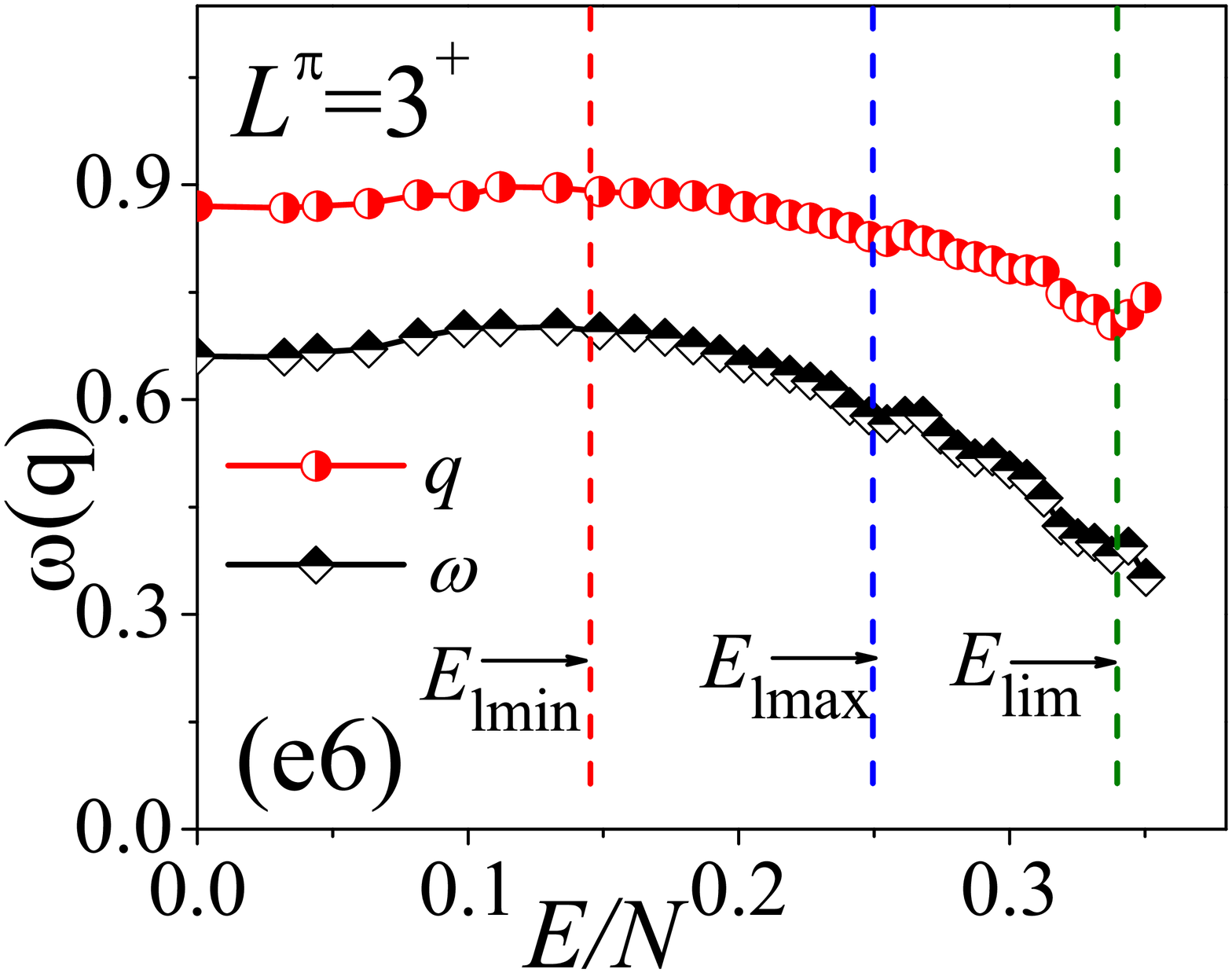}
\figcaption{The same as in Fig.~\ref{F2a} but for those
corresponding to the parameter point E.}\label{F2e}
\end{center}

As for the situation off the AW arc, one can find from
Fig.~\ref{F2e} that the potential curve at the point E is very
similar to the one at the point D. Both of them have three
stationary points lying in between $V_\mathrm{gmax}$ and
$V_\mathrm{gmin}$. However, the spectral fluctuations in the two
cases are very different and even evolve in the opposite directions.
First of all, the spectra at the point E are shown to be much more
chaotic than those on the arc. For example, the $P(S)$ statistics on
all the states with $L^\pi=2^+$ give $\omega\sim0$ and $q\sim0$ for
the point D but $\omega>0.6$ and $q>0.8$ for the point E. In
addition, the results given in the panels (e1), (e3) and (e5)
indicate that spectral fluctuations at the point E may increase with
the excitation energy until $E_{\mathrm{lmax}}$ and then turn to a
slow decrement. A notable point is that the $\omega(q)$ values for
$L^\pi=2^+$ exhibit a sudden enhancement around $E_\mathrm{lmin}$ as
seen in the panel (e3). All these features just reflect the effects
of the stationary points on the spectral fluctuations at the point
E. In contrast, the relatively smoother evolutions can be observed
if doing the statistical calculations from high to low energy as
given in the panels (e2), (e4) and (e6).

\begin{center}
\includegraphics[scale=0.165]{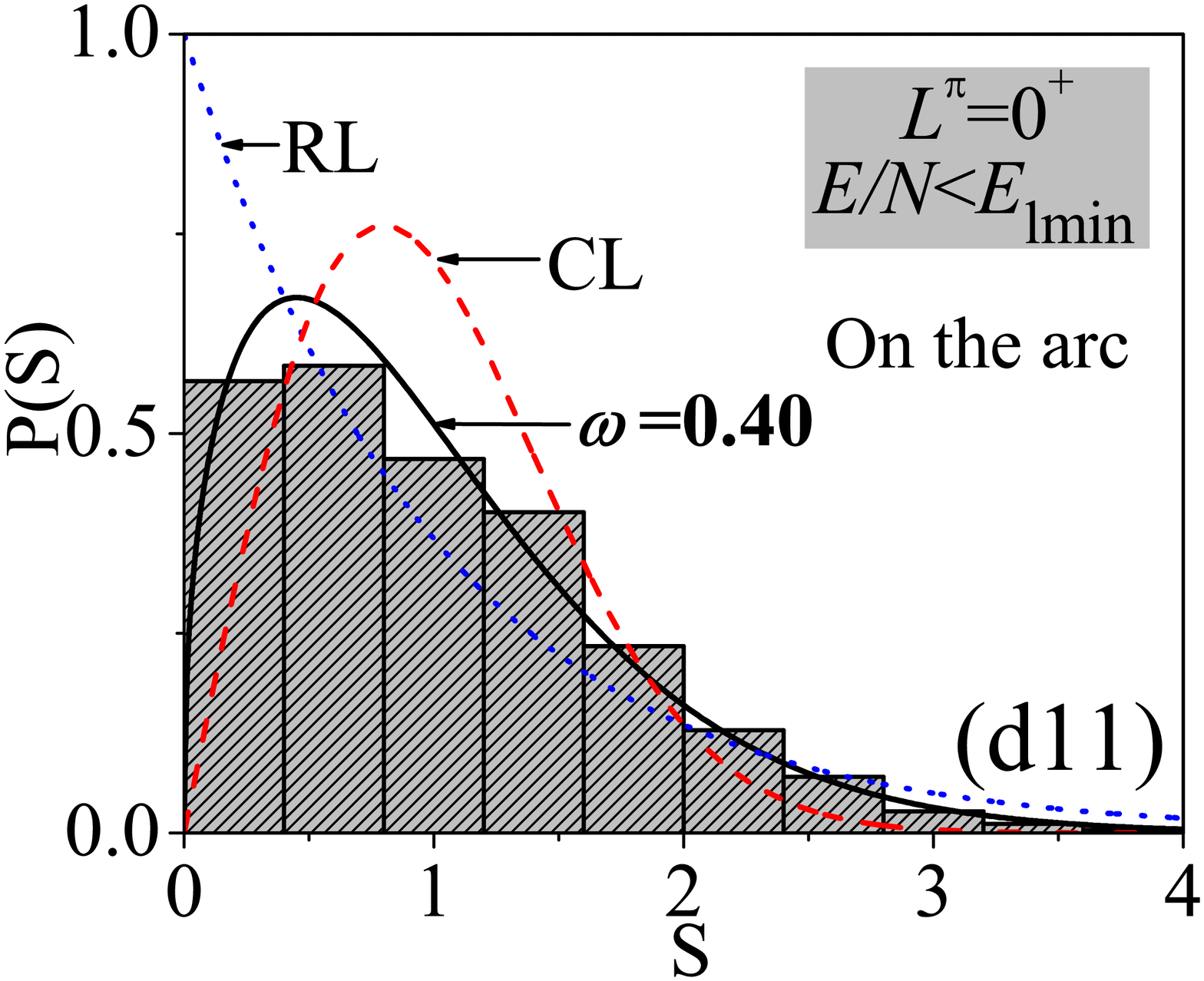}
\includegraphics[scale=0.165]{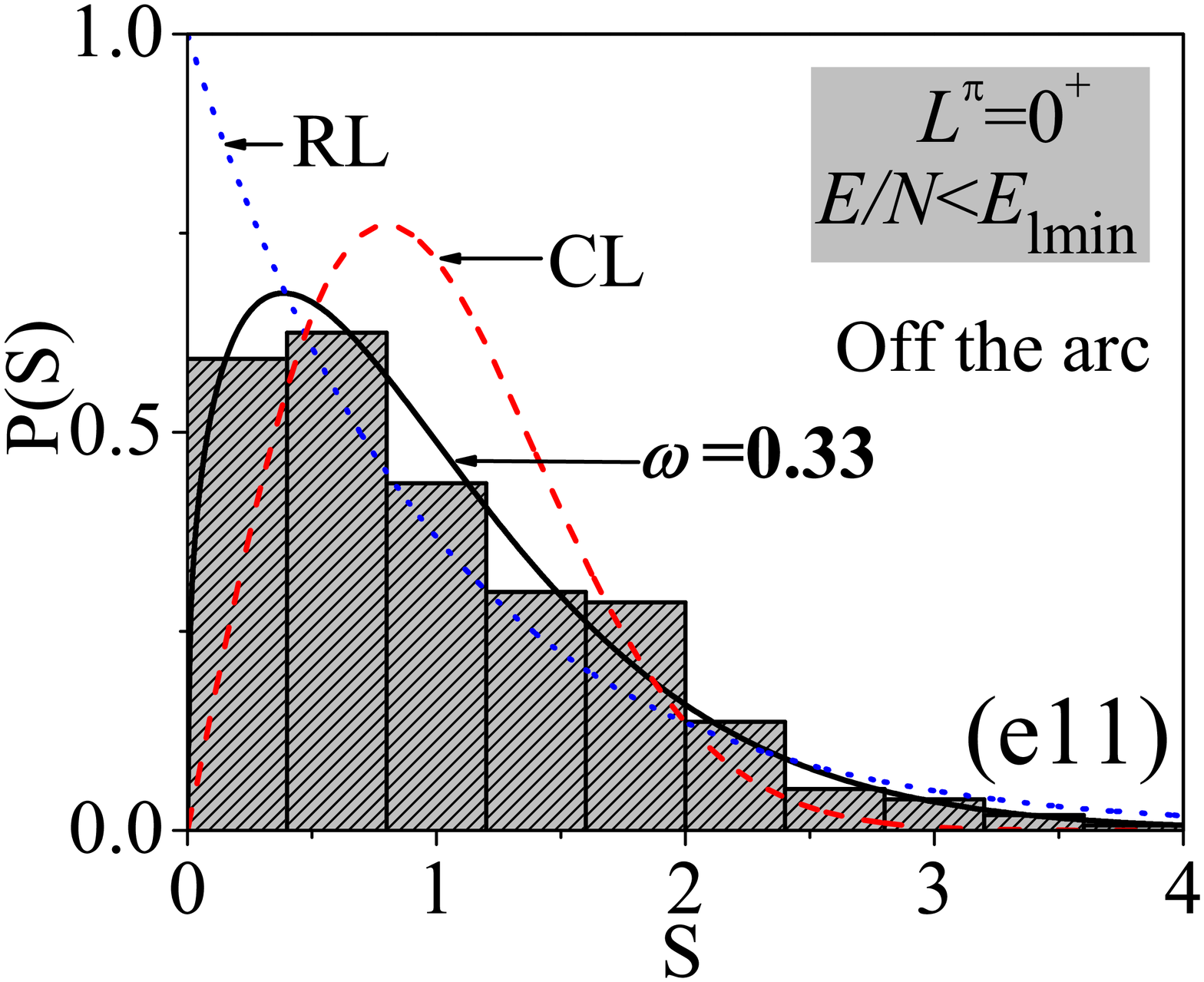}
\includegraphics[scale=0.165]{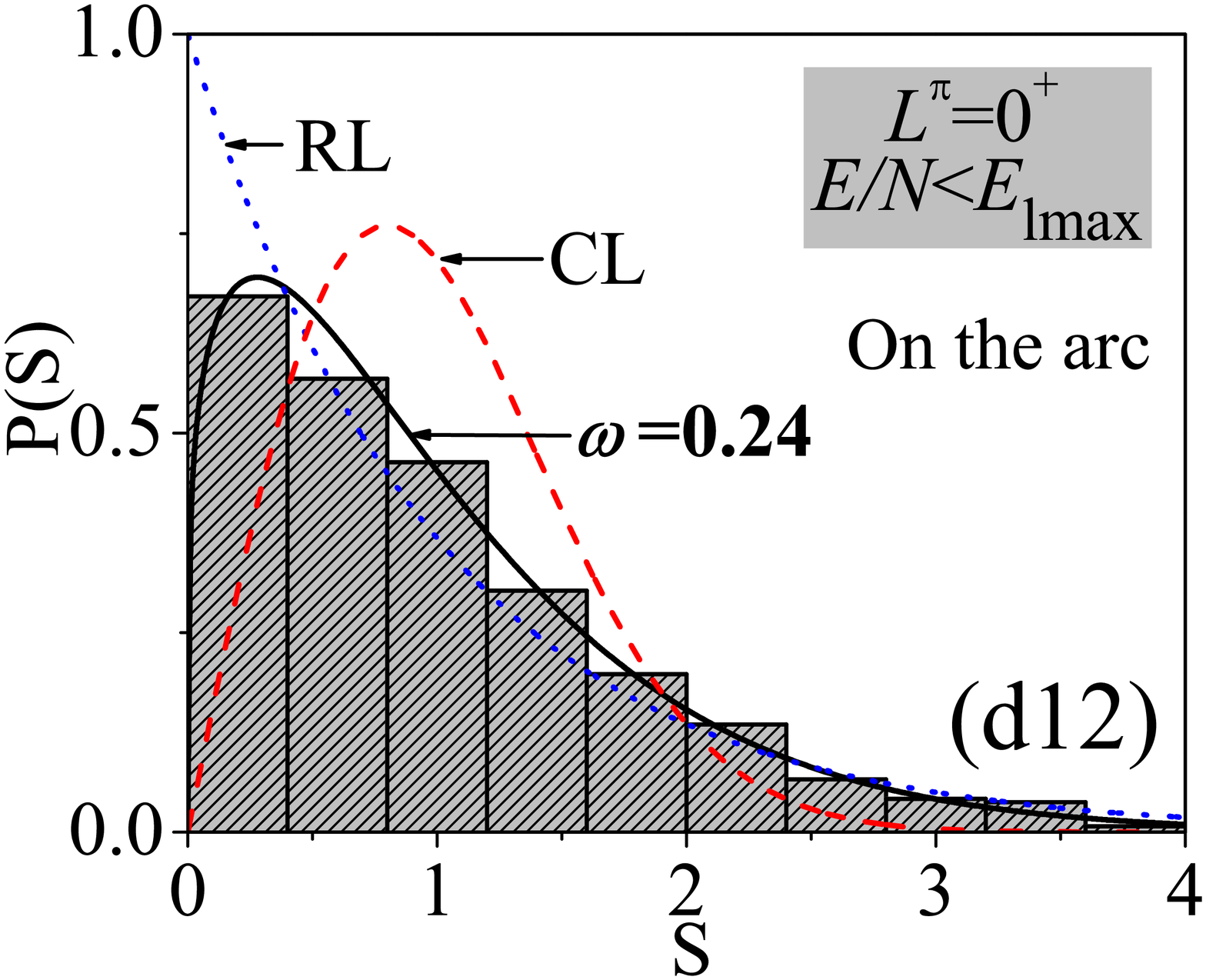}
\includegraphics[scale=0.165]{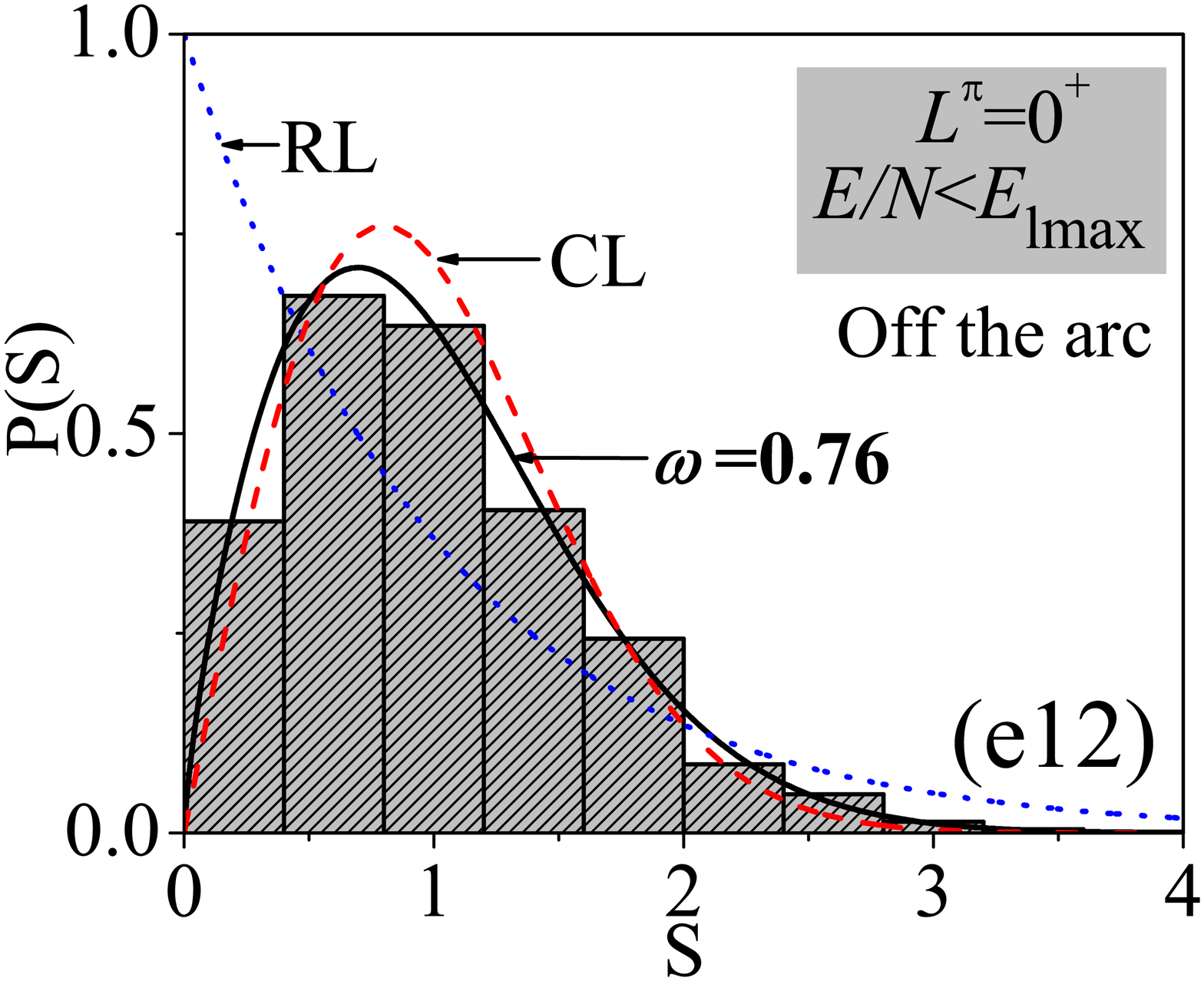}
\includegraphics[scale=0.165]{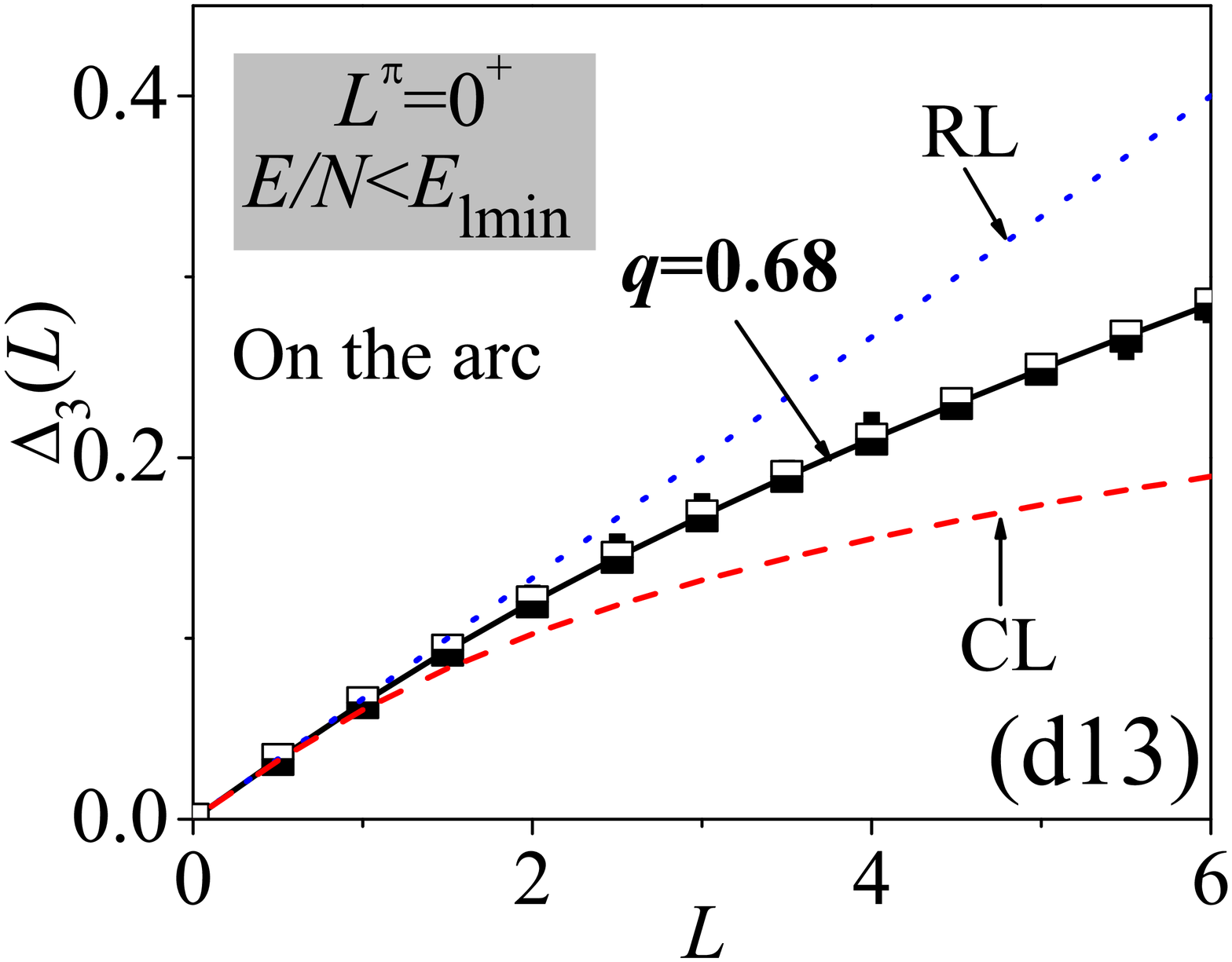}
\includegraphics[scale=0.165]{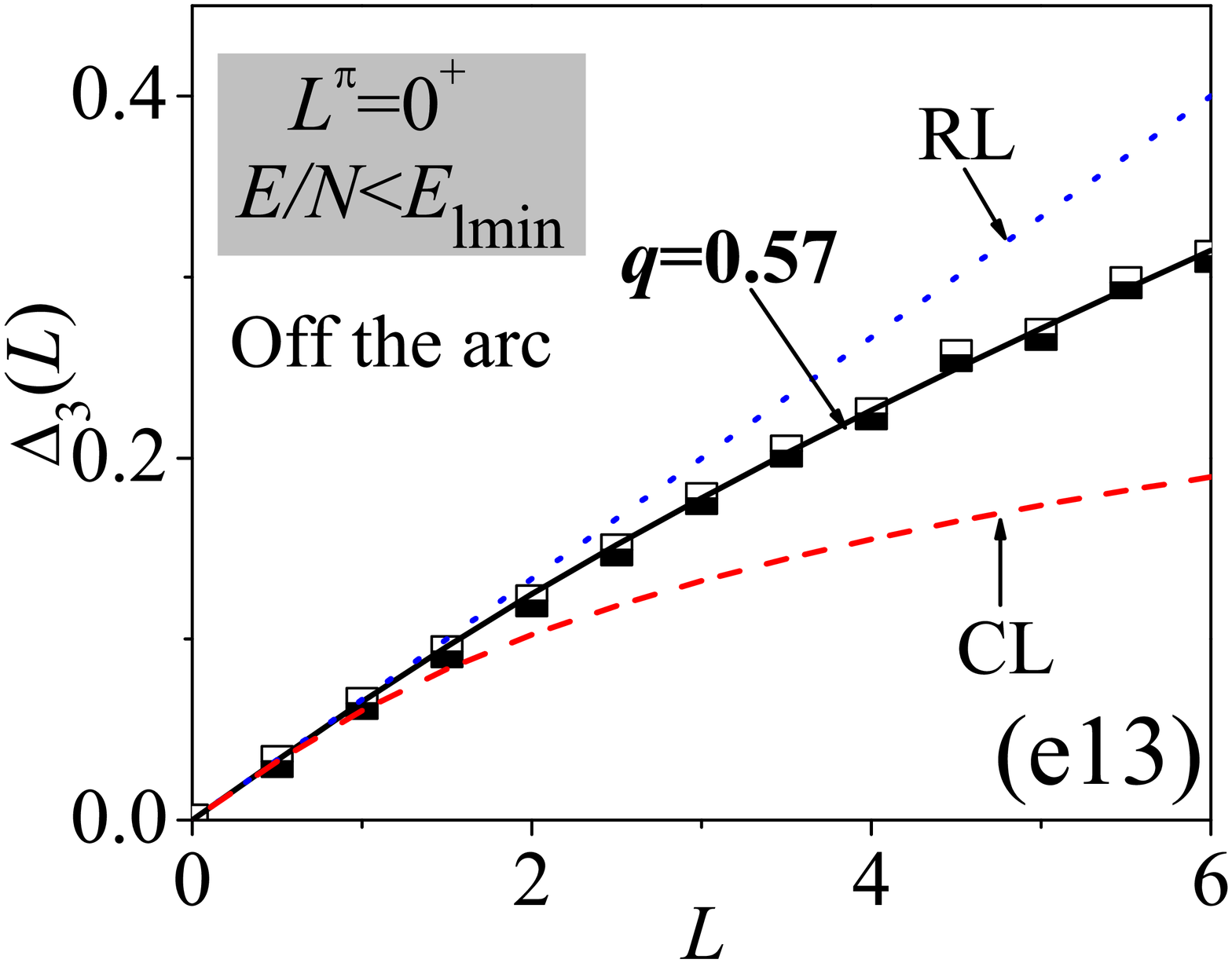}
\includegraphics[scale=0.165]{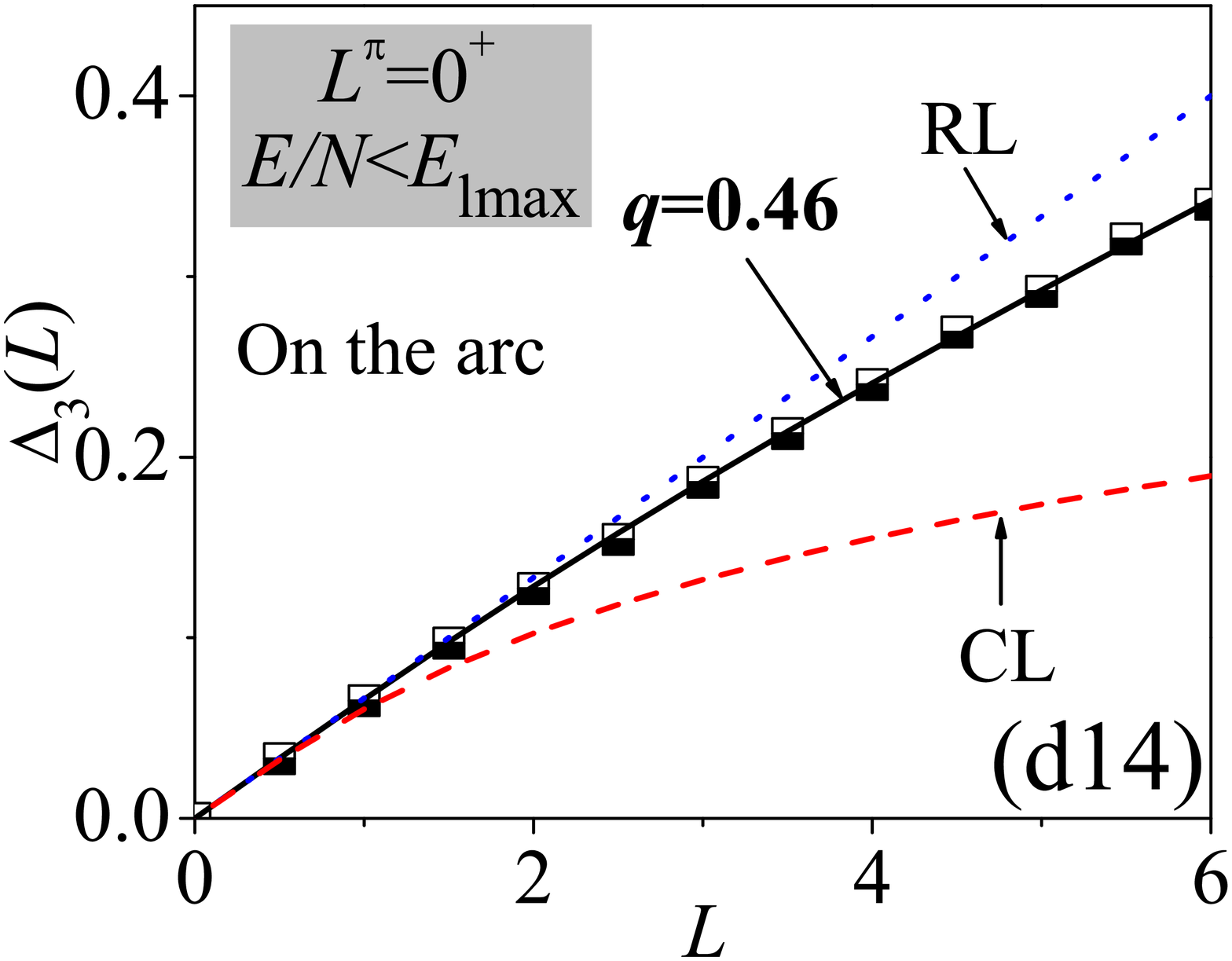}
\includegraphics[scale=0.165]{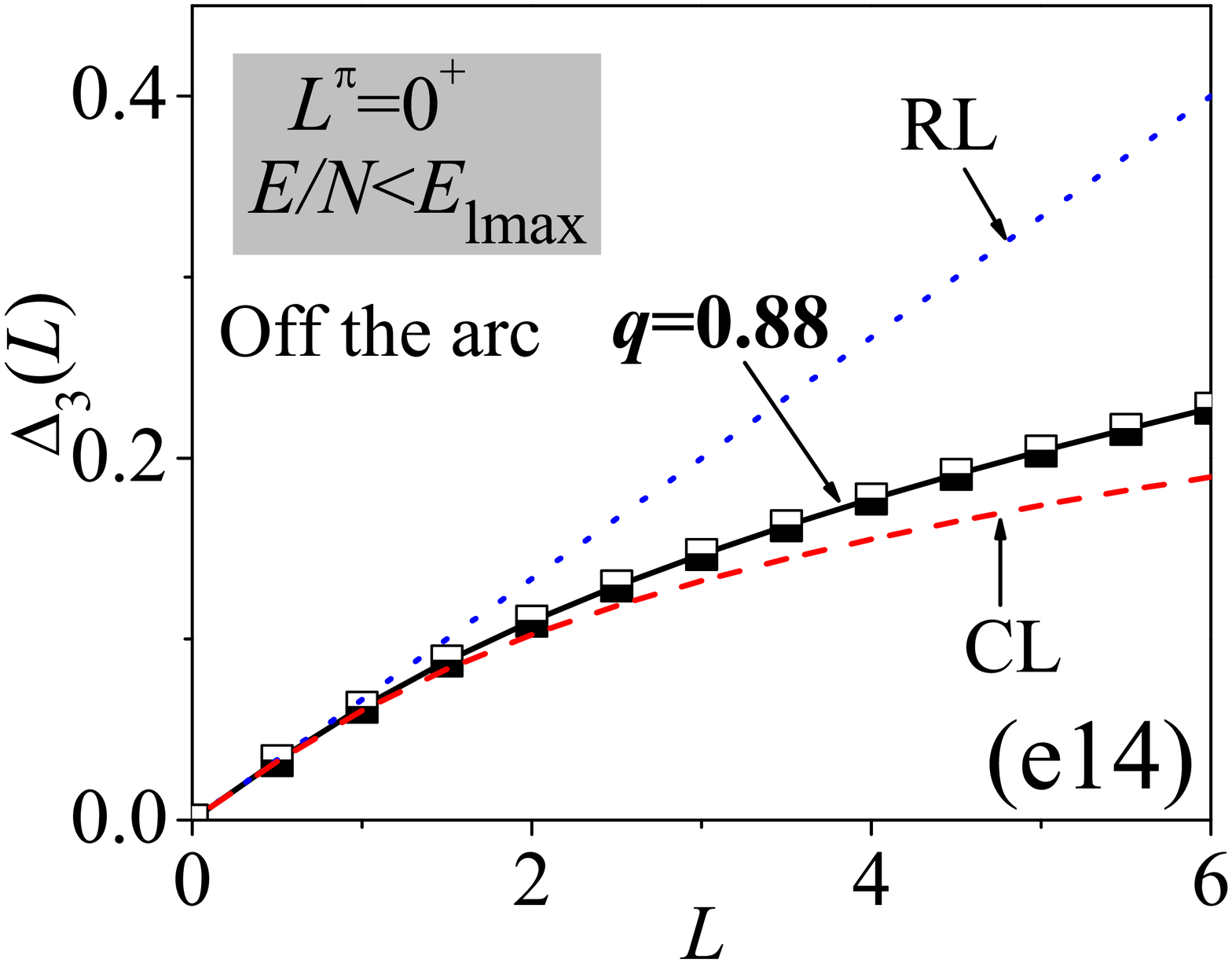}
\figcaption{To make a close comparison between the point D (on the
arc) and the point E (off the arc), the statistics for the $0^+$
levels bound below $E_\mathrm{lmin}$ and $E_\mathrm{lmax}$ in the
two cases are shown.}\label{F2de1}
\end{center}

\begin{center}
\includegraphics[scale=0.165]{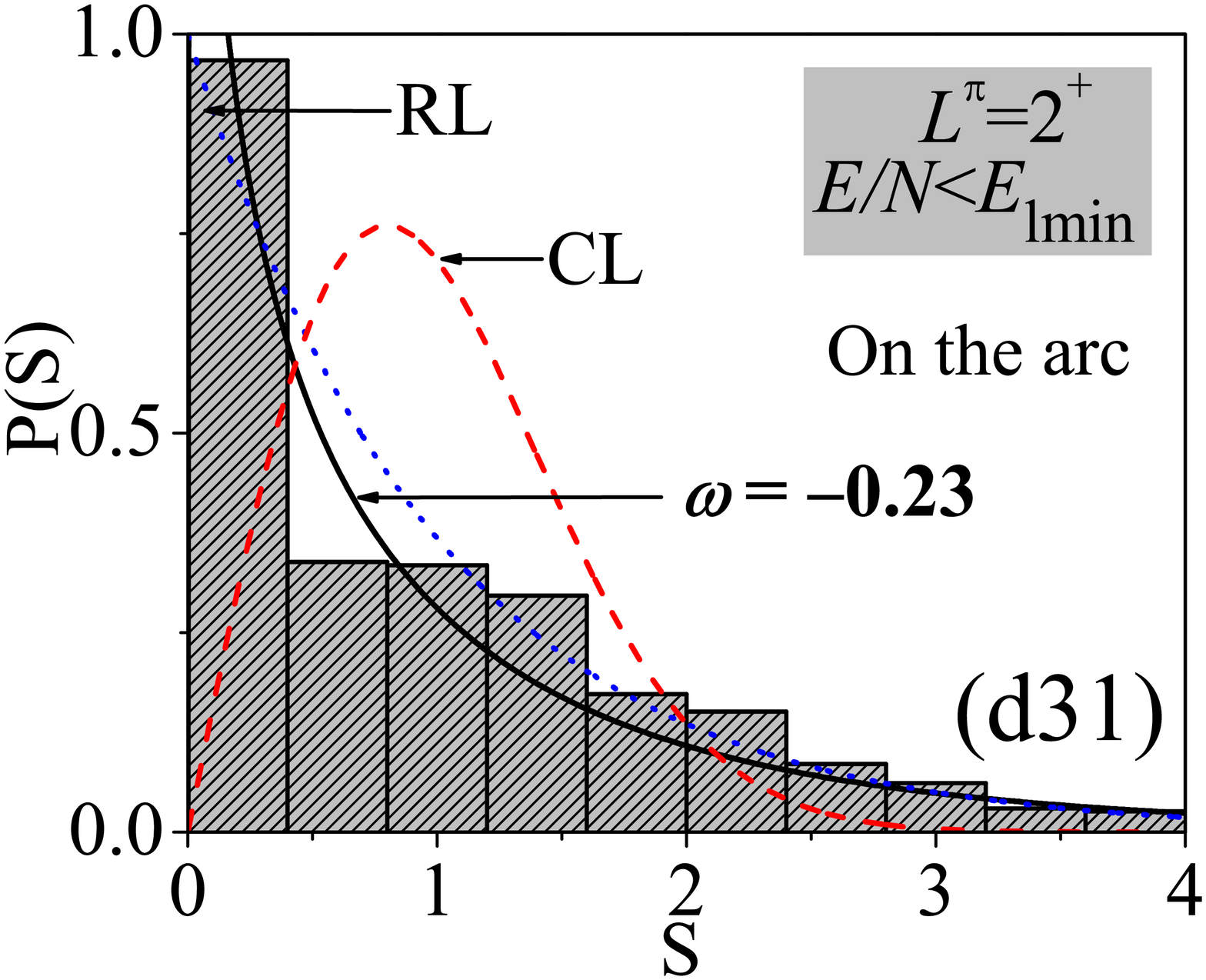}
\includegraphics[scale=0.165]{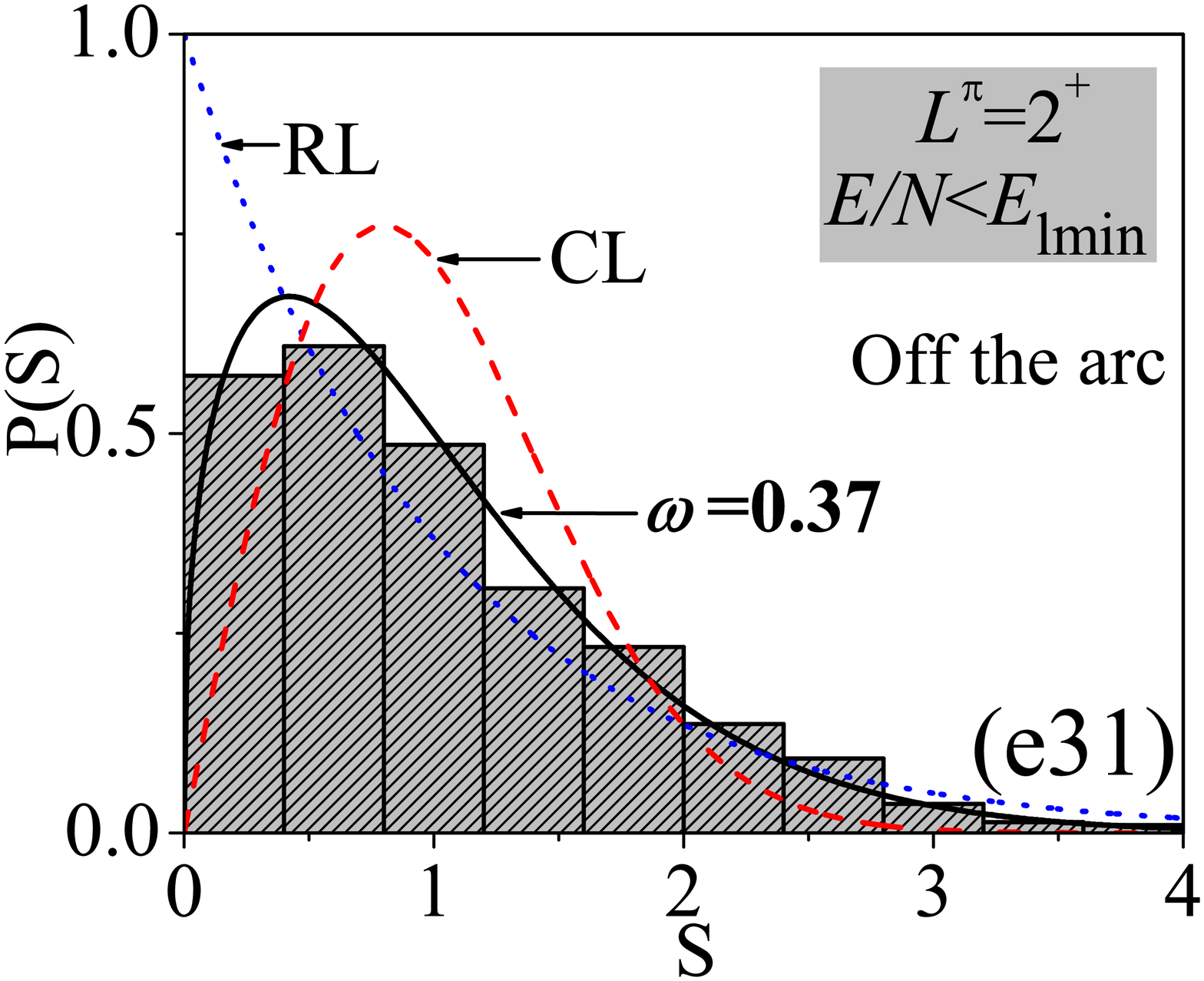}
\includegraphics[scale=0.165]{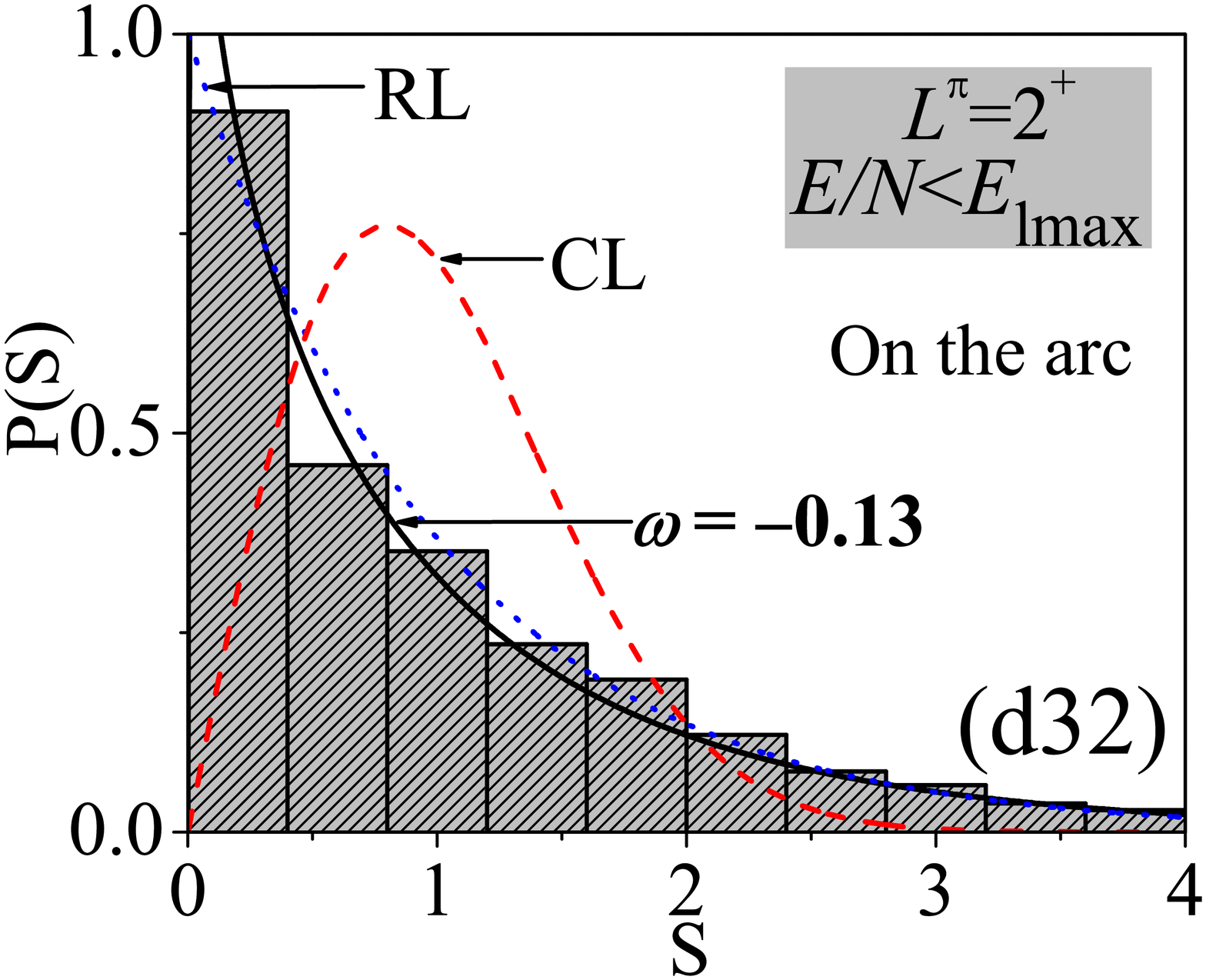}
\includegraphics[scale=0.165]{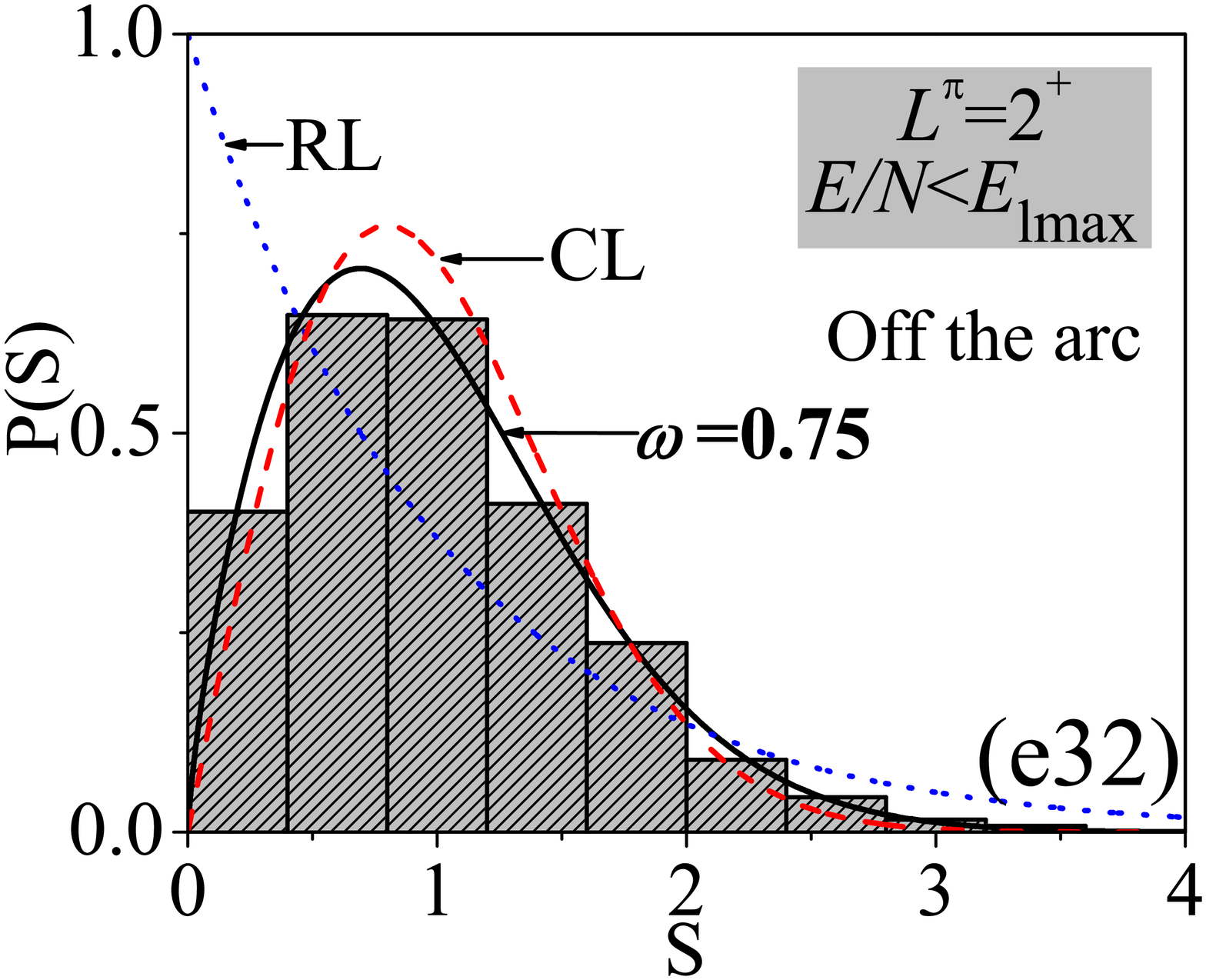}
\includegraphics[scale=0.165]{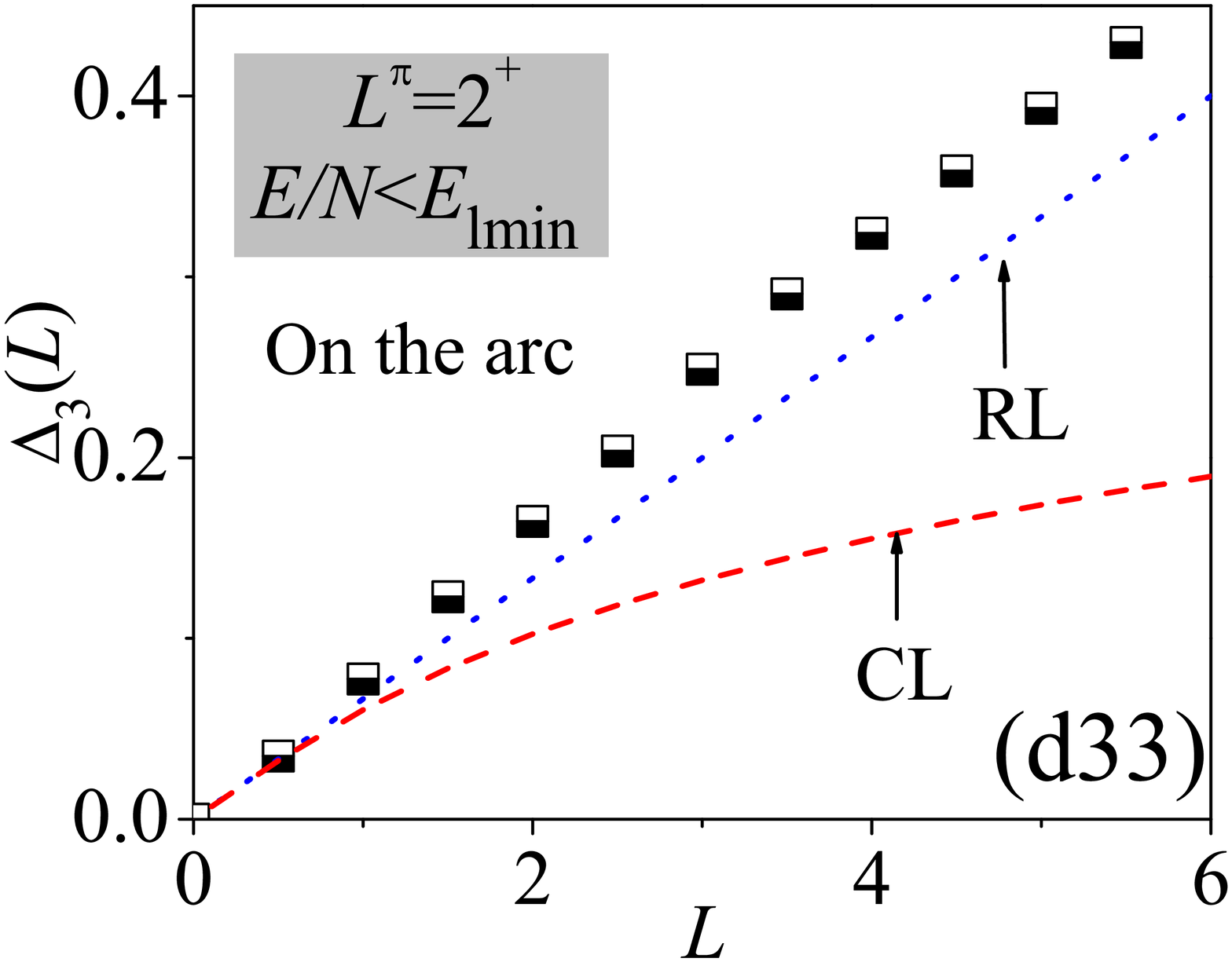}
\includegraphics[scale=0.165]{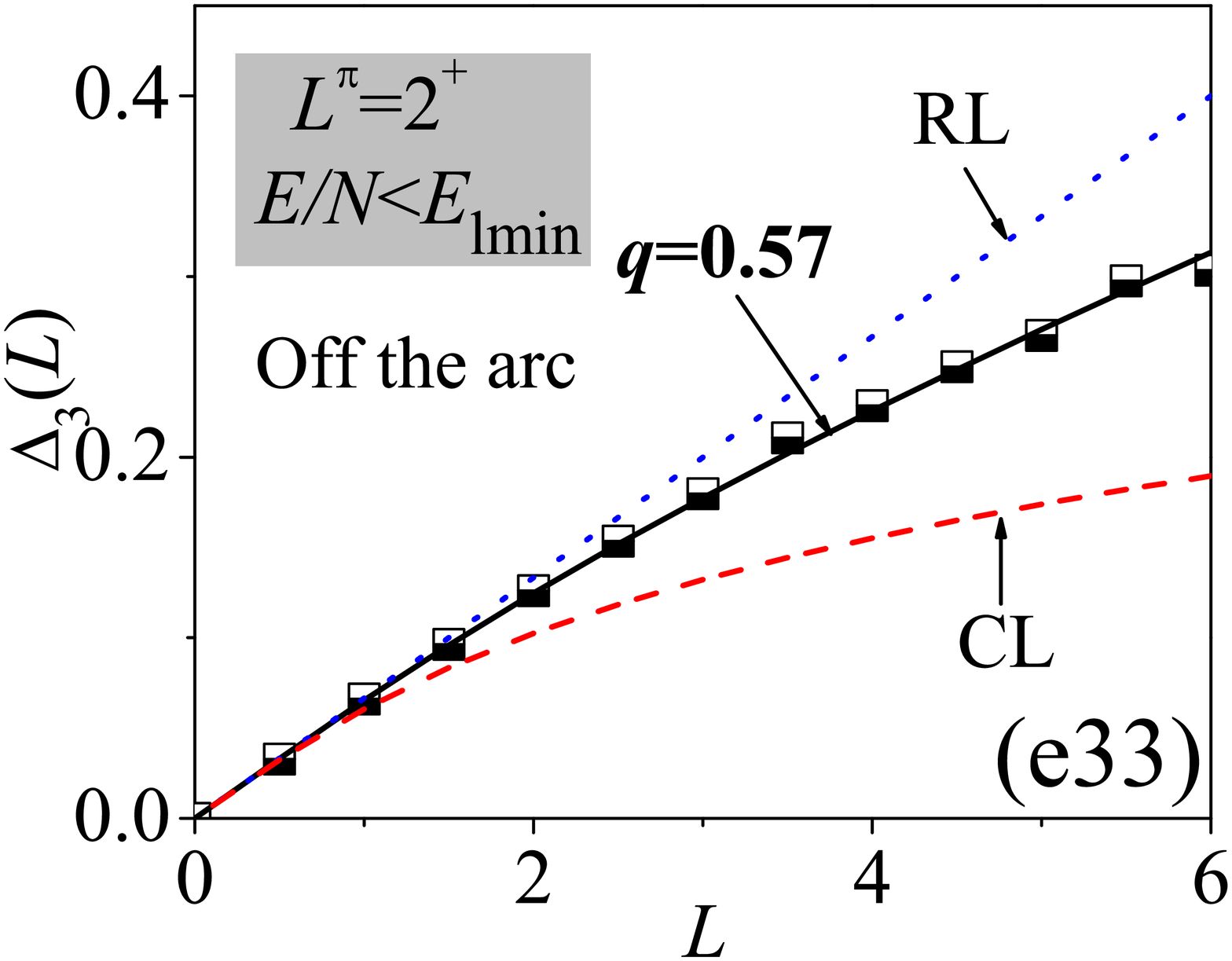}
\includegraphics[scale=0.165]{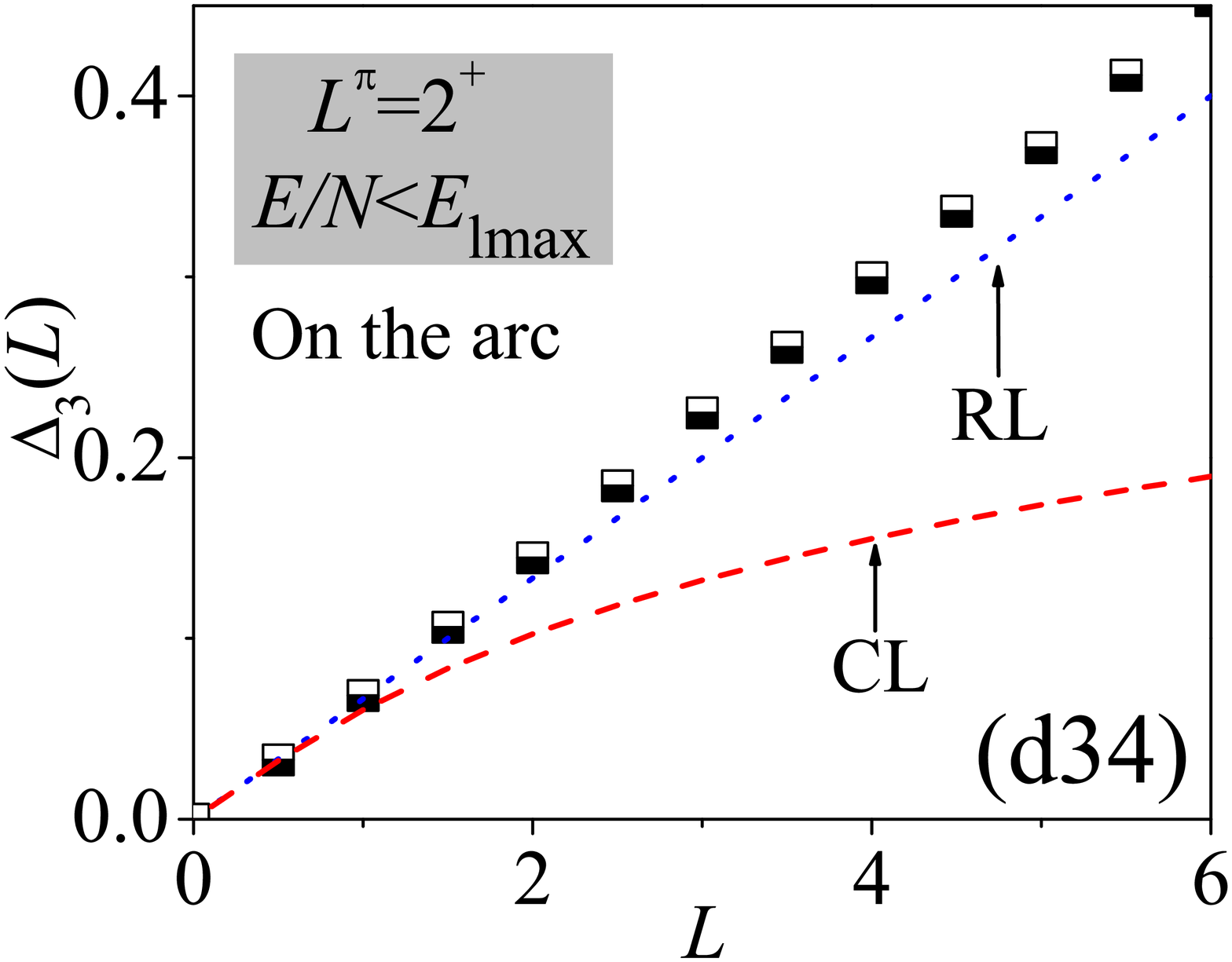}
\includegraphics[scale=0.165]{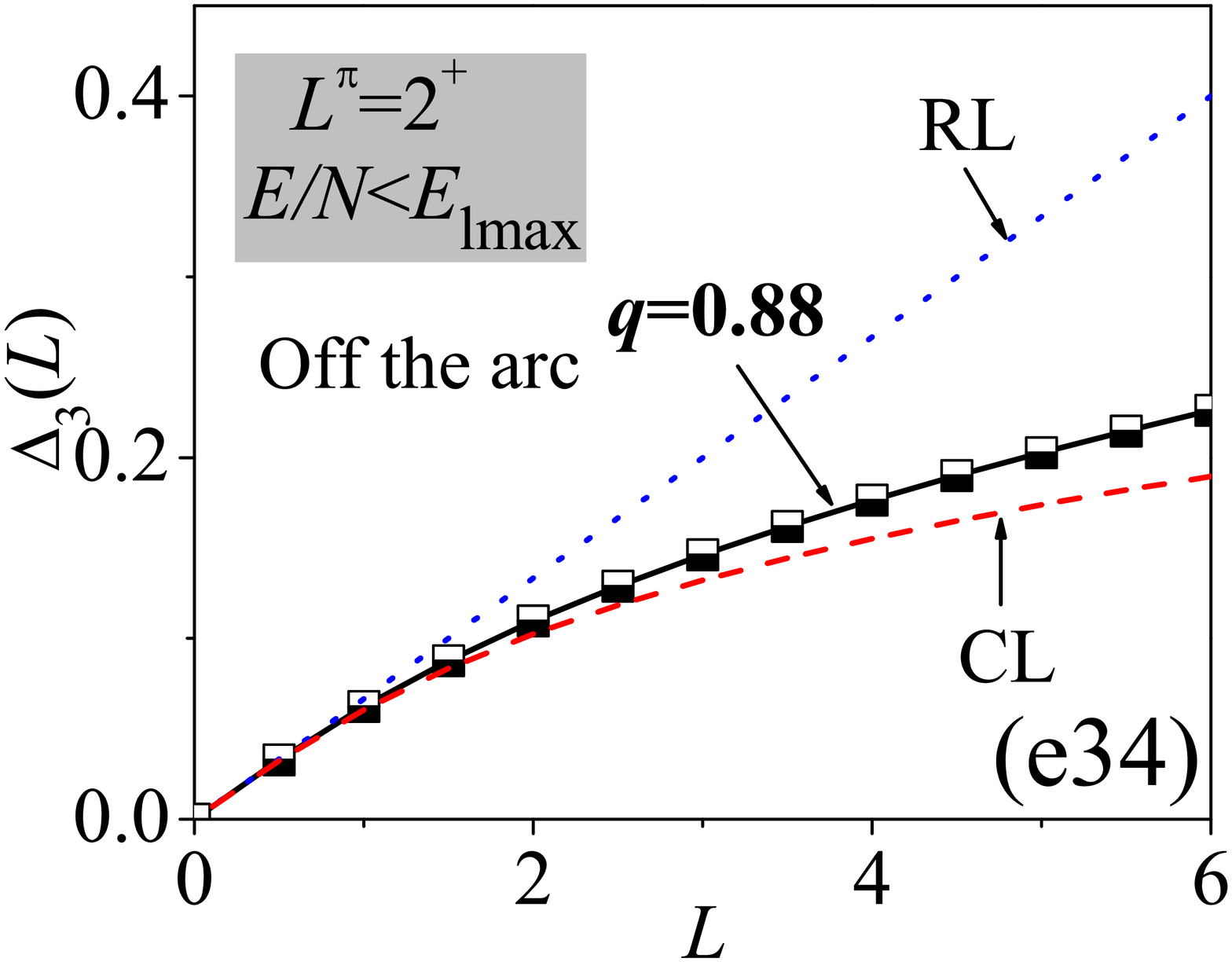}
\figcaption{The same as in Fig.~\ref{F2de1} but for the $2^+$
spectra.}\label{F2de3}
\end{center}

\begin{center}
\includegraphics[scale=0.165]{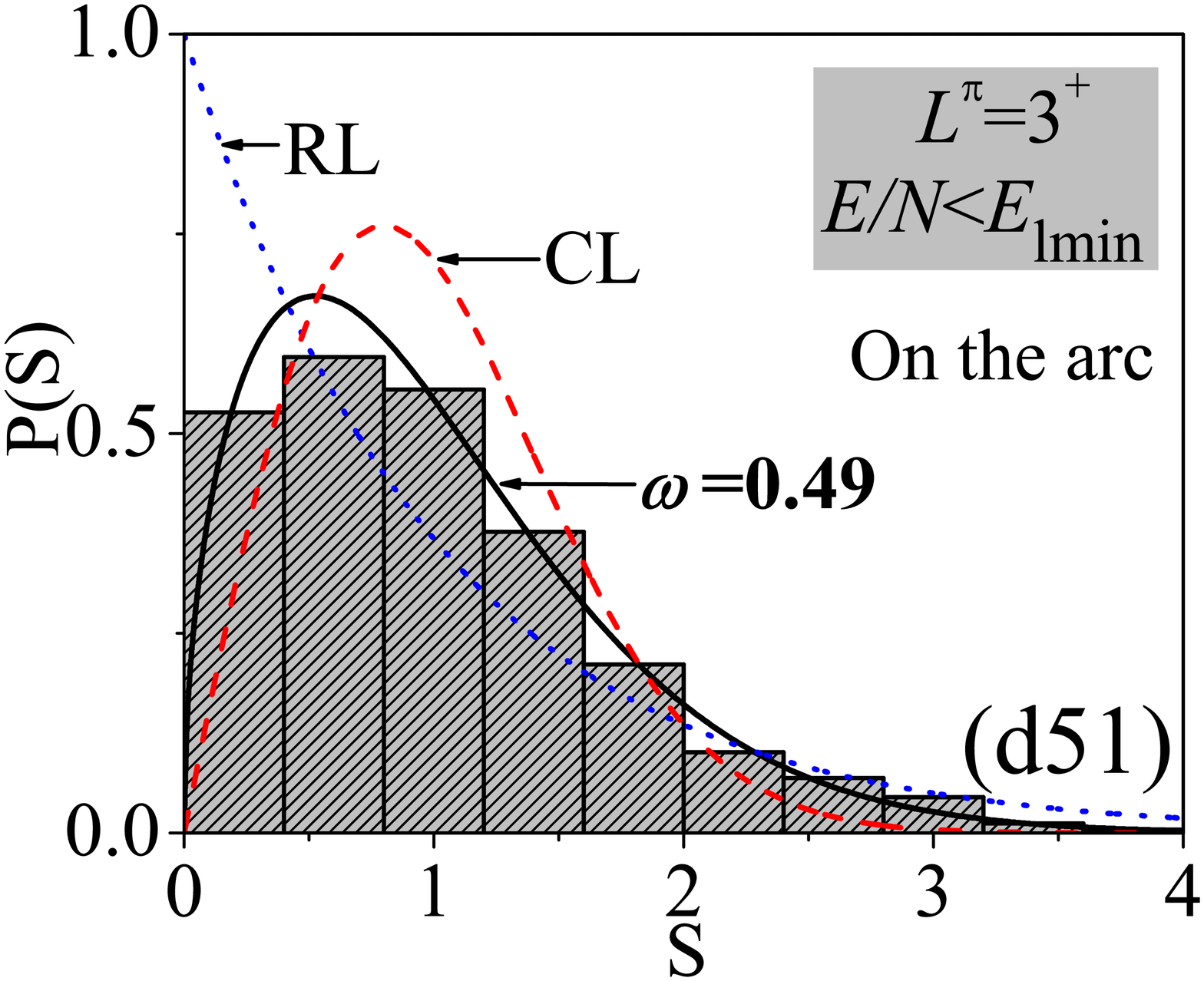}
\includegraphics[scale=0.165]{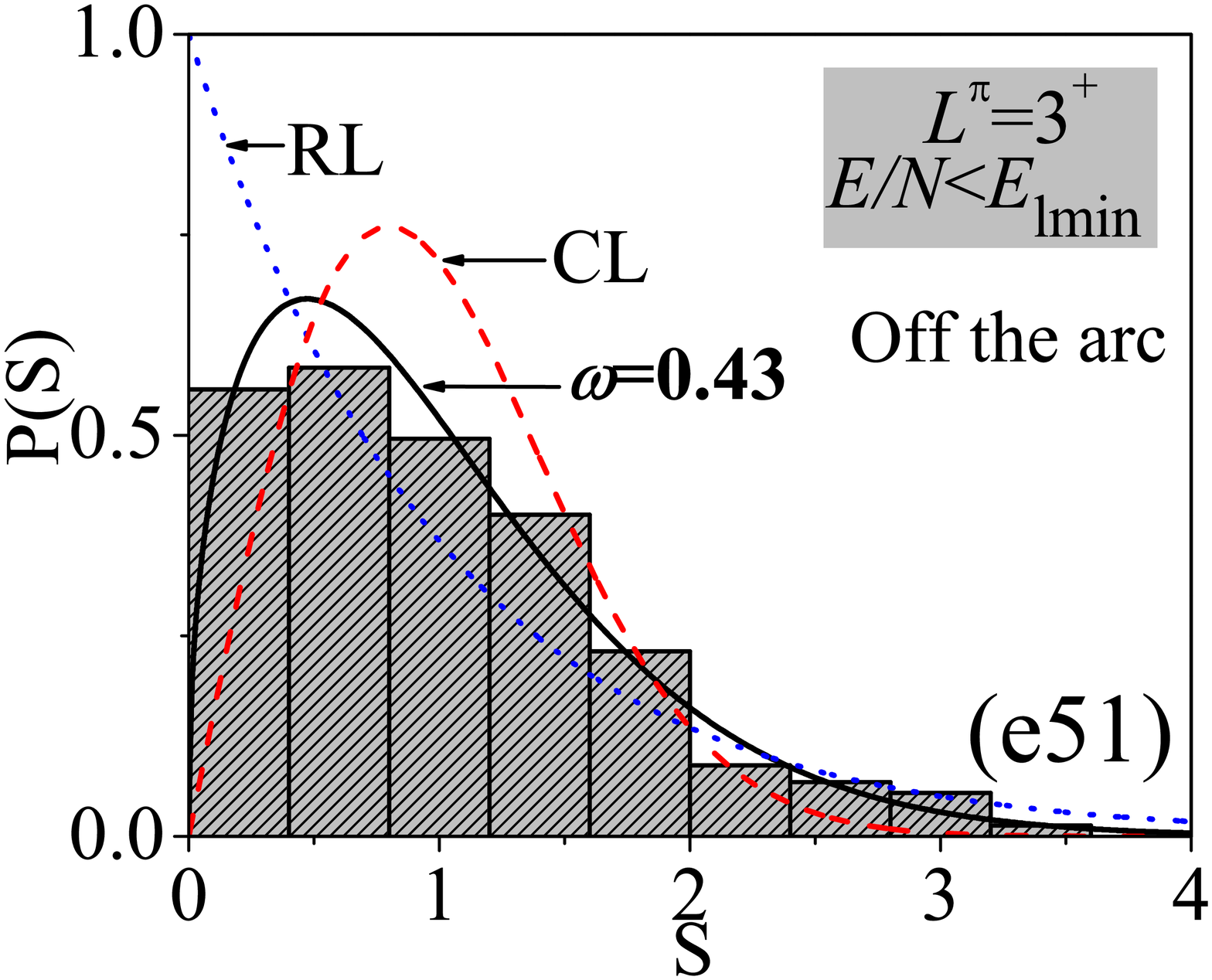}
\includegraphics[scale=0.165]{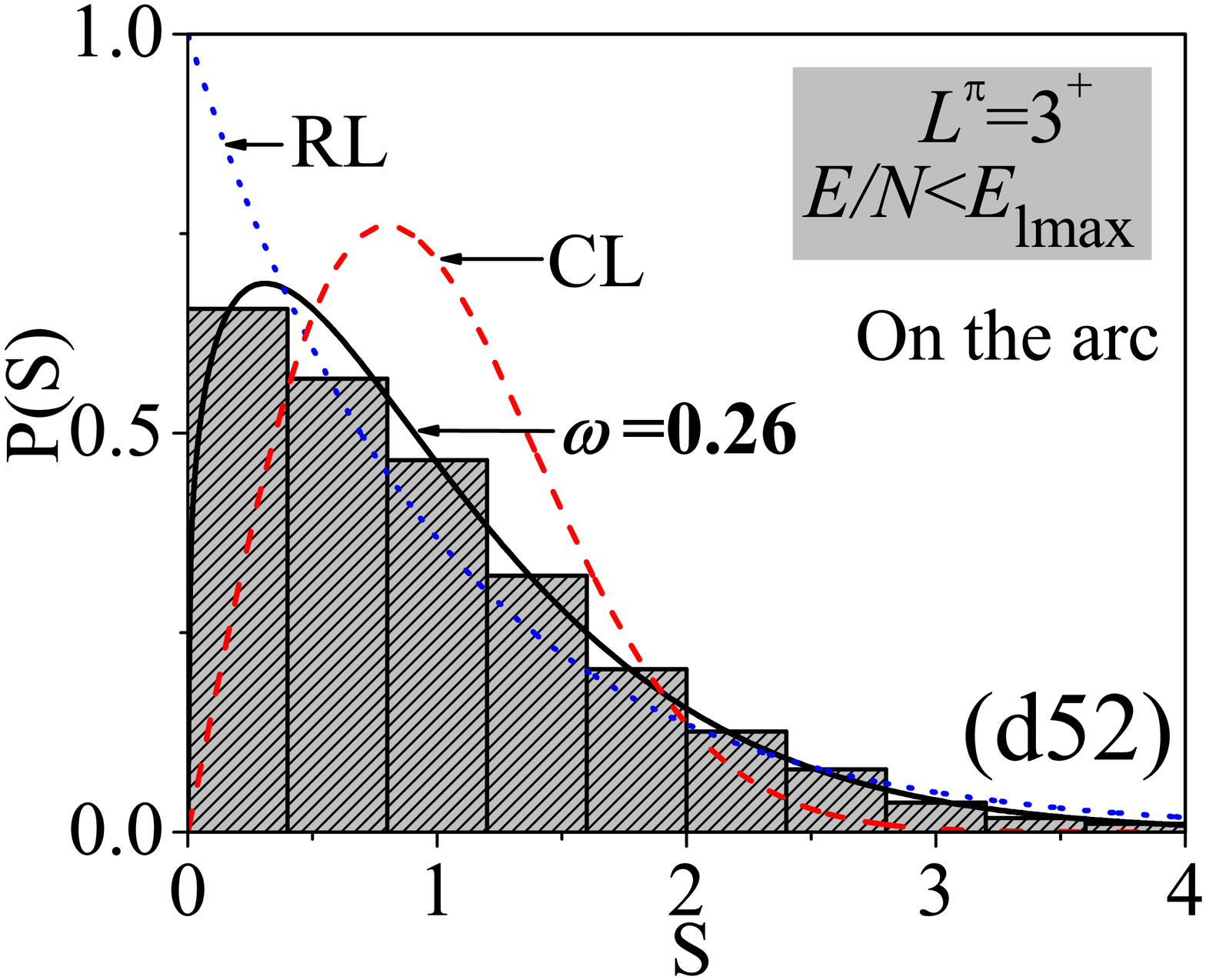}
\includegraphics[scale=0.165]{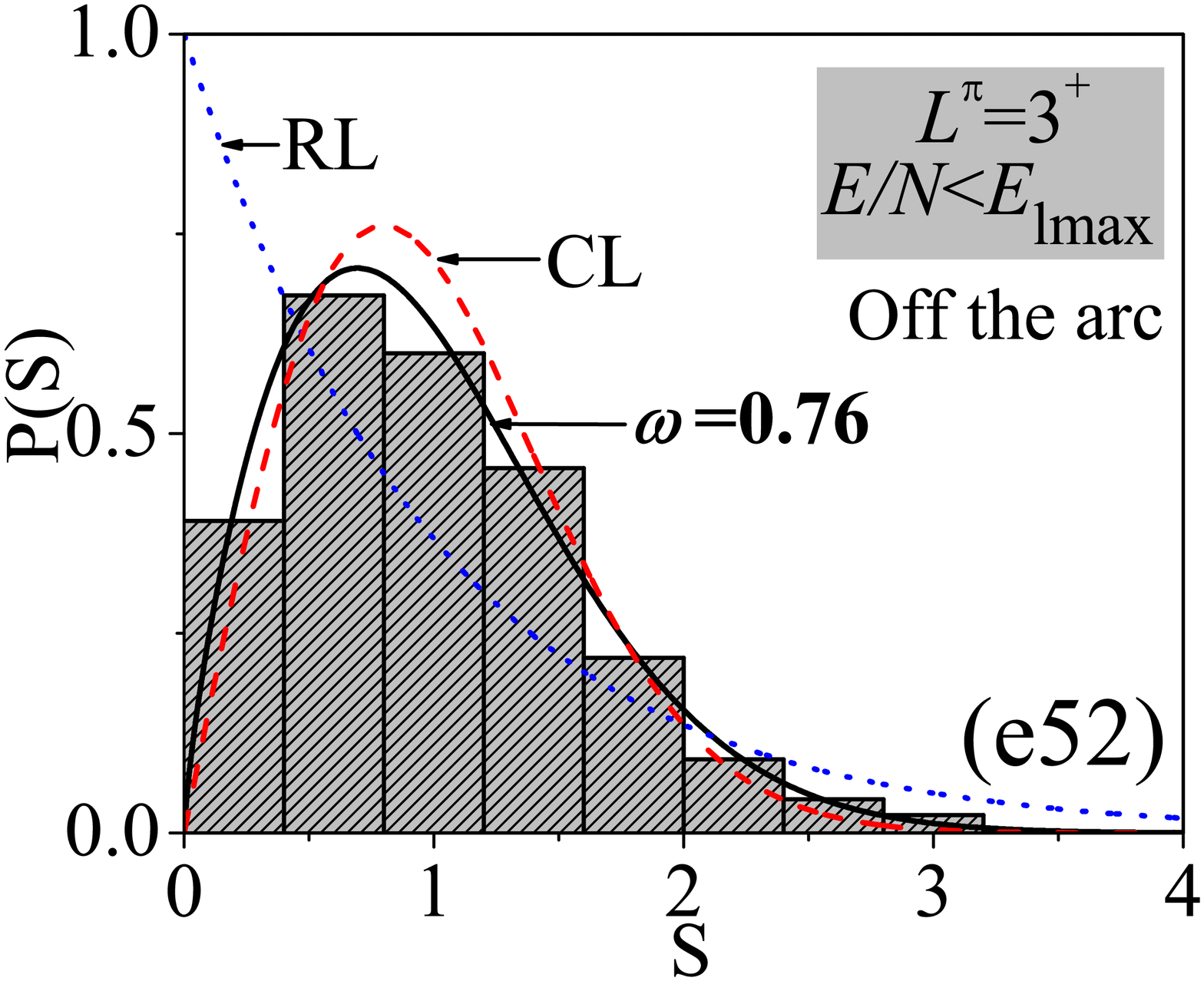}
\includegraphics[scale=0.165]{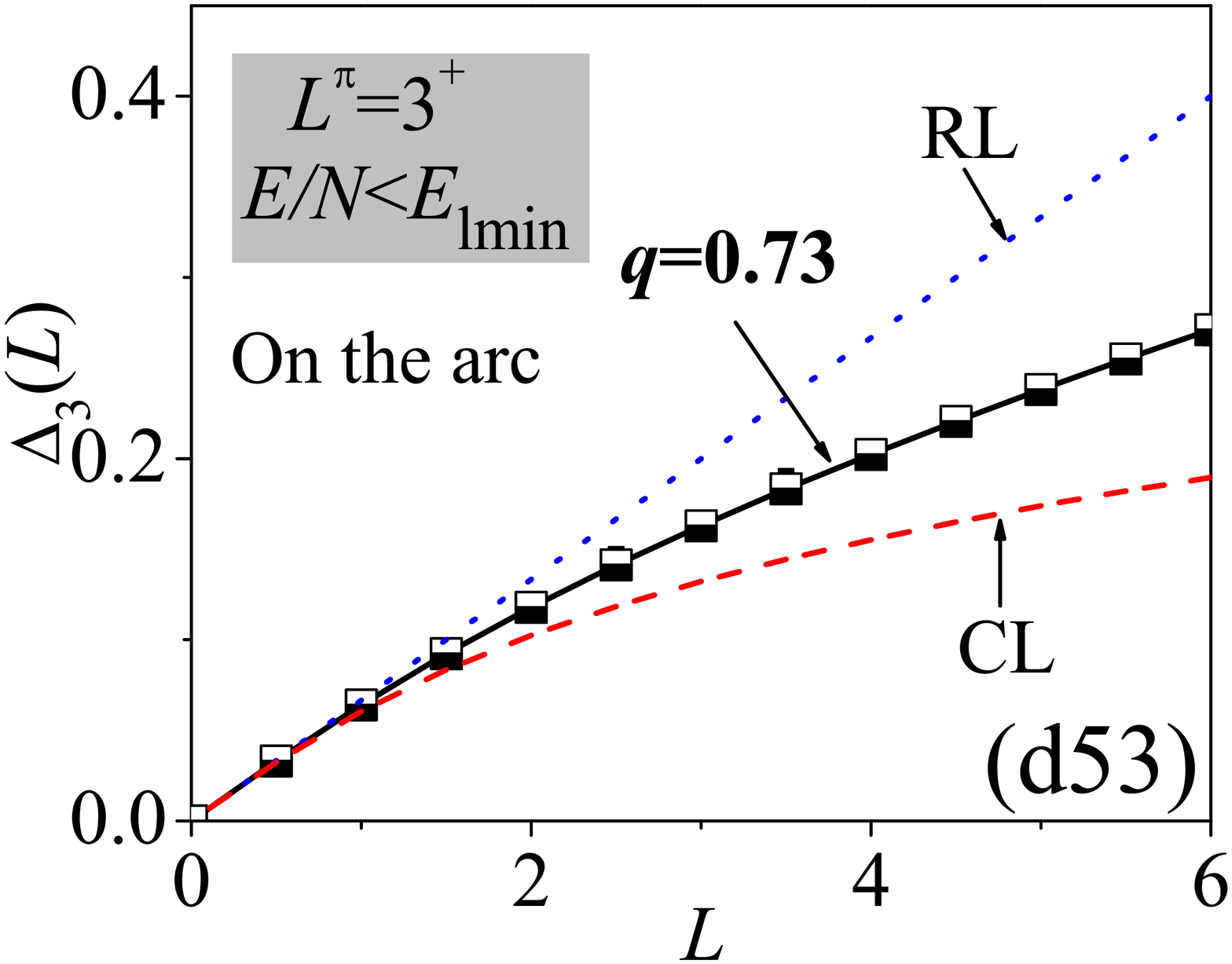}
\includegraphics[scale=0.165]{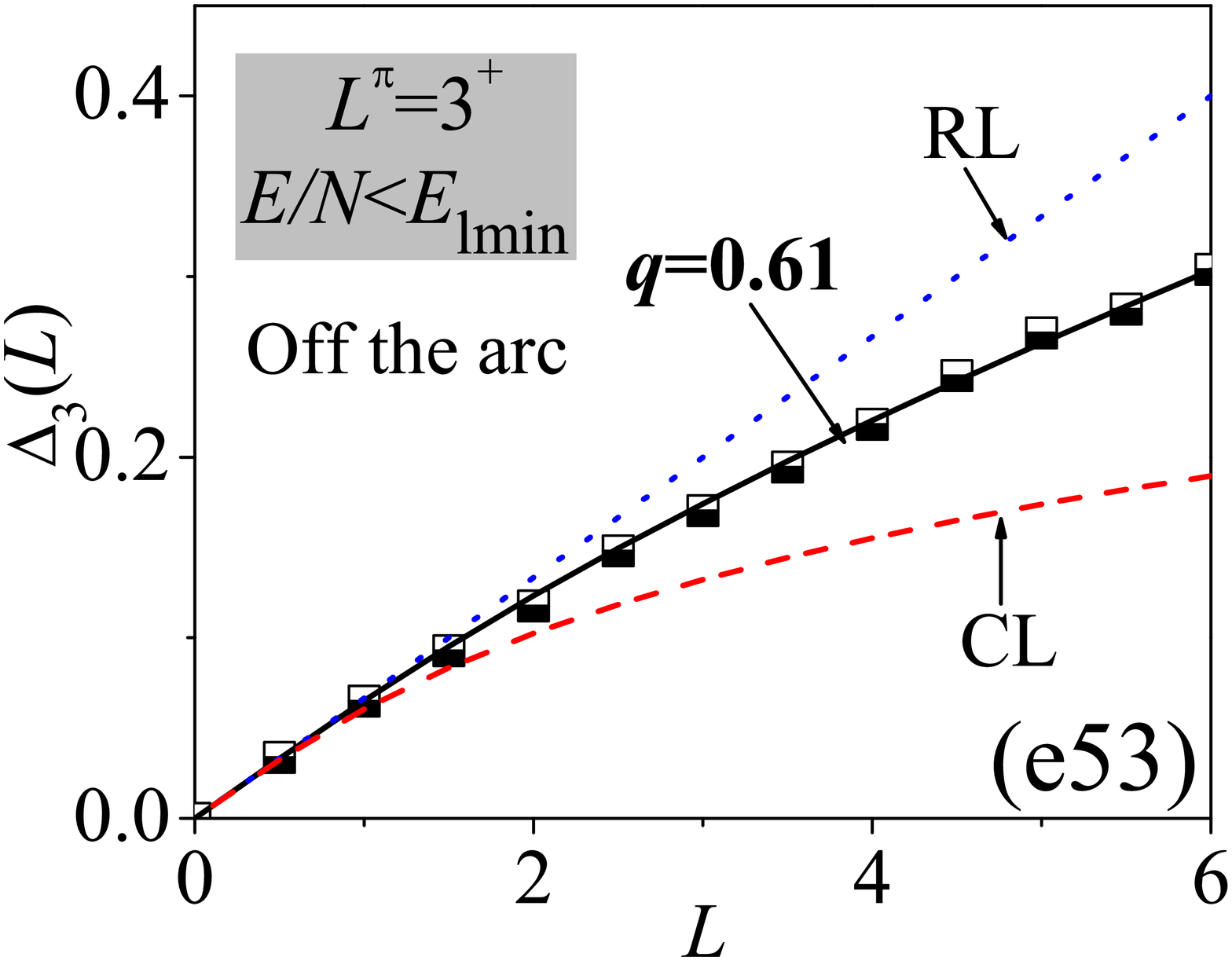}
\includegraphics[scale=0.165]{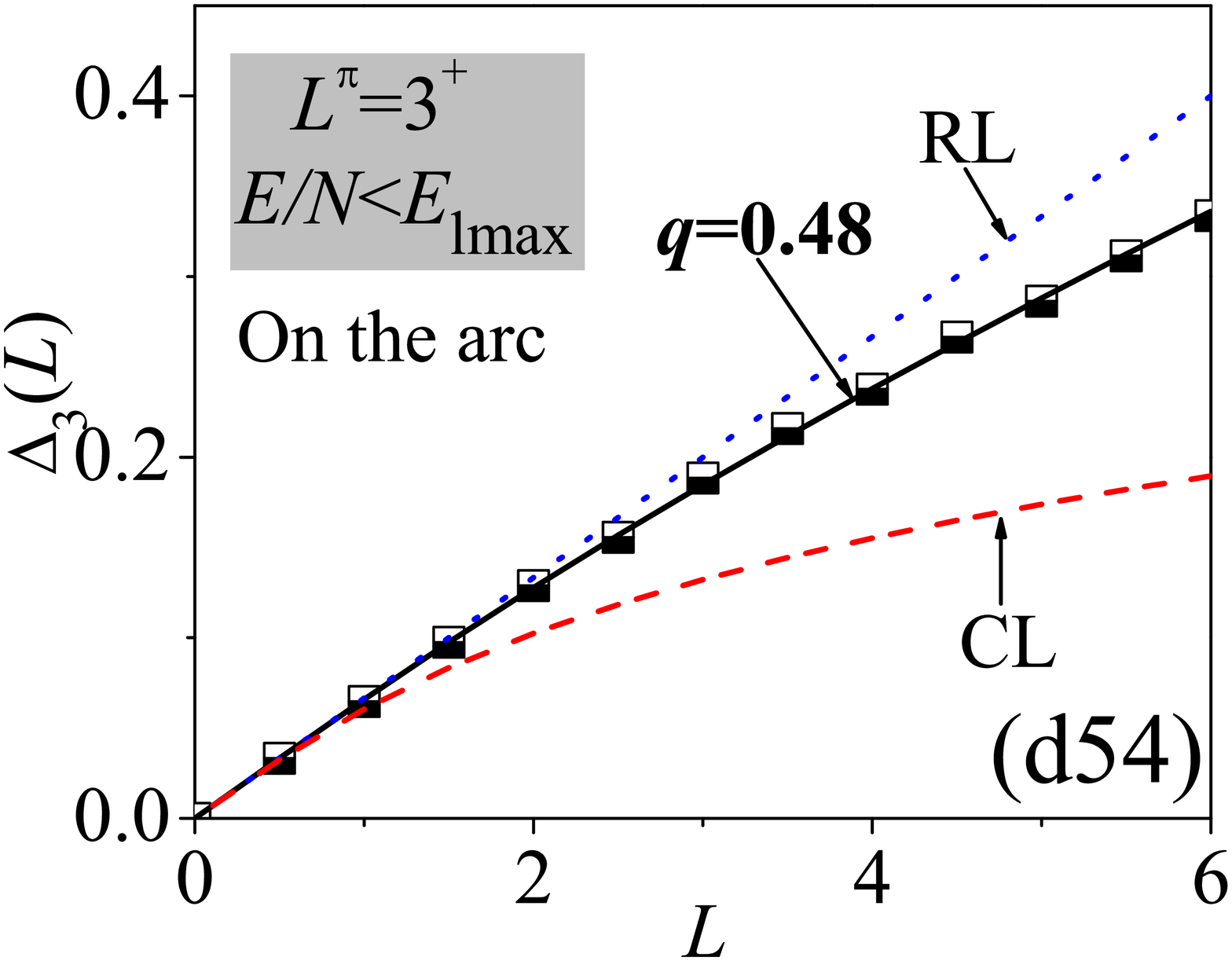}
\includegraphics[scale=0.165]{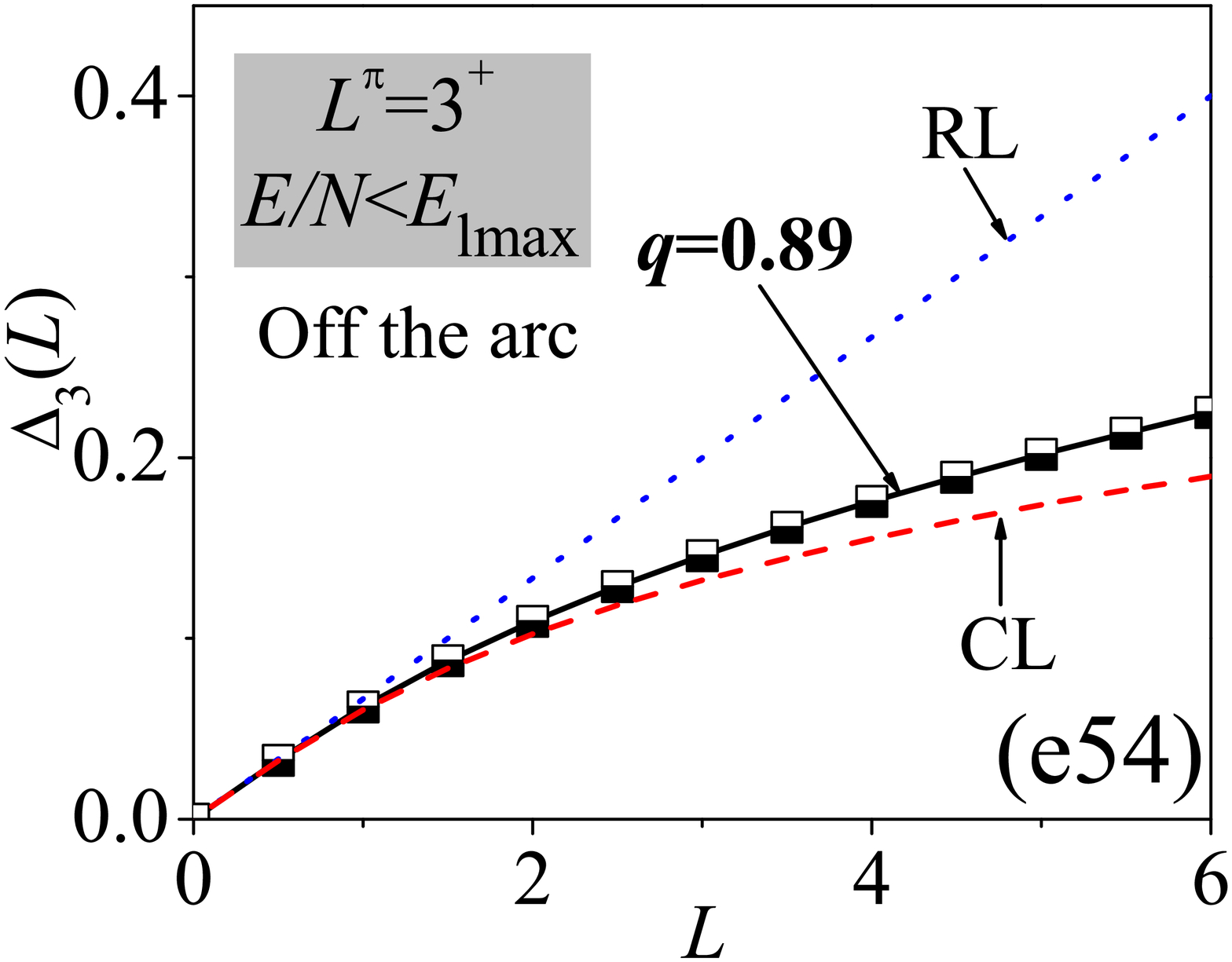}
\figcaption{The same as in Fig.~\ref{F2de1} but for the $3^+$
spectra.}\label{F2de5}
\end{center}

To make a closer comparison between the cases on and off the AW arc,
the concrete statistics on the levels below $E_\mathrm{lmin}$ and
$E_\mathrm{lmax}$ in the two cases are extracted out from
Fig.~\ref{F2d} and Fig.~\ref{F2e}. As shown in Fig.~\ref{F2de1}, the
results indicate that spectral fluctuations for $L^\pi=0^+$ below
$E_\mathrm{lmin}$ have no too much difference in between two cases
with $\Delta\omega=|\omega_1-\omega_2|=0.07$ and $\Delta
q=|q_1-q_2|=0.11$, where $\omega(q)_{1,2}$ represent the results in
the two cases. But the spectral fluctuations on the arc will
dramatically decrease if more states are involved in the statistics
by taking the energy cutoff $E_\mathrm{lmax}$. Instead, the ones off
the arc may become quite chaotic with $\omega=0.76$ and $q=0.88$.
Accordingly, the differences in between two cases come to the
greatest with $\Delta\omega=0.52$ and $\Delta q=0.42$, which are
even larger than the results involving all the $0^+$ levels into the
statistics, $\Delta\omega=0.48$ and $\Delta q=0.40$. This point also
agree with the classical analysis of the AW arc in \cite{Macek2007}
that the relative regularity of the arc is most significant just
around absolute energy $E_{\mathrm{cl}}=0$ corresponding to
$E_\mathrm{lmax}$. As seen from Fig.~\ref{F2de3}, the spectral
fluctuations for $L^\pi=2^+$ are very similar to those for
$L^\pi=0^+$ for the case off the arc with relatively small
$\omega(q)$ at the low energy cutoff and large $\omega(q)$ at the
high energy cutoff. In contrast, the $2^+$ spectrum on the AW arc
are shown to be over regular even at the high energy cutoff with a
negative $\omega$ value, which are actually consistent with those
shown in Fig.~\ref{F2d}(d3). As shown in Fig.~\ref{F2de5}, the
results for $L^\pi=3^+$ confirm that the regularity on the Arc will
be pronounced near $E_\mathrm{lmax}$, with the differences between
the two case reaching $\Delta\omega=0.50$ and $\Delta q=0.41$ at
this energy cutoff.

\begin{center}
\includegraphics[scale=0.25]{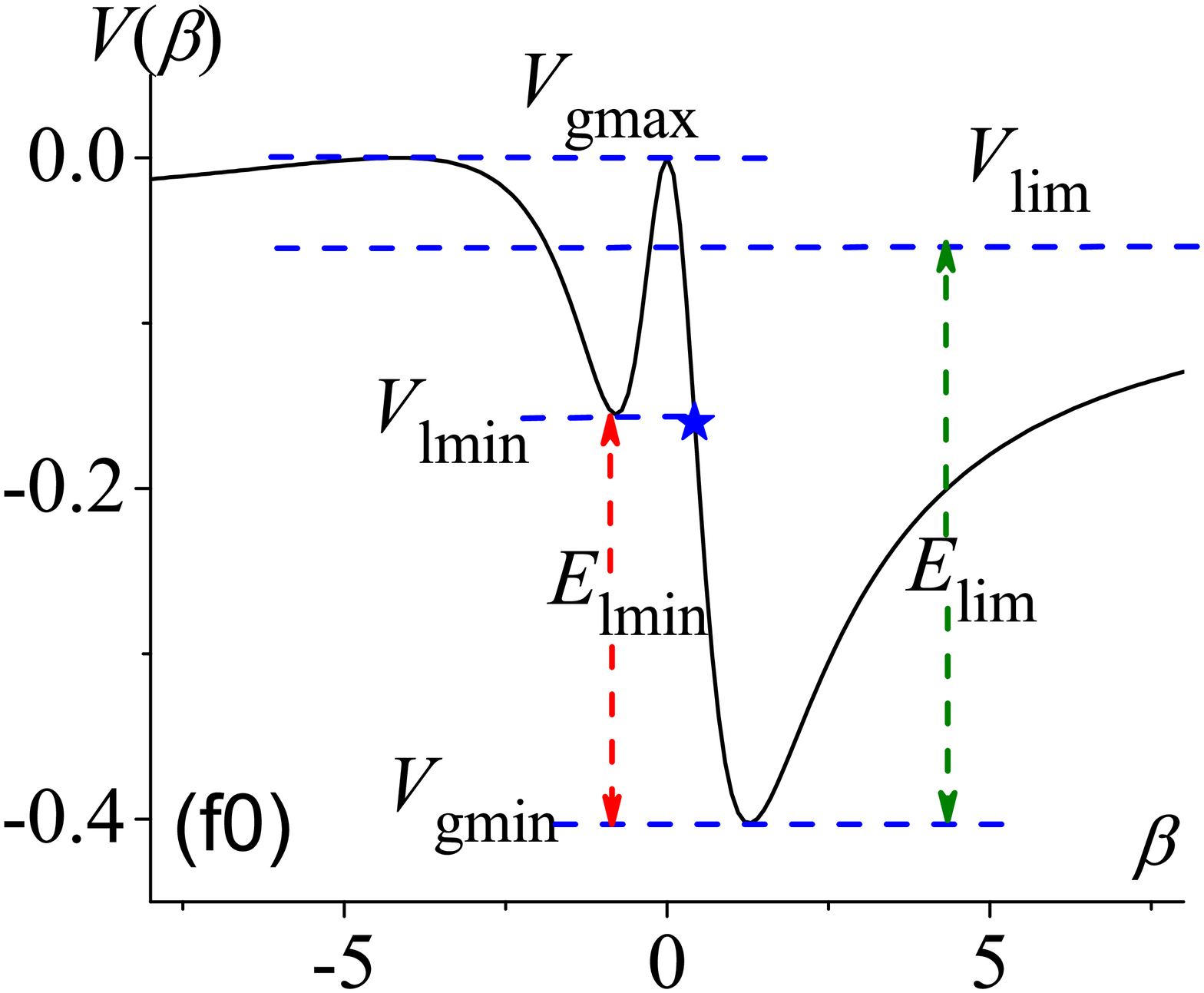}
\includegraphics[scale=0.16]{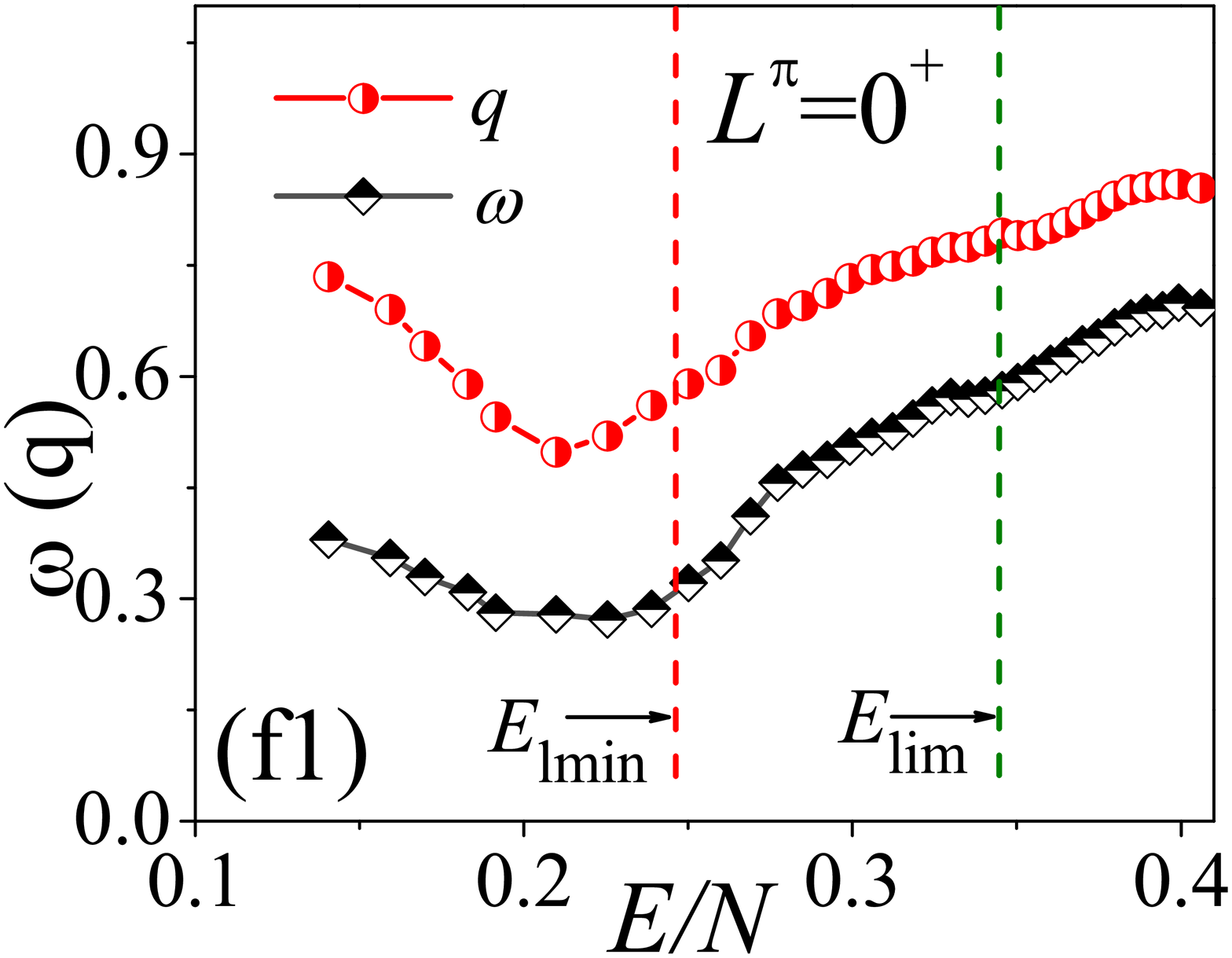}
\includegraphics[scale=0.16]{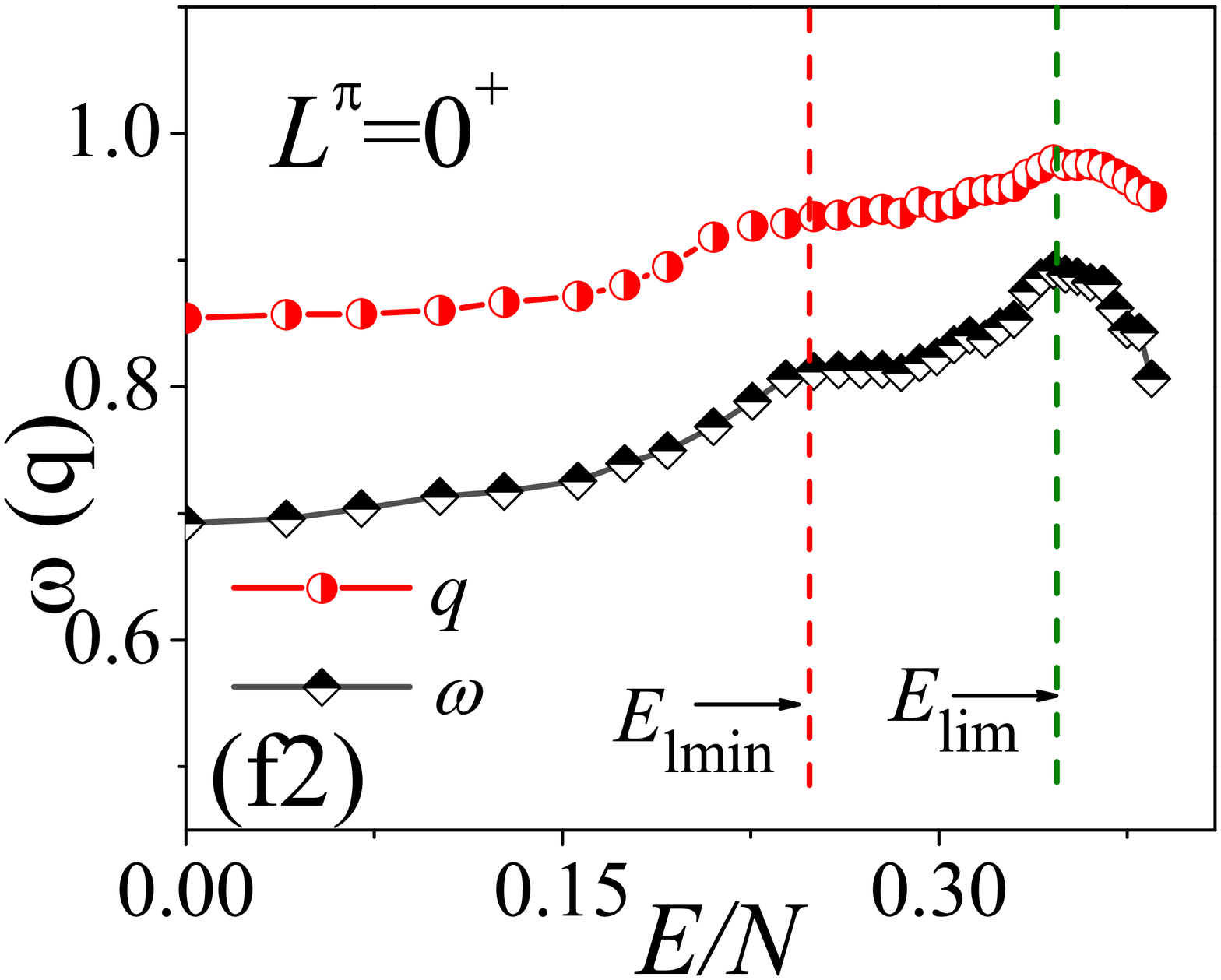}
\includegraphics[scale=0.16]{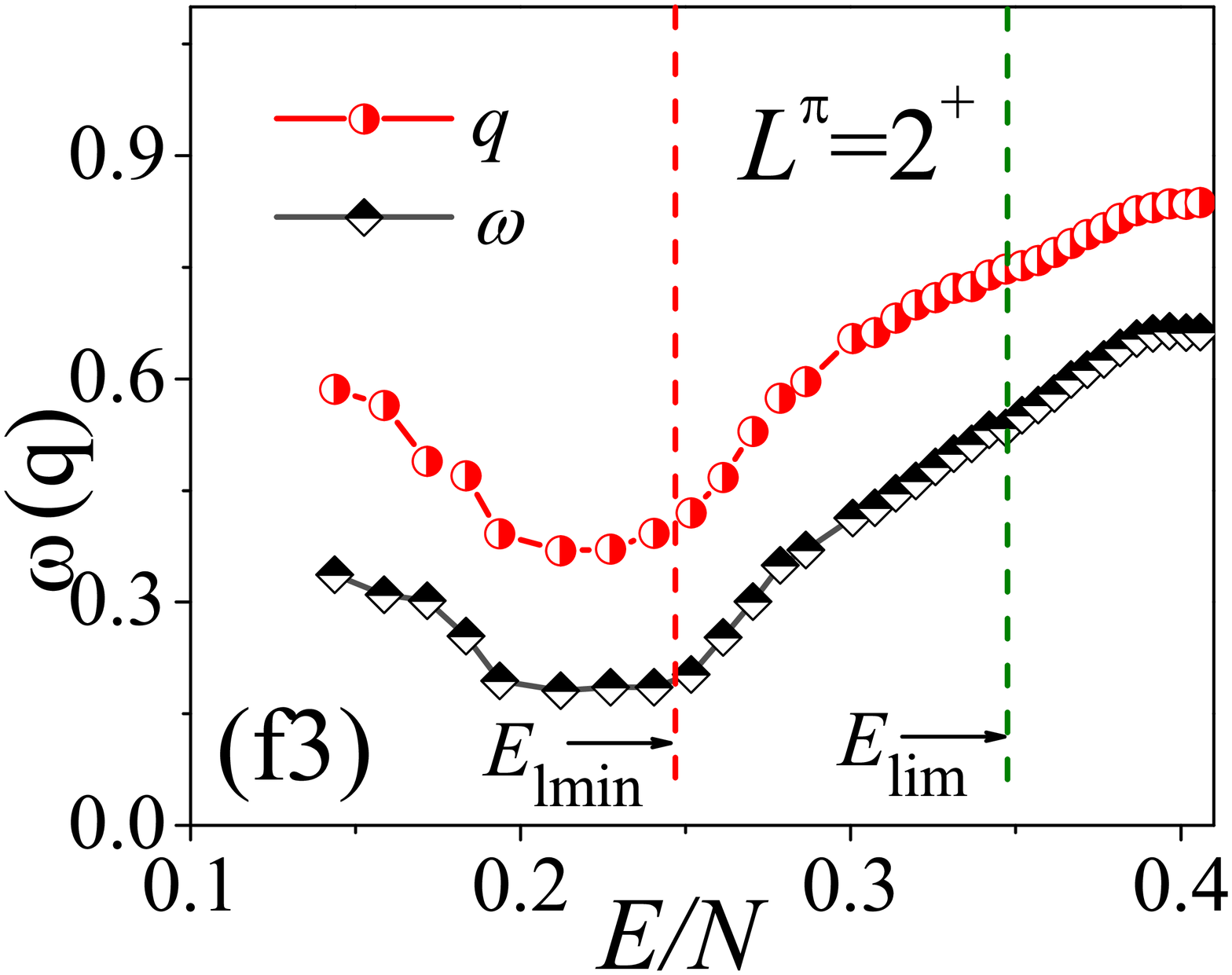}
\includegraphics[scale=0.16]{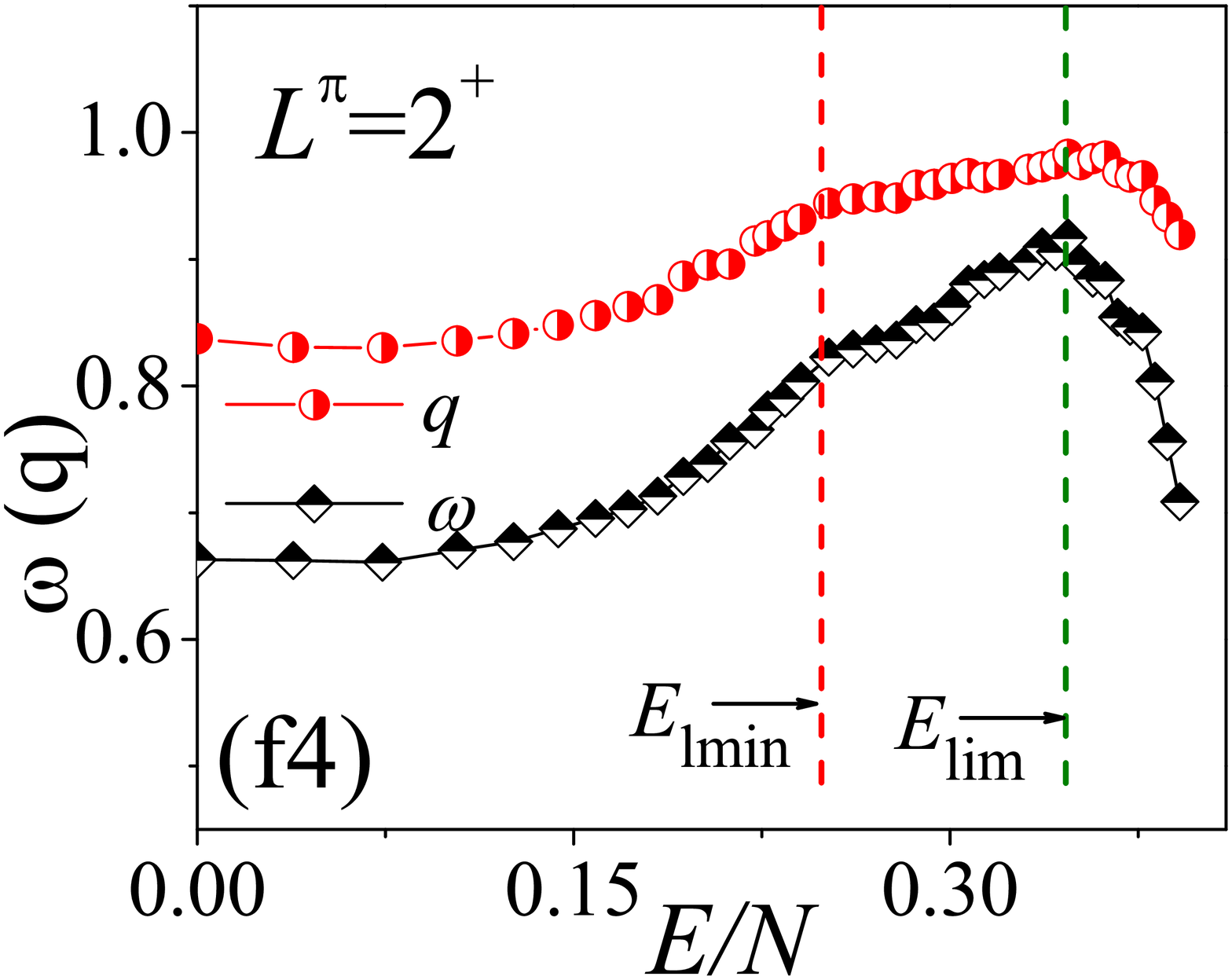}
\includegraphics[scale=0.16]{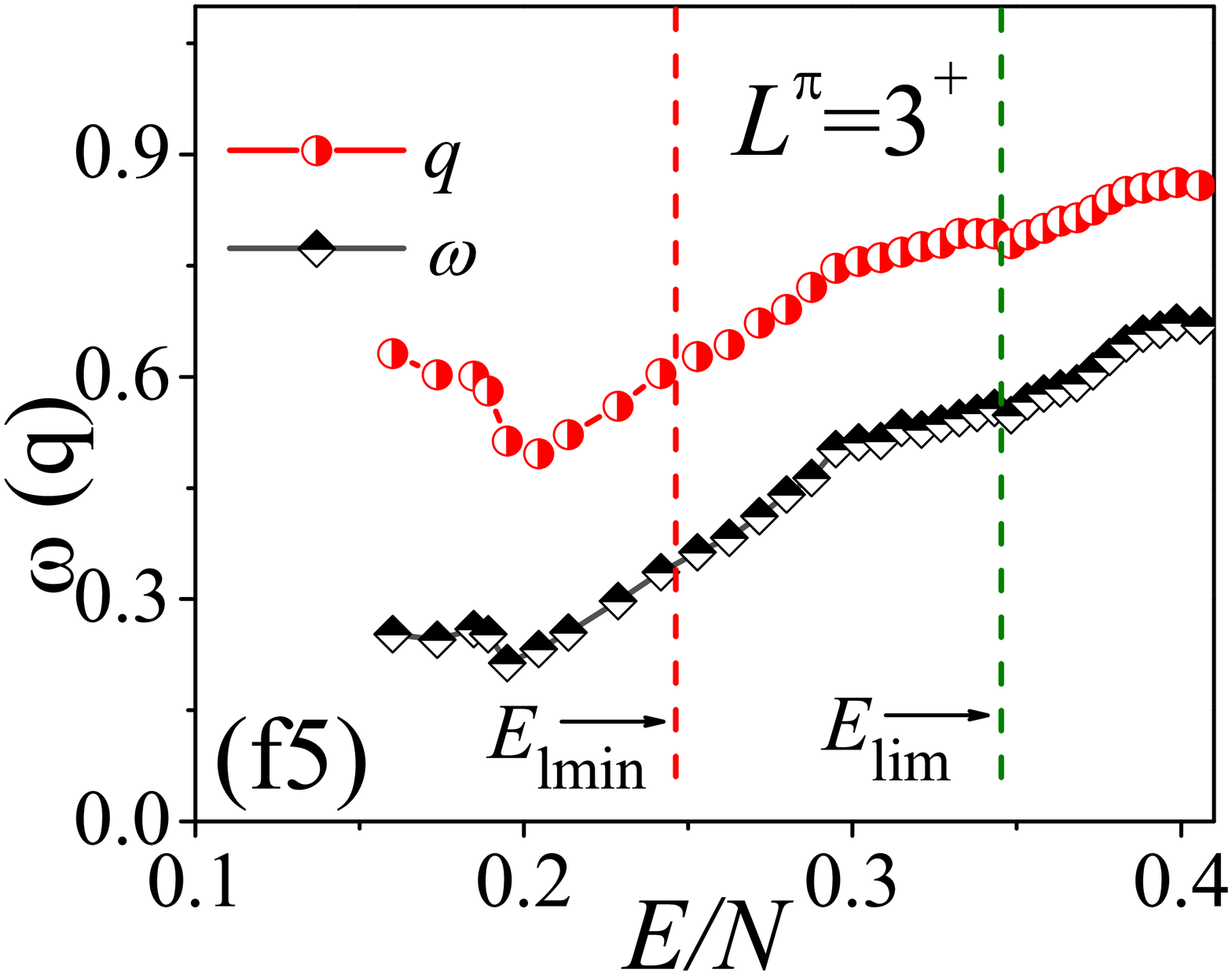}
\includegraphics[scale=0.16]{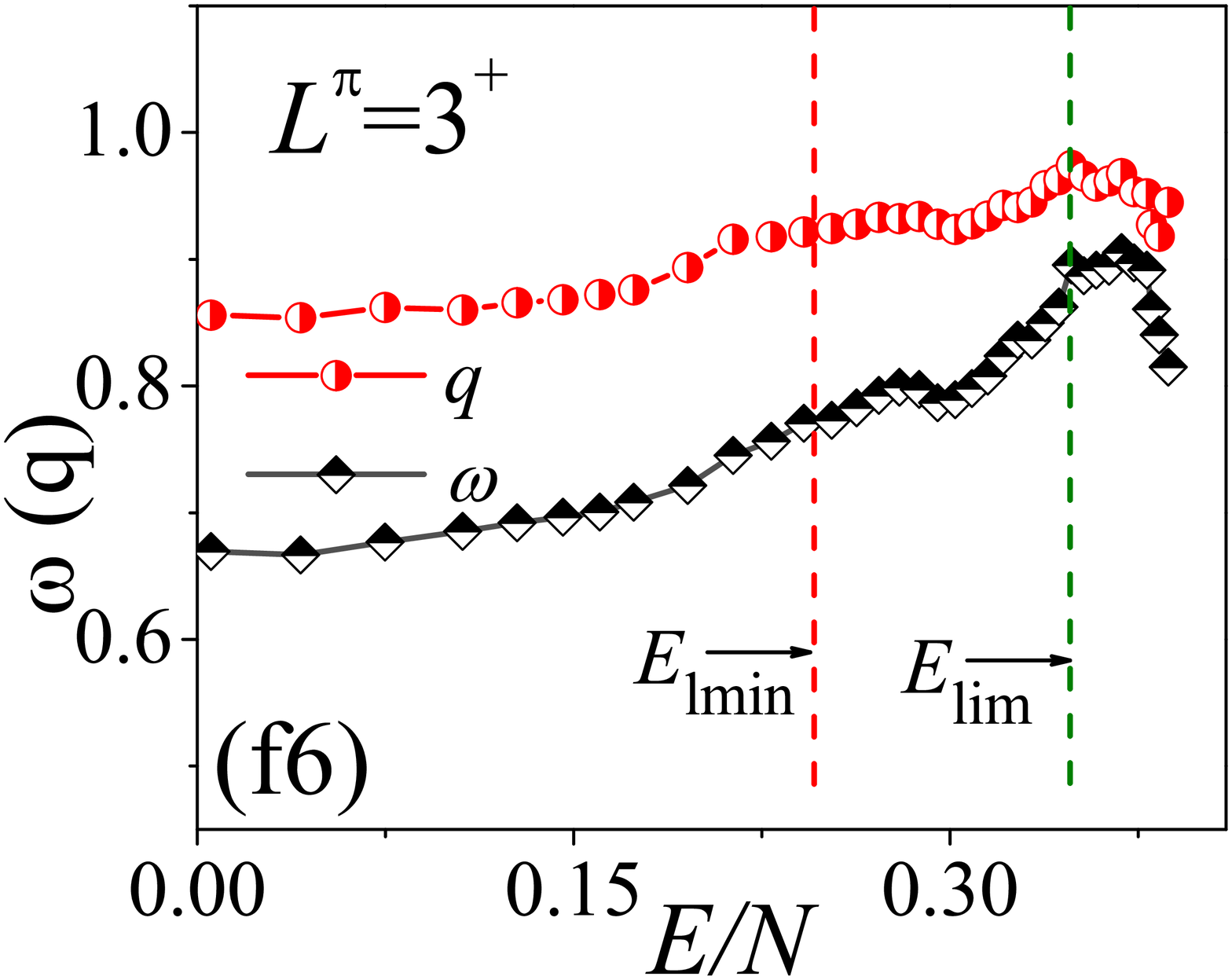}
\figcaption{The same as in Fig.~\ref{F2a} but for those
corresponding to the parameter point F.}\label{F2f}
\end{center}

\subsection{The SU(3)-O(6) transition}

The parameter point F represents a typical case in the SU(3)-O(6)
transition. As shown in the panel (f0), the potential curve at this
point has only two stationary points lying in between
$V_\mathrm{gmin}$ and $V_\mathrm{gmax}$ as its local maximal point
at $\beta=0$ may coincide with its global maximum. The spectra for
$L^\pi=0^+,~2^+,~3^+$ are all shown to be relatively chaotic with
$\omega>0.6$ and $q>0.8$ if involving all the levels into the
statistics for given spin. Like in the other cases, the spectral
fluctuations in this case are not uniform in energy and the
energy-dependence can be partially illustrated based on its
mean-field structure. For example, the $\omega(q)$ values as
functions of the excitation energy may reach their maxima near
$E_\mathrm{lim}$ as shown in the panels (f2), (f4) and (f6).
Influence of $E_\mathrm{lmin}$ on spectral fluctuations can be also
observed by seeing the panel (f3), where the results imply that the
spectral chaos will rapidly increase after this energy point. All
these just exhibit the effects of the stationary points on the
spectral fluctuations in this case.

\section{Summary}

In summary, the energy dependence of the spectral fluctuations in
the IBM and its connections to the mean-field structures have been
investigated for the cases across the U(5)-SU(3) GSQPT, near the AW
arc and in the SU(3)-O(6) transition. Two statistical measures, the
nearest neighbor level spacing distribution $P(S)$ and the
$\Delta_3$ statistics of Dyson and Mehta, are applied to inspect the
spectral fluctuations in each case. It is found that the spectral
fluctuations as a function of the energy cutoff may exhibit
different evolutional behaviors in different cases but their
behaviors are all shown to be closely related to the corresponding
mean-field structures. Specifically, most of the sudden changes in
the fluctuational evolutions can be attributed to the effects of the
stationary points particularly for the cases in the deformed area of
the phase diagram, where the ESQPT phenomena can be observed around
the same stationary points~\cite{Zhang2016}. These findings confirm
again the role of the mean-field structure in understanding excited
state properties. Another new finding is the appearance of negative
$\omega$ values indicating the approximate degeneracies in the $2^+$
spectrum on the AW arc. This may provide a statistical signature of
the approximate SU(3) symmetry along the AW
arc~\cite{Bonatsos2010,Bonatsos2011}, which, however, needs to be
further proved. The present large-$N$ analysis not only add new
information to the chaotic map of the IBM but also offer a reference
for the study of spectral fluctuations in the Bohr-Mottelson model
since the model may directly link the large-$N$ limit of the
IBM~\cite{IachelloBook87}. The discussions on spectral fluctuations
can be extended to discuss the fluctuations in other quantities such
as the $B(E\lambda)$ transitions~\cite{Karampagia2015} or to other
algebraic models including those for two-fluid
systems~\cite{Ramos2017,Bernal2008,Caprio2005,Caprio2011}, where the
stationary point and phase structures could be much richer than the
present case~\cite{Ramos2017,Caprio2005}. In addition, classical
measures~\cite{Alhassid1991I} can be also applied to reveal the
energy dependence of spectral fluctuations in the IBM. The related
work is in progress.

\end{multicols}

\begin{multicols}{2}

\end{multicols}

\clearpage

\end{CJK*}
\end{document}